\documentclass[letterpaper]{article}
\usepackage[margin=1.in]{geometry}
\usepackage{graphicx}
\usepackage{authblk}
\usepackage{amssymb}
\usepackage{amsthm}

\usepackage{color}
\usepackage{epstopdf}

\usepackage{subfigure}

\usepackage[acronym]{glossaries}

\newacronym{fsi}{FSI}{fluid-structure interaction}
\newacronym{dof}{DoF}{degrees of freedom}
\newacronym{eom}{EoM}{equations of motion}
\newacronym{curvib}{CURVIB}{curvilinear immersed boundary}
\newacronym{ib}{IB}{immersed boundary}
\newacronym{2d}{2D}{two-dimensional}
\newacronym{3d}{3D}{three-dimensional}
\newacronym{rao}{RAO}{response-amplitude operator}
\newacronym{viv}{VIV}{vortex induced vibration}
\newacronym{les}{LES}{large-eddy simulation}
\newacronym{tsr}{TSR}{tip-speed ratio}
\newacronym{bcs}{BCs}{boundary conditions}
\newacronym{rk4}{RK4}{fourth-order Runge-Kutta}
\newacronym{rk2}{RK2}{second-order Runge-Kutta}
\newacronym{dns}{DNS}{direct numerical simulation}
\newacronym{sgs}{SGS}{subgrid-scale}
\newacronym{hos}{HOS}{high-order spectral}
\newacronym{mpi}{MPI}{message-passing interface}
\newacronym{eno}{ENO}{essentially non-oscillatory}
\newacronym{weno}{WENO}{weighted essentially non-oscillatory}
\newacronym{pnw}{PNW}{Pacific Northwest}
\newacronym{jonswap}{JONSWAP}{Joint North Sea Wave Observation Project}
\newacronym{bem}{BEM}{boundary element method}
\newacronym{rans}{RANS}{Reynolds-averaged Navier-Stokes}
\newacronym{wec}{WEC}{wave energy converter}
\newacronym{safl}{SAFL}{Saint Anthony Falls Laboratory}
\newacronym{jpd}{JPD}{joint probability distribution}
\newacronym{msl}{MSL}{mean sea level}
\makeglossaries

\begin{document}

\title{Fluid-structure interaction simulation of floating structures interacting  with complex, large-scale ocean waves and atmospheric turbulence}

\author[1]{Antoni Calderer}
\author[1]{Xin Guo}
\author[1,2]{Lian Shen}
\author[3,*]{Fotis Sotiropoulos}

\affil[1]{Saint Anthony Falls Laboratory, University of Minnesota, 2 Third Avenue SE, Minneapolis, MN 55414, United States}
\affil[2]{Department of Mechanical Engineering, University of Minnesota, Minneapolis, MN 55455, USA}
\affil[3]{Department of Civil Engineering, College of Engineering and Applied Sciences, Stony Brook University, Stony Brook, NY}
\affil[*]{Corresponding author, email address: fotis.sotiropoulos@stonybrook.edu}
\setcounter{Maxaffil}{0}
\renewcommand\Affilfont{\itshape\small}


\maketitle

\begin{abstract}
We develop a numerical method for simulating coupled interactions of complex floating structures with large-scale ocean waves and atmospheric turbulence.  We employ an efficient large-scale model to develop offshore wind and wave environmental conditions, which are then incorporated into a high resolution two-phase flow solver with fluid-structure interaction (FSI). 
The large-scale wind-wave interaction model is based on the two-fluid dynamically-coupled approach of Yang and Shen (2011) \cite{yang_simulation_2011,yang_simulation_2011_1}, which employs a high-order spectral method for simulating the water motion and a viscous solver with undulatory boundaries for the air motion. The two-phase flow FSI solver, developed by Calderer, Kang, and Sotiropoulos (2014) \cite{calderer_fsi_2014}, is based on the level set method and is capable of simulating the coupled dynamic interaction of arbitrarily complex bodies with airflow and waves. The large-scale wave field solver is coupled with the near-field FSI solver by feeding into the latter waves via the pressure-forcing method of Guo and Shen (2009) \cite{guo_generation_2009}, which has been extended herein for the level set method. We validate the model for both simple wave trains and three-dimensional directional waves and compare the results with experimental and theoretical solutions. Finally, we demonstrate the capabilities of the new computational framework by carrying out large-eddy simulation of a floating offshore wind turbine interacting with realistic ocean wind and waves.
\end{abstract}

\section{Introduction}\label{sec:Introduction}
The potential seen in the ocean as an enormous supply of clean energy resource has motivated an increased attention in the scientific community towards \gls{fsi} problems involving waves and complex floating structures, such as \gls{wec} devices and offshore wind turbines, as further extension of previous studies on ship hydrodynamics and floating platforms in the petroleum industry. 
Such types of problems have generally been studied using simplified models assuming inviscid and irrotational flows as in \cite{kuhn_dynamics_2001,cheng_reliability_2002,veldkamp_influence_2005,jonkman_dynamics_2009,Gueydon2011Floating} dealing with floating wind turbines, or in \cite{de2010power, taghipour2008efficient, babarit2012numerical} applied to the study of \gls{wec}s. Potential flow models are accurate for simulating  problems with low amplitude motions and simplified non-breaking waves. However, offshore environments are often subject to more complex non-linear wave phenomena such as turbulence, wave-turbulence interaction, and wave overturning and breaking. Such cases in which viscosity plays an important role can be accurately represented by a Navier-Stokes solver in combination with turbulence models.

One of the main challenges in applying Navier-Stokes solvers to ocean wave problems is to deal with processes occurring at a disparate range of scales.  The study of the interaction of a floating structure with swells is an example. While swells interact with the wind flow and evolve for long distances requiring a large computational domain, the floating structure highly depends on flow motions at much smaller scales in the vicinity of the structure and requires the use of very fine meshes. One can reduce the high computational cost of viscous solvers and deal with multi-scale problems more efficiently by using a domain decomposition approach, limiting the application of the viscous solver to regions where the complexity of the flow requires it, and take advantage of other methods, such as the potential flow theory for treating the wave motion in the regions far away. 
Iafrati and Campana \cite{iafrati2003domain} used a domain decomposition approach to simulate \gls{2d} breaking water waves. In their work, a two phase Navier-Stokes solver based on the level set method was applied in the upper part of the domain, containing the free surface, where wave breaking occurs and viscous effects are most important, and an inviscid flow model was used in the lower part of the computational domain far from the free surface. 
In a later work, Colicchio et al. \cite{colicchio2006bem} developed another domain decomposition approach coupling also a level set based, two-phase, viscous flow solver, applied in a region where complex processes occur, with a potential flow solver applied in the regions with mild flow conditions. 
Both the methods of Iagrati et al. \cite{iafrati2003domain} and Colicchio et al. \cite{colicchio2006bem} are formulated in a \gls{3d} context but have only been applied to \gls{2d} problems.

A key aspect for developing such multi-scale methods is to choose an appropriate technique for transferring the far-field flow solution as input to the near-field solver. In particular, a major challenge in this regard is the approach for prescribing a specific large-scale wave environment as input into the \gls{3d} Navier-Stokes flow solver. The simplest and most obvious way is by directly specifying at the inlet boundary the velocity profile and surface elevation. For example, in the work of Colicchio et al.\ \cite{colicchio2006bem}, an algorithm for coupling a potential flow based \gls{bem} solver with a Navier-Stokes level set solver is presented. In Repalle et al.\ \cite{repalle_cfd_2007}, theoretical wave velocities are fed into a \gls{rans} model to simulate the wave run-up on a spar cylinder. In Xie et al.\ \cite{xie2016simulation} waves from a potential flow solver are directly prescribed as boundary conditions for a Navier-Stokes solver. In Christensen \cite{christensen2001sediment}, waves from a Boussinesq based model are incorporated to a Navier-Stokes solver. Generation of waves by specifying inlet boundary conditions can be problematic when strong reflected waves reach the inlet boundary. For example, Wei and Kirby \cite{wei_time_1995} demonstrated that, even if a generating-absorbing boundary condition is employed, large errors can accumulate and lead to inaccurate solutions after computation for long duration. 

An approach that can avoid the aforementioned difficulties when dealing with wave reflections is to employ an internal wave maker in combination with the use of sponge layers at the boundaries. The basic idea of internal wave generators is to apply an oscillatory force within an internal region of the domain, known as the source region. The force is introduced by adding a source/sink term either in the continuity or momentum equations. The internal wave generation method based on a mass source/sink was proposed by Lin and Liu \cite{lin_internal_1999}. They derived source function expressions based on the fact that the increase/decrease of mass in the source region contributes to the target wave generation. Given a submerged rectangular source region, their method was used to obtain expressions for the following wave cases: linear monochromatic waves, irregular waves, Stokes waves, cnoidal waves, and solitary waves. They demonstrated the accuracy of the method by comparing the results to analytical solutions. They also showed that the internal wave generator was not affected by the presence of reflected waves. Although this method has been widely used by many authors \cite{garcia_2d_2004,lara_rans_2006,lin_numerical_2007} for generating \gls{2d} waves, it has not been extended to the generation of \gls{3d} directional waves. A step further in the development of internal wave makers is to implement the source terms, not in the continuity equation but in the momentum equations. It is not obvious, however, how to derive a forcing term expression that can generate a free surface wave pattern with the specific target amplitude, because the free surface elevation is not a variable in the momentum equation. Such relation, however, can be directly established in exact form in the depth-integrated Boussinesq equations as shown by Wei et al. \cite{wei_generation_1999}. In their work \cite{wei_generation_1999}, source terms for the momentum equations were proposed for generating regular and irregular waves. The idea of Wei et al. \cite{wei_generation_1999} was later implemented by Choi and Yoon \cite{choi_numerical_2009} in a RANS turbulence model and by Ha et al. \cite{ha_internal_2011} in a \gls{les} model. In particular, the capabilities of the method to generate directional waves in a \gls{3d} basin were successfully demonstrated. All the above internal wave makers employ a fixed rectangular domain as source region that is located under the free surface. 
An alternative momentum source method is that of Guo and Shen \cite{guo_generation_2009}, in which the source region is not fixed but follows the motion of the free surface. 
This is equivalent of applying a surface pressure on the free surface, similar to the physical process of wave generation by wind forcing. An advantage of the forcing method of Guo and Shen \cite{guo_generation_2009} over the previous internal wave generators is that
it can be employed with more flexibility and in 3D directly. While it can be used in a similar manner as the other internal wave generation methods by choosing a source region and combining it with sponge layers at the boundaries, it can also be employed to generate, maintain, and suppress waves in simulations with periodic boundary conditions along the horizontal directions, as shown in their work \cite{guo_generation_2009}. Basically, in such type of simulations with periodic boundary conditions, the force is implemented in the entire domain of the free surface and no sponge layers are used near the boundaries.

The objective of this work is to develop a computational framework capable of simulating complex floating structures subject to the action of complex water wave fields such as those found in realistic offshore environments. To deal with the difficulty of large disparity of scales associated to the problem, we adopted a two-domain partitioned approach, i.e., a large-scale domain with periodic boundary conditions known as far-field domain in which the offshore ocean conditions are efficiently developed, and a reduced-scale domain with high grid resolution, known as near-field domain, where the floating structure is located. To connect the two domains, we adopted a loosely coupled approach by feeding the wind and wave fields that have been fully developed in the far-field domain to the near-field domain. We opted for a one-way loosely coupled approach, in contrast to a two-way strongly coupled approach, considering the computational cost and the fact that the far-field domain simulation is employed to generate large-scale wind and wave flows in which the presence of a single or several marine structures should only have local effects and not alter the ocean environmental conditions. The computational model that we adopted for the near-field domain is the \gls{fsi}-level set method of Calderer, Kang and Sotiropoulos \cite{calderer_fsi_2014} that has been demonstrated to predict with high accuracy the motion of complex bodies interacting non-linearly with the air-water interface and capture the complex flow features induced by the body-water-air interactions. As for the large scale far-field solver, we chose the wind-wave coupled approach of Yang and Shen \cite{yang_simulation_2011,yang_simulation_2011_1}, which employs an efficient potential-flow based wave solver using a spectral method for the water motion and a viscous solver with undulatory boundaries for the air motion.

To incorporate the far-field waves to the near-field domain, we employ a wave generation method by applying a pressure force on the free surface in form of source term in the momentum equation. This approach, known as pressure forcing method, was initially proposed by Guo and Shen \cite{guo_generation_2009} to generate, suppress, and maintain water waves in a computational approach where the free surface is treated with a sharp interface method. In the present work, we extend the pressure forcing method by adapting it to a diffused interface level set method. The pressure forcing method allows to generate \gls{3d} directional waves, as well as a broadband spectrum of waves by using superposition of various directional waves. Wave reflections at the lateral boundaries are suppressed by using a sponge layer method. The idea of the sponge layer method is to add a dissipation term in the momentum equation in the regions where waves are desired to be absorbed. The classic sponge layer method was proposed by Israeli and Orszag \cite{Israeli1981} and is based on the use of a viscous loss term known as the Darcy term. We have also considered an additional term accounting for inertial losses as in Choi and Yoon \cite{choi_numerical_2009}, which is meant to decrease fluctuations caused by wave breaking.

In what follows, we first present the respective governing equations and numerical methods of the near-field and far-field solvers. Subsequently, we present the numerical algorithm for coupling the two computational models in a loose manner and present and discuss the results of several validation test cases. Then we illustrate the potential of the computational framework by simulating an operational offshore floating wind turbine under realistic wind and wave conditions. Finally, we outline our findings and present the conclusions and future computational challenges to be addressed.

\section{The near-field flow solver}\label{sec:nearfield}
\subsection{The near-field two-phase flow solver}
The near-field model solves the spatially-filtered incompressible Navier-Stokes equations using a two-phase flow level set formulation. In this approach, a single set of equations is used for the two phases and the flow properties are variables adopting in each phase their corresponding values and smoothly transitioning over a thin layer of thickness $2\epsilon$ across the interface as follows: 
\begin{equation}\label{eq:density}
\rho\left(\phi\right)=\rho_{a}+\left(\rho_{w}-\rho_{a}\right)h\left(\phi\right),
\end{equation}
\begin{equation}\label{eq:viscosity}
\mu\left(\phi\right)=\mu_{a}+\left(\mu_{w}-\mu_{a}\right)h\left(\phi\right),
\end{equation}
where $\rho$ is the density, $\mu$ is the viscosity, $\phi$ is a signed distance function used for tracking the position of the interface with positive values in the water phase and negative values in the air phase, the subscripts $a$ and $w$ indicate the flow property in the air and water phases, respectively, and $h$ is the smoothed Heaviside function \cite{Osher2002Level} defined as
\begin{equation}\label{eq:heaviside}
h(\phi)=\left\{ \begin{array}{ll}
0 & \phi<-\epsilon,\\
\frac{1}{2}+\frac{\phi}{2\epsilon}+\frac{1}{2\pi}\sin(\frac{\pi \phi}{\epsilon}) & -\epsilon\leq\phi\leq\epsilon,\\
 1 & \epsilon<\phi.
 \end{array}\right.
 \end{equation}
Then the flow equations in generalized curvilinear coordinates and in tensor notation (repeated indices imply summation) read as follows ($i$, $j$, $k$, $l$ $=$
$1$, $2$, $3$)
\begin{equation}\label{eq:cont}
J\frac{\partial U^i}{\partial \xi^i}=0,
\end{equation}
\begin{eqnarray}\label{eq:mom}
\frac{1}{J}\frac{\partial U^j}{\partial t}&=&\frac{\xi_{l}^{i}}{J}\left(-\frac{\partial}{\partial \xi_j}\left(U^ju_l\right)+\frac{1}{\rho\left(\phi\right)Re}\frac{\partial}{\partial\xi^j}\left(\mu \left(\phi \right)\frac{\xi_{l}^{j} \xi_{l}^{k}}{J}\frac{\partial u_l}{\partial \xi^k} \right)
-\right. \nonumber\\
&&\left.
-\frac{1}{\rho\left(\phi\right)}\frac{\partial}{\partial \xi^j}\left(\frac{\xi_{l}^{j}p}{J}\right)-\frac{1}{\rho \left(\phi\right)}\frac{\partial \tau_{lj}}{\partial \xi^j}-\frac{\kappa}{\rho(\phi)We^2}\frac{\partial h(\phi)}{\partial x_j}+\frac{\delta_{i3}}{Fr^2}+S^w_i+S^s_i+S^{AL}_i\right),
\end{eqnarray}
where $\xi_l^i$ are the transformation metrics, $J$  is the Jacobian of the transformation, $u_i$ are the Cartesian velocity components, $U^i$ are the contravariant volume fluxes, $p$  is the pressure, $\tau_{li}$  is the \gls{sgs} tensor, $\delta_{ij}$ is the Kronecker delta,  $S^w_i$ is the source term for wave generation, $S^s_i$ is the source term for wave dissipation, $S^{AL}_i$ is the actuator line body force, $\kappa$ is the interface curvature, and $Re$, $Fr$, and $We$ are respectively the Reynolds, Froude, and Weber dimensionless numbers defined as
\begin{equation}\label{eq:dimensionless}
Re=\frac{U L \rho_{w}}{\mu_{w}},    Fr=\frac{U}{\sqrt{g L}},  We=U\sqrt{\frac{\rho_{w} L}{\sigma}},
\end{equation}
with $U$ being a characteristic velocity, $L$ a characteristic length, $g$ the gravity, and $\sigma$ the surface tension.

For simulating turbulent flows with \gls{les}, we use the dynamic Smagorinsky \gls{sgs} model of \cite{Germano1991Dynamic} in combination with a wall-layer modeling strategy \cite{cabot_approximate_2000, wang2002dynamic}. With the level set approach, the air-water interface is tracked by solving the following advection equation based on Osher and Sethian \cite{osher_fronts_1988}:
\begin{equation}\label{eq:levelset}
\frac{1}{J}\frac{\partial \phi}{\partial t}+U^j\frac{\partial \phi}{\partial \xi^j}=0.
\end{equation}
A mass conserving re-initialization equation is then solved to ensure proper conservation of mass within the two fluids as extensively described in Kang and Sotiropoulos \cite{kang_numerical_2012}. The parameters of the level set method for free surface tracking in cases involving moving bodies, such as the interface thickness $\epsilon$, the pseudo time step size, and the number of iterations of the re-initialization equation, are analyzed in \cite{calderer_fsi_2014}. 
 
The momentum equations (\ref{eq:mom}) are discretized using a second-order central differencing scheme for the pressure gradient, diffusion, advective, and \gls{sgs} terms, with the exception of the diffused region across the free surface interface where the third-order \gls{weno} scheme \cite{jiang_efficient_1996} is applied for the advective terms. 

The solution is advanced in time by using a second-order Crank-Nicholson scheme and the fractional step method. 
The level set equation (\ref{eq:levelset}) is discretized with a third-order \gls{weno} scheme in space, and \gls{rk2} scheme in time. The re-initialization equation uses a second-order \gls{eno} scheme \cite{sussman_improved_1998}.

\subsection{\Gls{fsi} of complex floating structures}
To simulate the dynamic motion of complex floating structures interacting with two-phase free surface flows we employ the level set-\gls{fsi}-\gls{curvib} method recently developed by Calderer, Kang, and Sotiropoulos \cite{calderer_fsi_2014}. In this method, a partitioned \gls{fsi} algorithm was adopted coupling the previously described two-phase flow solver with a structural model for predicting the six \gls{dof} dynamics of rigid bodies governed by the \gls{eom} derived from Newton's second law. For the sake of simplicity but without loss of generality, we briefly present the \gls{eom} for a single, rigid, elastically mounted, and damped body in a Lagrangian form and in principal axis as follows ($i=1,2,3$):
\begin{equation}
M\frac{\partial ^2 Y^i}{\partial t^2}+C\frac{\partial Y^i}{\partial t} +K Y^i=F^i_{f}+F^i_{e}
\label{eq:EOM1}
\end{equation}
\begin{equation}
J\frac{\partial ^2 \Theta^i}{\partial t^2}+C\frac{\partial \Theta^i}{\partial t} +K \Theta^i=M^i_{f}+M^i_{e}
\label{eq:EOM2}
\end{equation}
where $M$ and $J$ are respectively the mass and moment of inertia, $Y$ is the Lagrangian position describing the linear \gls{dof}, $\Theta$ is the position vector with the rotational \gls{dof}, $K$ is the spring stiffness coefficient, and $C$ is the damping coefficient. On the right hand side of the above equations, $F_f$ and $M_f$ are the forces and moments that the fluid imparts to the body and $F_e$ and $M_e$ are external forces and moments, respectively.

Equations (\ref{eq:EOM1}) and (\ref{eq:EOM2}) are integrated in time by first transforming them into a system of ordinary differential equations. The coupling between the flow field and the rigid body is achieved via imposed continuity of the velocity field at the fluid-body interface.  

The \gls{curvib} method, developed by Ge and Sotiropoulos \cite{ge_numerical_2007}, allows to deal with geometrically complex moving bodies with a sharp interface approach. An unstructured triangular mesh is used to discretize the body, which is then superposed on the underlying fluid mesh. The interfacing between the two domains is established by imposing the velocity boundary conditions of the body at the so called \gls{ib} nodes, located in the fluid domain in the immediate vicinity  of the structure. The velocity imposed at the IB nodes is reconstructed using interpolation along the wall-normal direction from the known values of velocity at the body and at the fluid nodes adjacent to the IB nodes. When the Reynolds number of the flow is relatively low and the \gls{ib} nodes fall inside the viscous sub-layer, the interpolation is done linearly. In contrast, when the Reynolds number is large and the grid resolution is not sufficiently fine to resolve the viscous sub-layer, the interpolation can be done using wall-layer modeling as described in  \cite{kang_highresolution_2011}.

\subsection{The actuator line model}\label{sec:actuatorline}
The method that we use for parameterizing a turbine rotor is the actuator line model proposed by S{\o}rensen and Shen \cite{sorensen2002numerical}. The basic idea behind the method is to subtract from the flow field an equivalent amount of momentum to that from a turbine rotor without the need to resolve the flow around its actual geometry. This effect is implemented by introducing a sink term on the right hand side of the momentum equation acting on those grid nodes that are located in the vicinity of the turbine rotor.

In the present actuator line method, the location of the turbine rotor is tracked by discretizing each of the blades in a Lagrangian manner with straight lines composed of several elements aligned with the radial direction. In each of the elements, the lift ($L$) and drag ($D$) forces are computed using the following expressions:
\begin{equation}\label{eq:act_line1}
L=\frac{1}{2}\rho C_L C u^2_{ref},
\end{equation}
\begin{equation}\label{eq:act_line2}
D=\frac{1}{2}\rho C_D C u^2_{ref},
\end{equation}
where $C_L$, $C_D$ are the lift and drag coefficients, respectively, taken from tabulated \gls{2d} airfoil profile data, $C$ is the chord length, and $u_{ref}$ is the incoming reference velocity computed as
\begin{equation}\label{eq:act_line3}
u_{ref}=(u_x,u_{\theta}-\Omega r)
\end{equation}
where $u_x$ and $u_{\theta}-\Omega r$ are the components of the velocity in the axial and azimuthal directions, respectively, $\Omega$ the angular velocity of the rotor, and $r$ the distance to the center of the rotor. 

The reference velocity at the line elements $u(X)$ can be calculated by using interpolation from the surrounding fluid nodes where the velocity is known. This is necessary as the nodes from the fluid mesh and line segments do not necessarily coincide. If we consider $X$ to be the coordinates of the actuator line nodes and $x$ the coordinates of the fluid mesh nodes, we can perform the interpolation using a discrete delta function in the following manner
\begin{equation}\label{eq:act_disk4}
u(X)=\sum_{N_D}u(x)\delta_h(x-X)V(x),
\end{equation}
where $\delta_h$ is a \gls{3d} discrete delta function, $V(x)$ is the volume of the corresponding fluid cell, and $N_D$ is the number of fluid cells involved in the interpolation.

The lift and drag forces, which have been computed at each of the line elements, can be transferred in a diffused manner into the flow domain through a source term $S^{AL}(x)$ in the momentum equation (\ref{eq:mom}) using the following equation:
\begin{equation}\label{eq:act_line4}
S^{AL}(x)=\sum_{N_L}F(X)\delta_h(x-X)A(x).
\end{equation}
where $N_L$ is the number of segments composing one of the actuator lines, $A(x)$ is the length of each segment, and $F(X)$ is the projection of $L$ and $D$ onto the Cartesian coordinates. 

For more details about the implementation of the present actuator line model, the reader is referred to Yang et al. \cite{yang2014large}. 

\section{The far-field flow solver}\label{sec:farfield}
The large-scale wind-wave model is based on the approach of Yang and Shen \cite{yang_simulation_2011, yang_simulation_2011_1}, which employs a potential flow based wave solver with a \gls{hos} method for the water motion, and a viscous solver with undulatory boundaries for the air motion. A brief description of the governing equations and numerical methods as well as the coupling algorithm is provided in this section.
\subsection{The high-order spectral method for simulating water waves}
To model the far-field wave field in a non-linear manner, we solve the potential flow wave problem formulated in the form of Zakharov \cite{Zakharov1968Stability} by applying the \gls{hos} method of Dommermuth and Yue \cite{Dommermuth1987HighOrder}. The kinematic and dynamic \gls{bcs} can be written as functions of the free surface elevation $\eta$ and the velocity potential $\Phi$ as follows
\begin{eqnarray}
\nabla^{2} \Phi = \frac{\partial^{2} \Phi}{\partial x_{i} \partial x_{i}} = 0,
\end{eqnarray}
\begin{eqnarray}
\frac{\partial \eta}{\partial t} + \frac{\partial \eta}{\partial x_{\alpha}} \frac{\partial \Phi^{s}}{\partial x_{\alpha}} - \left( 1 + \frac{\partial \eta}{\partial x_{\alpha}} \frac{\partial \eta}{\partial x_{\alpha}} \right) \left. \frac{\partial \Phi}{\partial x_{3}} \right|_{x_{3} = \eta} = 0,
\label{eq:FarField_Kinem}
\end{eqnarray}
\begin{eqnarray}
\frac{\partial \Phi^{s}}{\partial t} + \frac{\eta}{Fr^2} + \frac{1}{2} \frac{\partial \Phi^{s}}{\partial x_{\alpha}} \frac{\partial \Phi^{s}}{\partial x_{\alpha}} - \frac{1}{2} \left( 1 + \frac{\partial \eta}{\partial x_{\alpha}} \frac{\partial \eta}{\partial x_{\alpha}} \right) \left. \left( \frac{\partial \Phi}{\partial x_{3}} \right)^{2} \right|_{x_{3} = \eta} = - P_{a},
\label{eq:FarField_Dynam}
\end{eqnarray}
where $i = 1, 2, 3$, $\alpha = 1, 2$, and $\Phi^{s} = \left. \Phi \right|_{x_{3} = \eta}$ and $P_a$ are the velocity potential and air pressure at the water surface, respectively. Note that $x_1$ and $x_2$ correspond to the coordinate components along the horizontal directions and $x_3$ is that along the vertical direction. 

The following perturbation series of $\Phi$ expressed with respect to the wave steepness to order $M$ and the Taylor series expansion to the same order about the mean water level $x_3=0$ are used,
\begin{eqnarray}
\Phi \left( x_1, x_2, x_3, t \right) = \sum^{M}_{m = 1}\Phi^{(m)}(x_1,x_2,x_3),
\end{eqnarray}
\begin{eqnarray}
\Phi^{s} \left( x_1, x_2, t \right) = \sum^{M}_{m = 1} \sum^{M - m}_{l = 0} \left. \frac{\eta^{l}}{l!} \frac{\partial^{l} \Phi^{\left( m \right)}}{\partial x_{3}^{l}} \right|_{x_{3}^{l} = 0}
\end{eqnarray}
and each $\Phi^{(m)}$ can be expressed with $N$ modes using an eigenfunction expansion,
\begin{eqnarray}
\Phi^{(m)} \left(  x_1,x_2,x_3, t \right) = \sum^{N}_{n = 1} \Phi^{(m)}_{n}(t)\Psi_n(x_1,x_2,x_3), \;  z_3\leq 0,  
\end{eqnarray}
For deep water waves, $\Psi_n$ has the expression 
\begin{eqnarray}
\Psi_{n}( x_1, x_2) = \exp \left( \sqrt{k_{1n}^2+k_{2n}^2}\cdot x_3    +\mathbf{\imath} (k_{1n}\cdot x_1 + k_{2n}\cdot x_2) \right),  
\end{eqnarray}
where $\imath=\sqrt{-1}$, and $k_{1n}$ and $k_{2n}$ are the components of the wavenumber vector $k$ in the $x_1$ and $x_2$ directions, respectively. Later in this paper, we use $k_{x}$ to refer to $k_{1}$ and $k_{y}$ to refer to $k_{2}$.

Periodic \gls{bcs} are imposed in the horizontal directions, which allow the \gls{hos} method to use an efficient spacial discretization scheme based on a pseudo-spectral method. The equations (\ref{eq:FarField_Kinem}) and (\ref{eq:FarField_Dynam}) are advanced in time with the \gls{rk4} scheme. An extensive description of this approach with validations and applications was presented in \cite{Dommermuth1987HighOrder} and in \cite{mei2005theory}.

\subsection{The \gls{les} method for the air field}

To simulate the air flow over water waves in the far-field model, we employ the method of Yang, Meneveau, and Shen \cite{yang2013dynamic}, which is an extension to \gls{les} of the initial \gls{dns} approach of Yang and Shen \cite{yang_simulation_2011}. We solve the following filtered incompressible Navier-Stokes equations governing the flow of a single phase fluid ($i,j=1,2,3$)
\begin{eqnarray}
\displaystyle \frac{\partial \widetilde{u}_{i}}{\partial t} + \frac{\partial \left( \widetilde{u}_{i} \widetilde{u}_{j} \right)}{\partial x_{j}} = - \frac{1}{\rho} \frac{\partial \widetilde{P}}{\partial x_{i}} - \frac{\partial \tau_{ij}}{\partial x_{j}},
\label{eq:Farfield_Mom}
\end{eqnarray}
\begin{eqnarray}
\displaystyle \frac{\partial \widetilde{u}_{i}}{\partial x_{i}} = 0.
\label{eq:Farfield_Cont}
\end{eqnarray}
where $\widetilde{u}_{i}$ are the filtered velocity components, $\widetilde{P}$ is the filtered dynamic pressure, and $\tau_{ij}$ is the \gls{sgs} stress tensor modeled by the scale-dependent Lagrangian dynamic model \cite{Bou-Zeid2005}. 
Note that the viscous term in equation (\ref{eq:Farfield_Mom}) has been neglected due to the consideration of high Reynolds number and the negligible effect of viscosity at the resolved scales. A wall-layer model is used to account for the viscous effects at the bottom free surface boundary \cite{yang2013dynamic}. 
A boundary fitted grid is employed and adapts to the motion of the free surface which is seen by the air domain as an undulatory boundary (see \cite{yang_simulation_2011} for details of the coordinate transformation mapping). The geometry and the velocity of the bottom boundary are prescribed from the \gls{hos} simulation. The air flow can be driven either by applying a constant pressure gradient in the stream-wise direction or by applying a shear stress at the top boundary, and periodic \gls{bcs} are considered in the horizontal directions.

For the spacial discretization in the horizontal directions, a Fourier-series-based pseudo-spectral method is used taking advantage of the fact that the \gls{bcs} are periodic. In the vertical direction, a second-order finite difference scheme is used. A semi-implicit fractional-step method is applied to advance the governing equations (\ref{eq:Farfield_Mom}) and (\ref{eq:Farfield_Cont}) \cite{yang2013dynamic}. 

\subsection{The \gls{les}-\gls{hos} coupling algorithm}
\label{sec:leshos}
The \gls{les} and \gls{hos} models are dynamically coupled by following an iterative procedure. First the \gls{hos} method is used to advance the wave to the next time step $(n+1)$ under the forcing of air pressure $P_{a}^{n}$ on the wave surface. The free surface elevation $\eta^{n+1}$ and surface velocity $u_s^{n+1}$ can then be imposed as Dirichlet BC at the bottom boundary of the \gls{les} model. The air flow can then be advanced to time step $(n+1)$ by solving the described \gls{les} model, and a new value of surface pressure $P_a^{n+1}$ is computed to continue the simulation.
\section{Coupling the near-field solver with the far-field solver}\label{sec:coupling}

We propose two far-field/near-field coupling approaches to take full advantage of the numerical expedience of the far-field method for large-scale wind-wave simulations and the advanced capabilities of the near-field solver for simulating air-water-body interactions. 

In Approach 1, the two solvers are loosely coupled in time. The wind flow from the far field can be incorporated directly into the near-field solver by prescribing at each time step the instantaneous air velocity at the inlet boundary. The process for incorporating the wave field involves the following two steps: (1) extract the energy and phases of surface waves from the far-field model by performing a Fourier analysis, and (2) incorporate the resulting far-field waves into the near-field domain by applying the surface forcing method of Guo and Shen \cite{guo_generation_2009} that has been appropriately adapted to the level set method in the present study. This approach is aimed to simulate a single or multiple floating structures. 

In Approach 2, the wind-wave simulation at a particular instant in time for which the flow is fully developed is used as the initial condition for the air-water-body simulation, i.e., the wind-wave simulation is restarted using the \gls{fsi}-level set method with the addition of the floating structures. The wind field is directly prescribed in the whole air domain of the \gls{fsi}-level set method using interpolation and the wave field is initialized with the pressure forcing method of Guo and Shen \cite{guo_generation_2009}. In this approach periodic boundary conditions are used in the air-water-body simulation. This approach is suited for simulating infinite arrays of floating structures.

The major difference in Approach 2 compared with Approach 1 is the way that the pressure forcing method is applied in time and space.  While in Approach 1 the distributed pressure is continuously applied in time during the entire simulation, in Approach 2 the forcing method is only applied at the very initial time. The area of application of the pressure, which is referred to as the source region, also differs between the two approaches. While in Approach 1 the source region corresponds to a rectangular band as illustrated in Figs.\ \ref{fig:wave_scketch_2d} and \ref{fig:wave_scketch_2d_NF}, in Approach 2 it spans the whole free surface.

The forcing method of \cite{guo_generation_2009} relies on the linearized Cauchy-Poisson problem that allows to relate a given free surface pressure field to its leading-order free surface elevation response (details of the Cauchy-Poisson problem can be found in \cite{mei2005theory}). For simplicity but without loss of generality we write the pressure-elevation relationship in \gls{2d} form as
\begin{equation}\label{eq:cauchy}
\eta(x,t)=- \frac{1}{2 \pi}\int_{-\infty}^{\infty}{dk_x \exp [i k_x x]}\int_{0}^{t}{P_a(k_x,\tau) {Fr}^2 \omega \sin[\omega(t-\tau)] d\tau},
\end{equation}
where $\omega$ is the wave angular frequency, and $P_a$ is the free surface pressure field. 

\subsection{Approach 1: Inlet-outlet \gls{fsi} simulation}
\label{sec:WaveGeneration}

\begin{figure}[h!bt]
	\centering
	\includegraphics[trim=0.0cm 0.0cm 0.0cm 0.0cm,clip,width=0.9\textwidth]{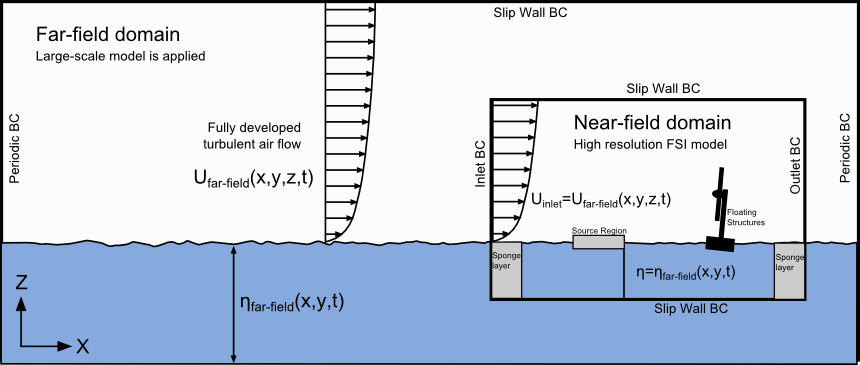}
	\caption{Schematic description of the near-field/far-field coupling approach 1. The inlet velocity in the near-field domain is prescribed from the far-field velocity field, and the wave field is imposed through applying the pressure forcing method in the source region. Air-water-body interactions can be studied by placing the body between the source region and the outlet wall sponge layer.}
\label{fig:wave_scketch_2d}
\end{figure}

\begin{figure}[h!bt]
    \centering
    \subfigure[Lateral view]{
        \label{subfig:wave_scketch_2d_NF_a}
        \centering
        \includegraphics[trim=0.0cm 0.0cm 0.0cm 0.0cm,clip,width=0.7\textwidth]{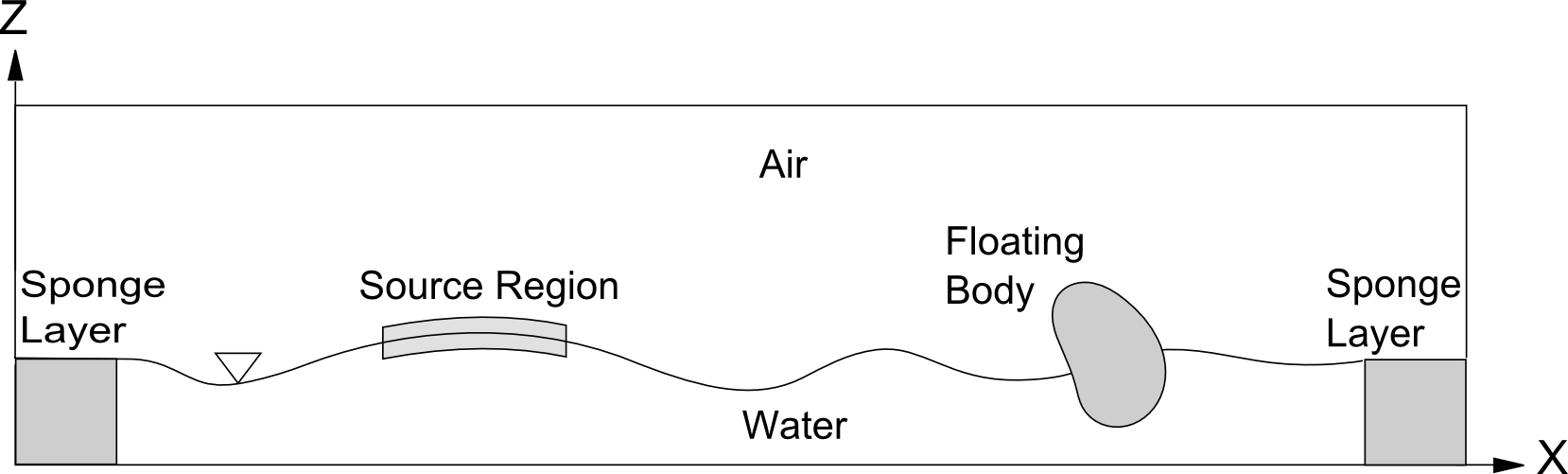}%
    }
    \subfigure[Top view]{
        \label{subfig:wave_scketch_2d_NF_b}		
        \centering
        \includegraphics[trim=0.0cm 0.0cm 0.0cm 0.0cm,clip,width=0.7\textwidth]{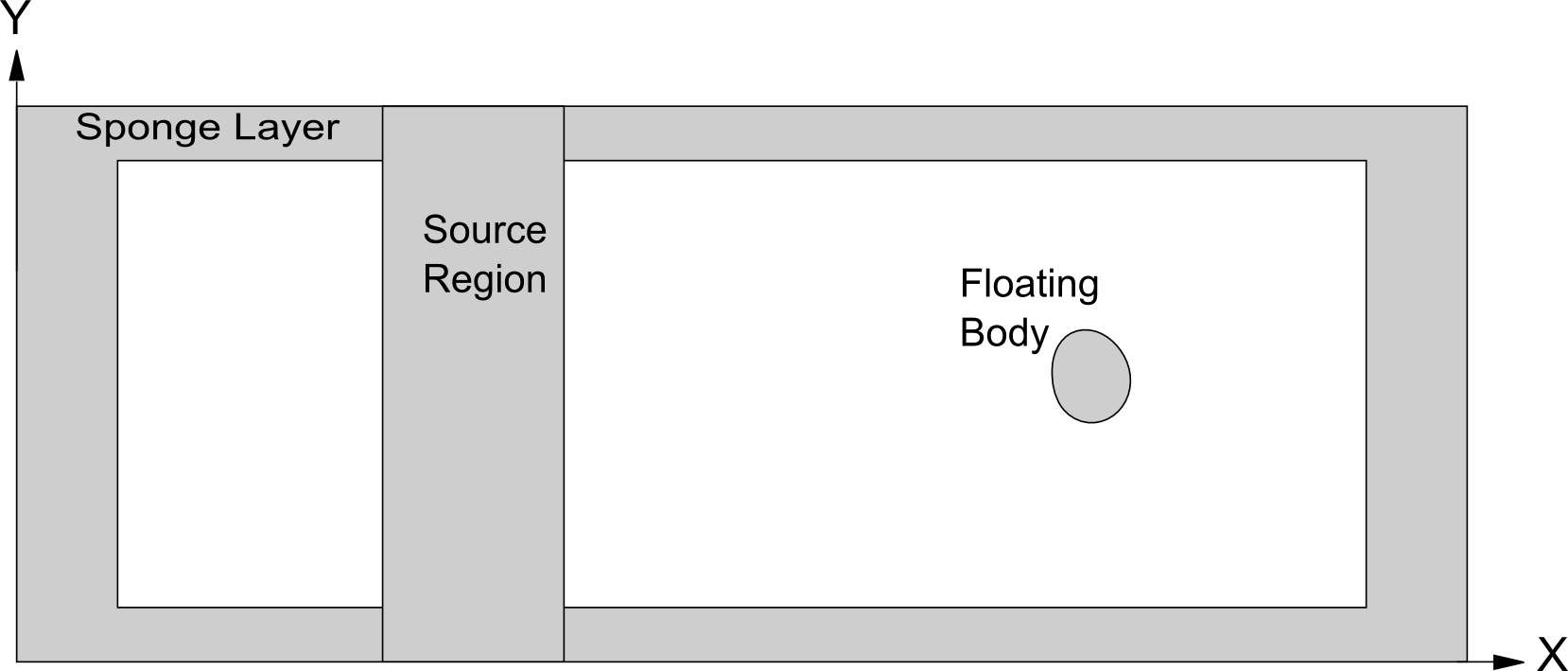}%
    }
	\caption{Schematic description of the near-field domain when using the far-field/near-field coupling approach 1. The source region for wave generation, which is applied on the free surface along a span-wise rectangular band, generates \gls{3d} directional waves propagating away from the center of application. When the waves reach the side walls the wave energy is dissipated using the sponge layer method to suppress wave reflections. The floating structure subject to wave interactions is located between the source region and the outlet sponge layer.}
\label{fig:wave_scketch_2d_NF}
\end{figure}

A schematic description of the overall procedure in Approach 1 is depicted in Figs.\ \ref{fig:wave_scketch_2d} and \ref{fig:wave_scketch_2d_NF}. After the wind and wave fields from the far-field simulation are fully developed, the far-field solution can be fed to the near-field domain as described in this section.

\subsubsection{\Gls{2d} wave generation}

\begin{figure}[h!bt]
\centering
\subfigure[Nodal force]{
\label{subfig:wave_scketch_F}
\centering
\includegraphics[trim=0.0cm 0.0cm 0.0cm 0.0cm,clip,width=.4\textwidth]{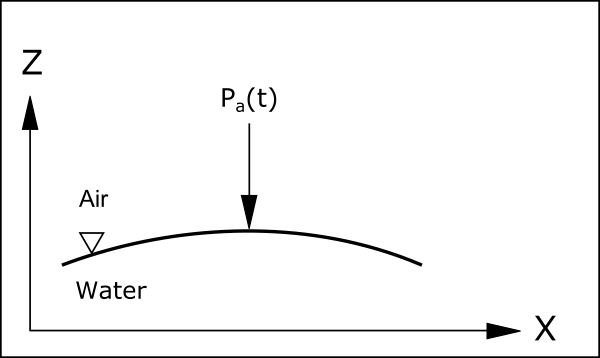}%
}
\subfigure[Distributed pressure]{
\label{subfig:wave_scketch_Ps}		
\centering
\includegraphics[trim=0.0cm 0.0cm 0.0cm 0.0cm,clip,width=.4\textwidth]{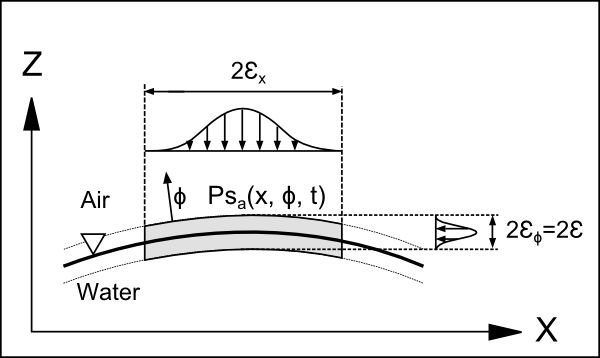}%
}
\caption{Schematic description  of the force applied on the free surface for wave generation when using the far-field/near-field coupling approach 1. The nodal force shown in (a) is implemented with the distributed manner shown in (b).}\label{fig:wave_scketch_force}
\end{figure}

Using equation (\ref{eq:cauchy}) one can obtain a progressive monochromatic wave of amplitude $A$ and frequency $\omega$ adopting the following form 
\begin{equation}\label{eq:elevation2d}
\eta(x,t)=A \cos (k_x x - \omega t),
\end{equation}
and surface velocity
\begin{equation}\label{eq:elevation2d_vel}
\eta_t(x,t)=A \omega \sin (k_x x - \omega t),
\end{equation}
by applying the following free surface force corresponding to an oscillatory nodal force applied at $x=0$
\begin{equation}\label{eq:FS_force}
P_a(t)=P_0  \sin (\omega t),
\end{equation}
where $P_0$, described below, is a coefficient that depends on the wave and fluid characteristics.

The application of a nodal force $P_a(t)$ generates progressive waves, which propagate symmetrically with respect to the point of application $x=0$. The numerical implementation of the nodal force on the free surface is not straightforward. It is important to consider that the interface is moving in time and the point of application of the force does not necessarily coincide with a grid coordinate point of the fixed Eulerian fluid mesh. Also, in the context of the present level set free surface tracking method the interface is diffused, adding additional numerical difficulties when applying the nodal force. For these reasons, we propose to diffuse the nodal force over both a distance $\epsilon_x$ along the $x$ direction and a distance $\epsilon_\phi$ along the free surface normal direction as follows 
\begin{equation}\label{eq:FS_force_smoothed}
Ps_a(x,\phi,t)=P_0\delta(x,\epsilon_x) \delta(\phi,\epsilon_{\phi})sin(\omega t),
\end{equation}
where $\delta$ is a distribution function defined as 
\begin{equation}\label{eq:delta_func}
 \delta (\alpha,\beta) = \left\{ \begin{array}{rl}
 \frac{1}{2\beta}\left[1+\cos\left(\frac{\pi \alpha}{\beta}\right)\right] 
 &\mbox{ if $-\beta<\alpha<\beta$} \\
 0 
 &\mbox{ otherwise.}
 \end{array} \right.
\end{equation} 

We choose the value of $\epsilon_{\phi}$  to be equal to the smoothing thickness of the free surface interface $\epsilon$.  The smoothing distance $\epsilon_x$ can in principle adopt any value comprised in the interval $(0,L/2)$ with $L$ being the wavelength, although we observed better accuracy using $\epsilon_x\approx L/2$.

In order to determine the value of $P_0$ in equation (\ref{eq:FS_force}), we use the wave energy flux concept. The energy flux $Q_{Ps_a}$ induced by the distributed pressure $Ps_a$ is set to be equal to the theoretical wave energy flux $Q_{theory}$ desired. The theoretical wave energy flux is written as 
\begin{equation}\label{eq:wave_pot1}
Q_{theory}=E c_g,
\end{equation}
were $E$ is the mean wave energy density defined as $E=\frac{1}{2}\rho g A^2$ and $c_g$ is the wave group velocity that using the dispersion relation can be written as $c_g=\sqrt{g/k}/2$.

The energy flux induced by the smoothed pressure force is computed as follows
\begin{equation}\label{eq:wave_pot2}
Q_{Ps_a}=\frac{1}{T}\int_{0}^{T}{\int_{-\epsilon_x}^{\epsilon_x}{\int_{-\epsilon_{\phi}}^{\epsilon_{\phi}}{Ps_a(x,\phi,t)\rho(\phi)\eta_{t}(x,t)}d\phi dx}dt}.
\end{equation}
 
Substituting (\ref{eq:elevation2d_vel}) and  (\ref{eq:FS_force_smoothed}) into (\ref{eq:wave_pot2}) and considering $\rho(\phi)=0.5\left(\rho_a+\rho_w\right)+0.5\left(\rho_a-\rho_w\right)\sin(0.5 \pi \phi / \epsilon_{phi})$, we write the wave energy flux induced by the force as
\begin{equation}\label{eq:wave_pot3}
Q_{Ps_a}=-\frac{P_0 A \omega}{4 \epsilon_{x}}
\frac{\pi^2}{k (\pi^2 - \epsilon_{x}^2 k^2)}
sin(k \epsilon_x)
\frac{\rho_a+\rho_w}{2},
\end{equation}
with the theoretical energy flux in equation (\ref{eq:wave_pot1}),
\begin{equation}\label{eq:wave_P0}
 P_0=A 
 \frac{g^2}{\omega ^2} 
 \frac{2 \rho_w}{\rho_a+\rho_w}
 \frac{\epsilon_x}{f(\epsilon_x,k_x)},
\end{equation}
where $f(\epsilon_x,k_x)$ is 
 \begin{equation}\label{eq:surfacepressure5} 
 f(\epsilon_x,k_x)=\frac{\pi^2}{k_x \left(\pi^2-\epsilon_x^2k_x^2 \right)}\sin(k_x\epsilon_x).
 \end{equation}

The distributed force in the source region is introduced in the code in form of a source term in the filtered momentum equations as follows
\begin{equation}\label{eq:wave_source_2d}
S^w_i(x,t)=n_i(\phi)P_0\delta(x,\epsilon_x)\delta(\phi,\epsilon_\phi)sin(\omega t)
\end{equation} 
where $n_i$ denotes the normal direction of the free surface.
The source region always follow the motion of the free surface adapting to its topological shape.

Typical values adopted in this work for $\epsilon_x$ are in the order of half wavelength, and for $\epsilon_\phi$ between 3 and 6 grid sizes.
By applying superposition principles, the above method can be applied to generate complex wave fields with multiple wave frequencies as demonstrated in the results section.

\subsubsection{\Gls{3d} directional wave generation}

The wave generation method can be extended to \Gls{3d} directional waves defined by the following free surface elevation
\begin{equation}\label{eq:surfacepressure2}
\eta(x,y,t)=A \cos (k_x x + k_y y -\omega t + \theta),
\end{equation} 
where $k_x$ and $k_y$ are the components of the wavenumber vector and $\theta$ is the wave phase.
In this case the constructed forcing term reads as follows 
\begin{equation}\label{eq:surfacepressure1}
S_i(x,y,t)=n_i(\phi)P_0\delta(x,\epsilon_x)\delta(\phi,\epsilon_\phi)sin(\omega t-k_y y-\theta),
\end{equation} 
where $P_0$ is a coefficient that depends on the wave and fluid characteristics,
 \begin{equation}\label{eq:surfacepressure4}
 P_0=A \frac{g^2}{\omega ^2} \frac{\epsilon_x}{f(\epsilon_x,k_x)} \frac{2 \rho_w}{\rho_a+\rho_w}\frac{k_x}{\left( k_x^2+k_y^2\right)^{1/2}},
 \end{equation}
$\delta$ is the distribution function as defined in (\ref{eq:delta_func}), and $f(\epsilon_x,k_x)$ is given in (\ref{eq:surfacepressure5}).
 

\subsubsection{Sponge layer}
The method used in this work for dissipating the energy of the waves near the boundaries of the computational domain is the sponge layer method expressed in the form proposed by \cite{choi_numerical_2009} as follows
\begin{equation}\label{eq:sponge}
S^s_i(x,y,t)=-\left[ \mu C_0 u_i+\rho C_1 u_i \left| u_i\right|  \right]\frac{\exp \left[ \left(\frac{x_s-x}{x_s} \right)^{n_s}\right]-1}{\exp(1)-1} \mbox{,  for } (x_0-x_s)\leq x \leq x_0,
\end{equation}
where $x_0$ denotes the starting coordinate of the source region, $x_s$ is the length of the source region, and $C_0$, $C_1$, and $n_s$ are coefficients determined empirically \cite{choi_numerical_2009}.
 

 
\subsubsection{Air flow coupling}

\begin{figure}[h!bt]
	\centering
	\subfigure[Far-field vertical grid storage]{%
        \includegraphics[trim=-1.0cm 0.0cm -1.0cm 0.0cm,clip,width=0.30\textwidth]{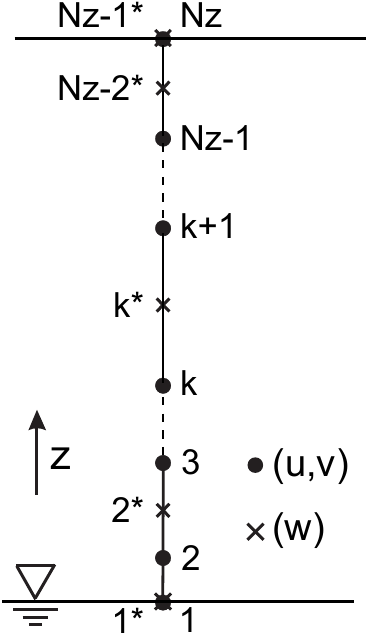}%
		\label{subfig:coupling_grid_scketch_a}%
	}%
	\subfigure[Near-field interpolation]{%
        \includegraphics[trim=-1.0cm -0.4cm -1.0cm 0.0cm,clip,width=0.45\textwidth]{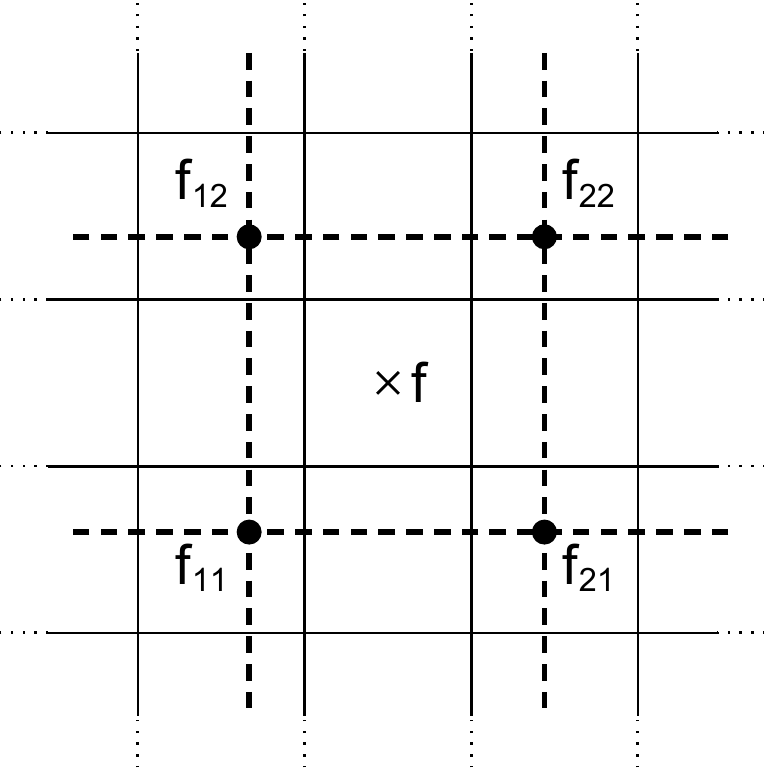}%
		\label{subfig:coupling_grid_scketch_b}%
	}
	\caption{Schematic description of the far-field and near-field grid data storage. (a) Location of the points where the three velocity components are stored on the far-field mesh in the vertical direction. (b) Superposition of the far-field mesh, in dashed line, over the near-field mesh, indicated in solid line.}
	\label{fig:coupling_grid_scketch}
\end{figure}

The air flow from the far-field solver is directly prescribed on the inlet plane of the near-field solver at every time step. This process requires special considerations as the far-field method uses a boundary-fitted approach for which the mesh continuously deforms adapting to the free surface elevation. In contrast, the fluid mesh in the near-field method is purely Eulerian and its grid points are kept fixed during the simulation. Because the grid nodes of the far-field source plane and the near-field inlet plane do not coincide, we perform a bi-linear interpolation.

The storage of the velocity components in the far-field domain is particular. While along the horizontal directions all three velocity components, $u$, $v$, and $w$ are stored at the grid points, along the vertical direction, $v$ and $u$ are stored at the regular grid points and $w$ at the grid midpoint as illustrated in Fig. \ref{fig:coupling_grid_scketch}a.
  
The far-field plane corresponding to the last cross section of the computational domain is exported to a data file, which is read by the near-field solver later. Once the far-field velocity at a particular time step is available, it is prescribed at the inlet plane of the near-field domain using bi-linear interpolation. The velocity components in the inlet-plane of the near-field domain are stored in Cartesian form at the cell centers. The bi-linear interpolation for obtaining the velocity components $f$ (see Fig.\ \ref{fig:coupling_grid_scketch}) is as follows:
%
%
%
%

%
\begin{eqnarray}\label{eq:bilinear}
f(x,y) & \approx &
 \frac{1}{(x_2-x_1)(y_2-y_1)} \left[
 (x_2-x)(y_2-y)f_{11} +
\right. \nonumber\\
&&\left. 
 +(x-x_1)(y_2-y)f_{21} +
 (x_2-x)(y-y_1)f_{12} +
 (x-x_1)(y-y_1)f_{22} 
 \right],
\end{eqnarray}
where $f_{ij}$ indicate the far-field velocity components.


\subsection{Approach 2: Periodic \gls{fsi} simulation}
\label{sec:approach2_method}

In this approach the wave generation method is a direct extension of the $\delta$-function method, described in \cite{guo_generation_2009}, to the level set method. Equation (\ref{eq:cauchy}) can be used to relate a surface pressure with the following free surface elevation, corresponding to a progressive wave, expressed as a sum of two standing waves,
\begin{equation}\label{eq:wave_appr2_1}
\eta(x,t)=A \cos (k x -\omega t + \theta)=A \sin(\omega t - \theta) \cos (k x)+A \cos(\omega t - \theta) \sin (k x).
\end{equation} 

The above free surface elevation can be accomplished in a sharp free surface interface method by applying an instantaneous pressure field on the entire undisturbed free surface. As proposed in \cite{guo_generation_2009}, the pressure field is smeared over a period of time equal to $2\Delta$ to facilitate its numerical implementation by using a smoothed $\delta$ function as follows 
\begin{equation}\label{eq:wave_appr2_2}
P_a(x,t)=
-\frac{A}{Fr^2 \omega}\delta(t,\Delta)\cos(k x)
+\frac{A}{Fr^2 \omega}\delta \left( t- \frac{\pi}{2\omega},\Delta\right)\sin(k x),
\end{equation} 
with the smoothed $\delta$ function given in equation (\ref{eq:delta_func}). As indicated by \cite{guo_generation_2009}, $\Delta$ adopts a value comprised between $4dt$ and $10\%$ of the wave period.

To extend the pressure force to a smoothed interface approach, we can distribute $P_a(x,t)$ as follows:
 \begin{equation}\label{eq:wave_appr2_3}
P_\phi(x,t,\phi)=B\delta(\phi)P_a(x,t),
\end{equation} 
where $B$ is a coefficient to ensure that the energy flux $P_{P_\phi}$ induced by  $P_\phi(x,t,\phi)$ is equivalent to the energy flux $P_{P_a}$ induced by $P_a(x,t)$. The energy fluxes are defined as follows
\begin{equation}\label{eq:wave_appr2_4}
P_{P_a}=\frac{1}{2T\Delta}\int_{0}^{L}{\int_{-\Delta}^{\Delta}{P_a(x,t)\rho_w\eta_t(x,t)}}dtdx,
\end{equation} 
\begin{equation}\label{eq:wave_appr2_5}
P_{P_\phi}=\frac{1}{2T\Delta}\int_{-\epsilon}^{\epsilon}{\int_{0}^{L}{\int_{-\Delta}^{\Delta}{P_a(x,t)\rho(\phi)\eta_t(x,t)}}dt dx d\phi},
\end{equation} 
By setting (\ref{eq:wave_appr2_4}) equals to (\ref{eq:wave_appr2_5}), $B$ becomes
\begin{equation}\label{eq:wave_appr2_6}
B=\frac{2\rho_w}{\rho_w+\rho_a}.
\end{equation}
%
\section{Results}\label{sec:results}
In this section, we employ the proposed far-field/near-field coupling algorithm to simulate a number of water wave cases. We first test the ability of the pressure forcing method (Approach 1) to generate monochromatic waves in a \gls{2d} rectangular channel and a \gls{3d} basin. Subsequently, we evaluate the far-field/near-field coupling algorithm by incorporating various \gls{3d} directional wave cases, initially originated in the far-field domain, to the near-field domain using both coupling approaches. Finally, we illustrate the capability of the present computational framework by applying it to simulate a floating wind turbine platform interacting with offshore waves and wind.
\subsection{Forcing method validation case: monochromatic waves}
\label{sec:monoch}
\begin{table}[t]
  \caption{Description of the parameters of the six test cases of monochromatic waves}\
\centering
\begin{tabular}{c c c c c}
\hline
\textbf{Wave Case} & \textbf{$L [m]$} & \textbf{$A [m]$} & \textbf{$A k_x$} & \textbf{$\epsilon_x$}\\
\hline
1   &   0.6   &   0.005  & 0.05  & 0.3\\
2   &   1.2   &   0.01   & 0.05  & 0.6\\
3   &   2.4   &   0.02   & 0.05  & 1.2\\
4   &   1.2   &   0.0019  & 0.01  & 0.6\\
5   &   1.2   &   0.0095  & 0.05  & 0.6\\
6   &   1.2   &   0.0191  & 0.10  & 0.6\\
\hline
\end{tabular}
\label{tab:monowaves}
\end{table}
\begin{table}[t]
  \caption{Description of the five grids employed in the monochromatic wave cases}\
\centering
\begin{tabular}{c c c c}
\hline
\textbf{Grid} & \textbf{Grid size} & \textbf{Streamwise spacing} & \textbf{Minimum vertical spacing [m]}\\
\hline
1   &    $301\times180$   &   $L/7.5$  & $0.005$\\
2   &    $601\times180$   &   $L/15$  & $0.005$\\
3   &    $1200\times180$   &   $L/30$  & $0.005$\\
4   &    $601\times 140$   &   $L/15$  & $0.01$\\
5   &    $601\times 100$   &   $L/15$  & $0.02$\\
\hline
\end{tabular}
\label{tab:monowaves_grid}
\end{table}

\begin{figure}[h!bt]
	\centering
	\subfigure[Time $t=4s$]{
		\includegraphics[trim=0.1cm 0.0cm 0.1cm 0.0cm,clip,width=.7\textwidth]{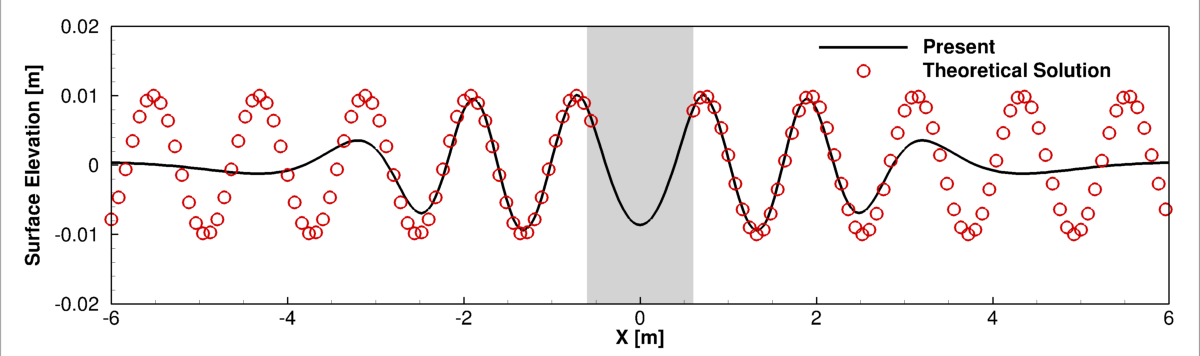}
		\label{subfig::monowavestime_a}
	}
	\subfigure[Time $t=6s$]{
		\includegraphics[trim=0.1cm 0.0cm 0.1cm 0.0cm,clip,width=.7\textwidth]{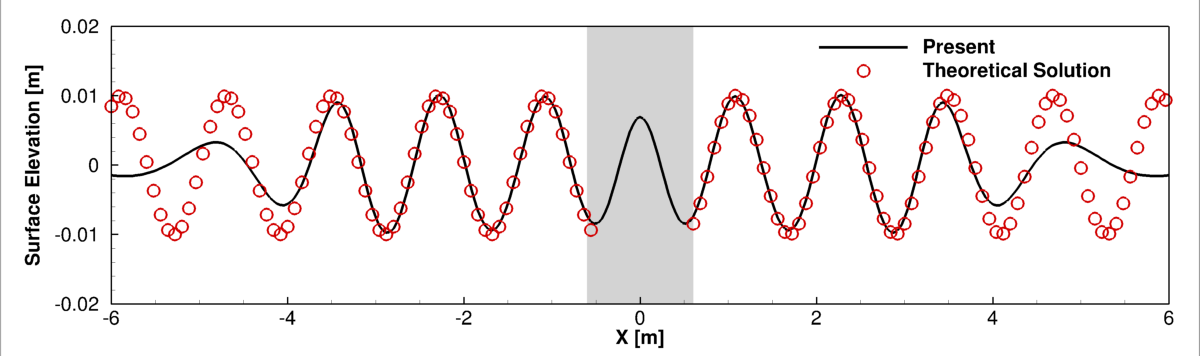}
		\label{subfig::monowavestime_b}
	}
	\subfigure[Time $t=8s$]{
		\includegraphics[trim=0.1cm 0.0cm 0.1cm 0.0cm,clip,width=.7\textwidth]{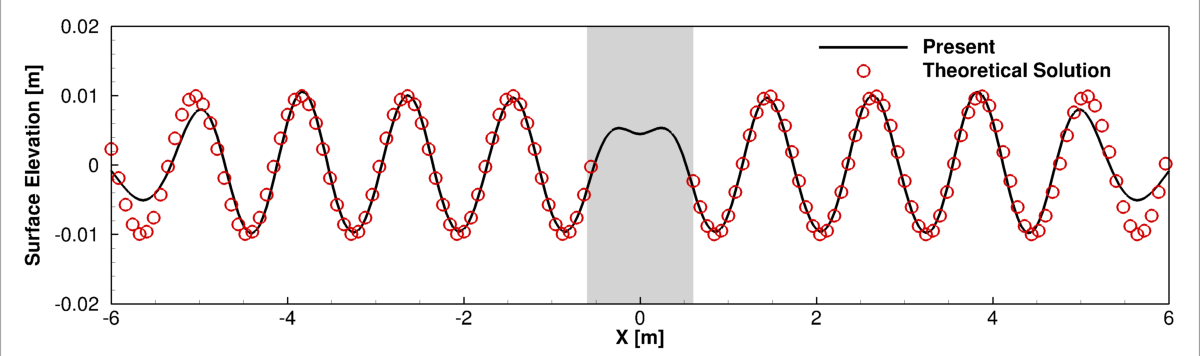}
		\label{subfig::monowavestime_c}
	}
	\caption{Generation of monochromatic waves. Free surface elevation at three instances in time for wave case 2. The time step used is $0.002s$. The grey shaded area represents the source region.}\label{fig:monowavestime}
\end{figure}
\begin{figure}[h!bt]
\centering
\subfigure[$X = 1.92 m$]{
\label{subfig::monowavesvel_a}
\centering
\includegraphics[trim=0.0cm 0.0cm 0.0cm 0.0cm,clip,height=.3\textheight]{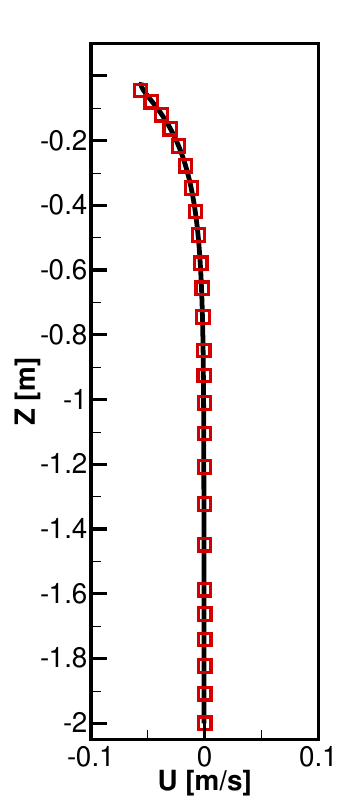}%
\includegraphics[trim=0.0cm 0.0cm 0.0cm 0.0cm,clip,height=.3\textheight]{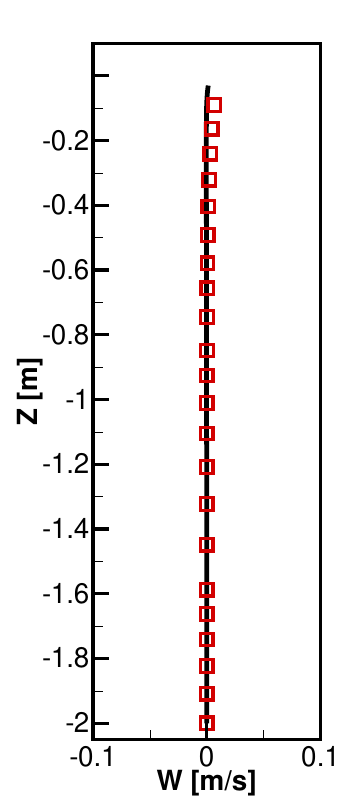}%
}
\subfigure[$X = 2.20 m$]{
\label{subfig::monowavesvel_b}		
\centering
\includegraphics[trim=0.0cm 0.0cm 0.0cm 0.0cm,clip,height=.3\textheight]{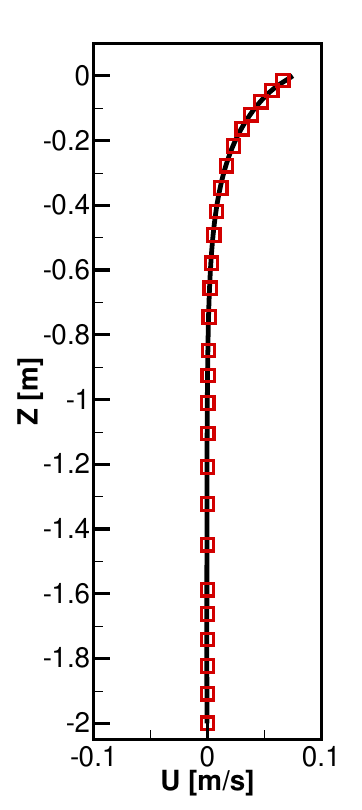}%
\includegraphics[trim=0.0cm 0.0cm 0.0cm 0.0cm,clip,height=.3\textheight]{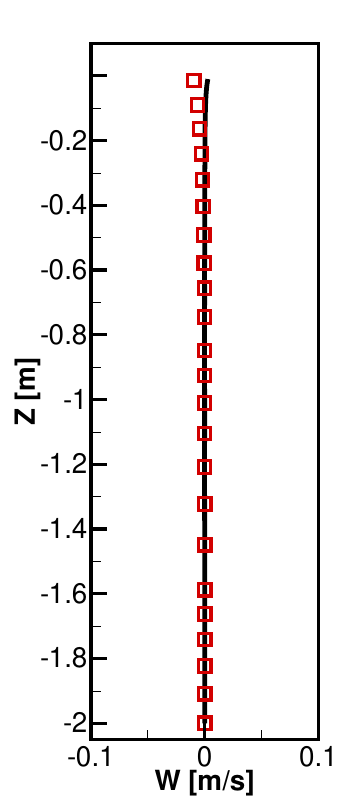}%
}
\subfigure[$X = 2.52 m$]{
\label{subfig::monowavesvel_c}
\centering
\includegraphics[trim=0.0cm 0.0cm 0.0cm 0.0cm,clip,height=.3\textheight]{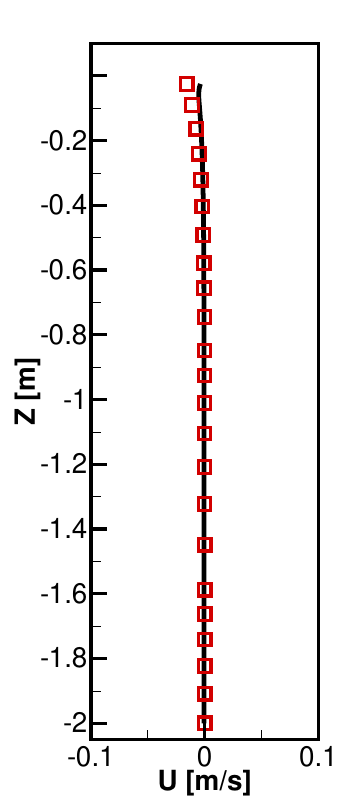}%
\includegraphics[trim=0.0cm 0.0cm 0.0cm 0.0cm,clip,height=.3\textheight]{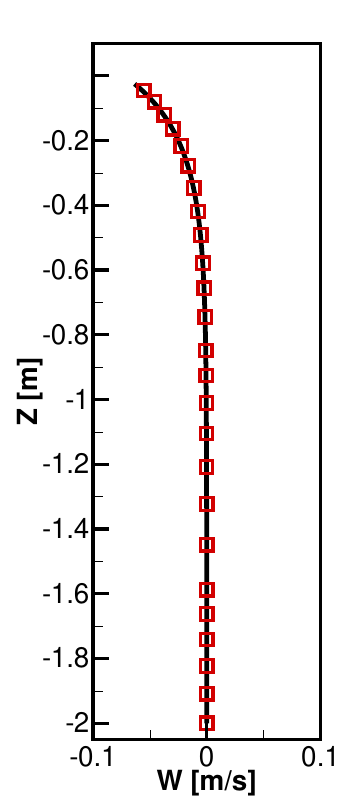}%
}
\subfigure[$X = 2.84 m$]{
\label{subfig::monowavesvel_d}
\centering
\includegraphics[trim=0.0cm 0.0cm 0.0cm 0.0cm,clip,height=.3\textheight]{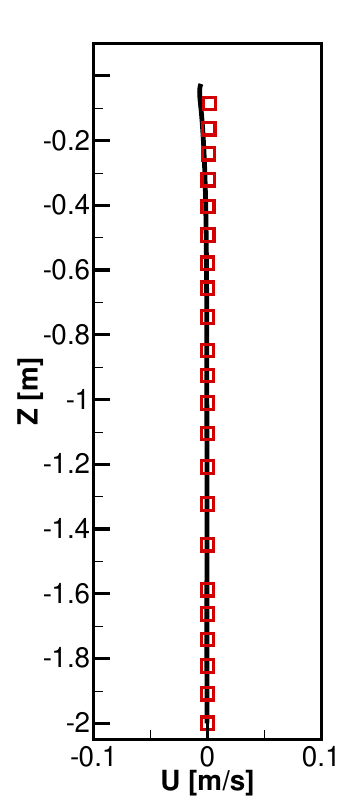}%
\includegraphics[trim=0.0cm 0.0cm 0.0cm 0.0cm,clip,height=.3\textheight]{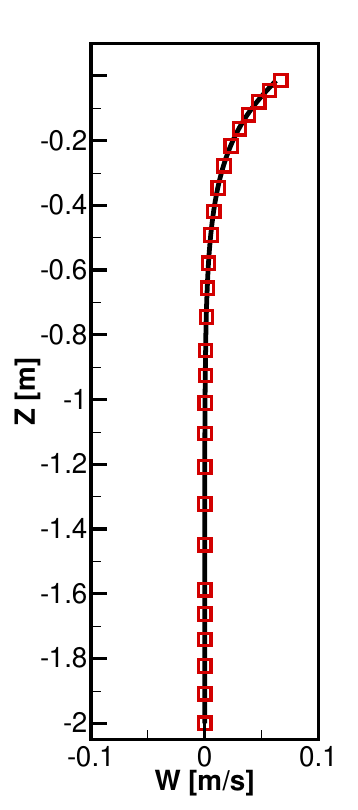}%
}
\subfigure[Position of the wave profiles]{
\label{subfig::monowavesvel_e}
\centering
\includegraphics[trim=0.0cm .5cm 0.0cm .7cm,clip,width=.35\textwidth]{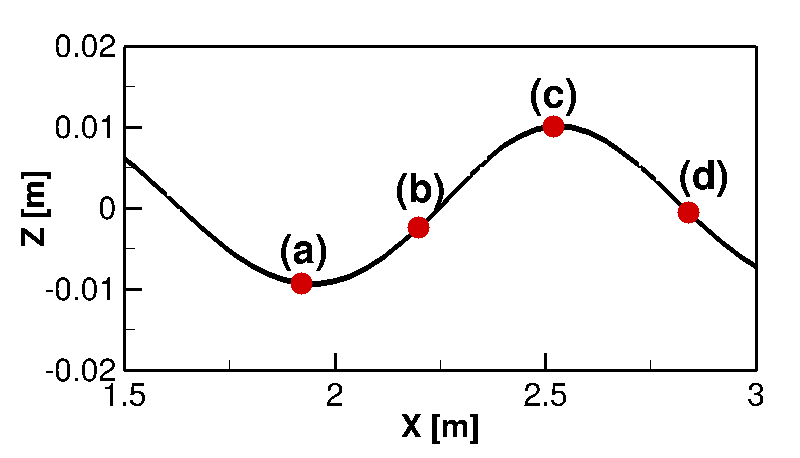}%
}	
\caption{Generation of monochromatic waves. Vertical profiles of horizontal velocity $u$ and vertical velocity $v$ at different streamwise locations for wave case 2 ($L=1.2 m$). Solid lines represent the present computed solution and circles the analytical solition from linear wave theory. The profiles correspond to time $t=14s$ and the time step used is $0.002s$.}\label{fig:monowavesvel}
\end{figure}
\begin{figure}[h!bt]
\centering
\subfigure[Grid refinement in the stream-wise direction]{
\label{subfig:monochrom_refin_a}
\centering
\includegraphics[trim=0.1cm 0.0cm 0.1cm 0.0cm,clip,width=.7\textwidth]{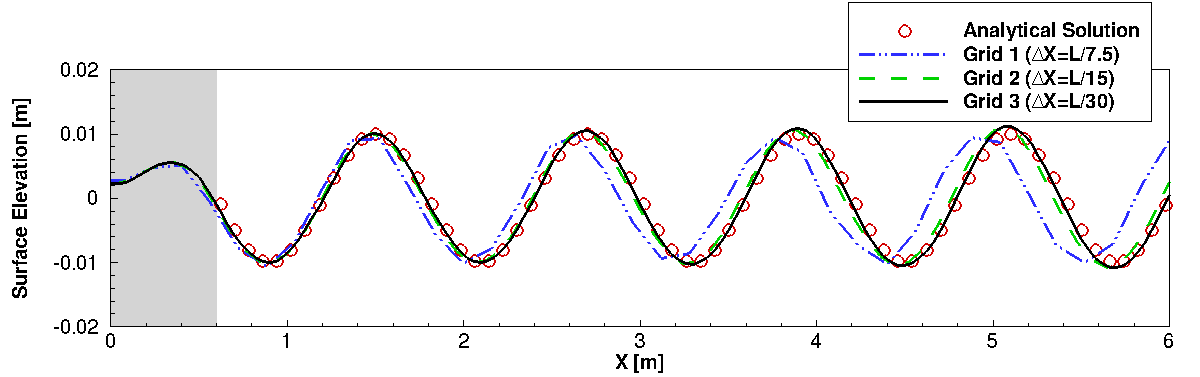}%
}
\subfigure[Grid refinement in the vertical direction]{
\label{subfig:monochrom_refin_b}		
\centering
\includegraphics[trim=0.1cm 0.0cm 0.1cm 0.0cm,clip,width=.7\textwidth]{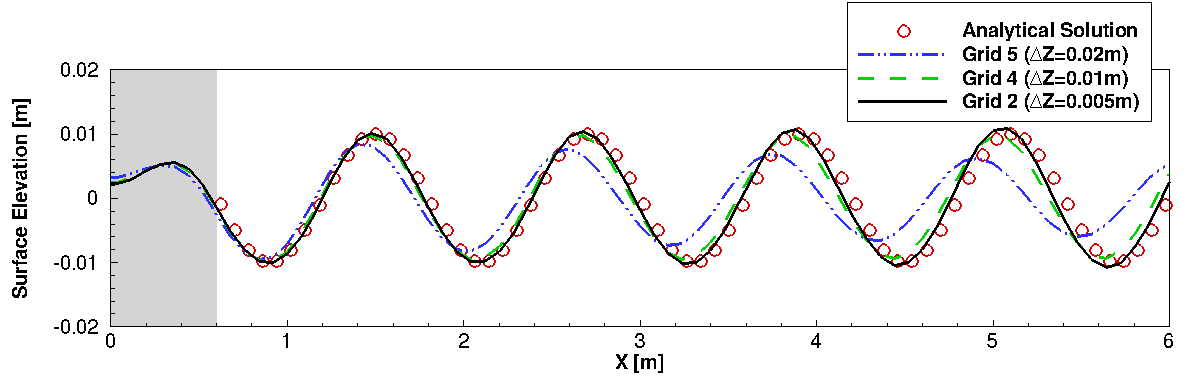}%
}
\caption{Generation of monochromatic waves. Free surface elevation for wave case 2 using different grids. The results correspond to time $t=16 s$ and the time step used is $0.002 s$. The grey shaded area represents the source region.}\label{fig:monochrom_refin}
\end{figure}
\begin{figure}[h!bt]
	\centering
	\subfigure[Wave case 1: $L=0.6 m$]{
		\includegraphics[trim=0.1cm .3cm 0.1cm 0.5cm,clip,width=.7\textwidth]{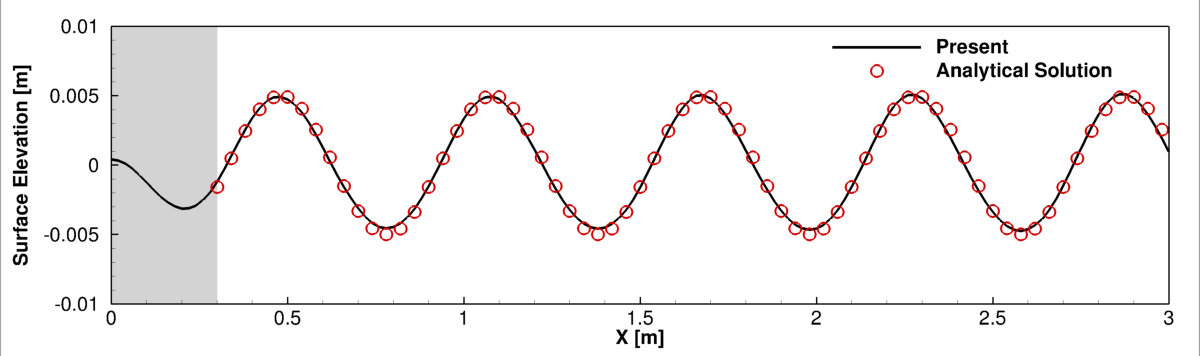}
		\label{subfig::monowavesnum_a}
	}
	\subfigure[Wave case 2: $L=1.2 m$]{
		\includegraphics[trim=0.1cm 0.3cm 0.1cm 0.5cm,clip,width=.7\textwidth]{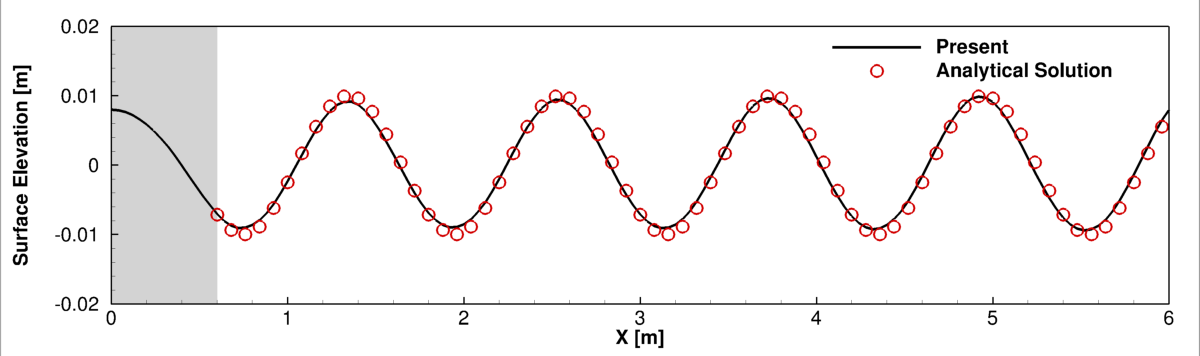}
		\label{subfig::monowavesnum_b}
	}
	\subfigure[Wave case 3: $L=2.4 m$]{
		\includegraphics[trim=0.1cm 0.3cm 0.1cm 0.5cm,clip,width=.7\textwidth]{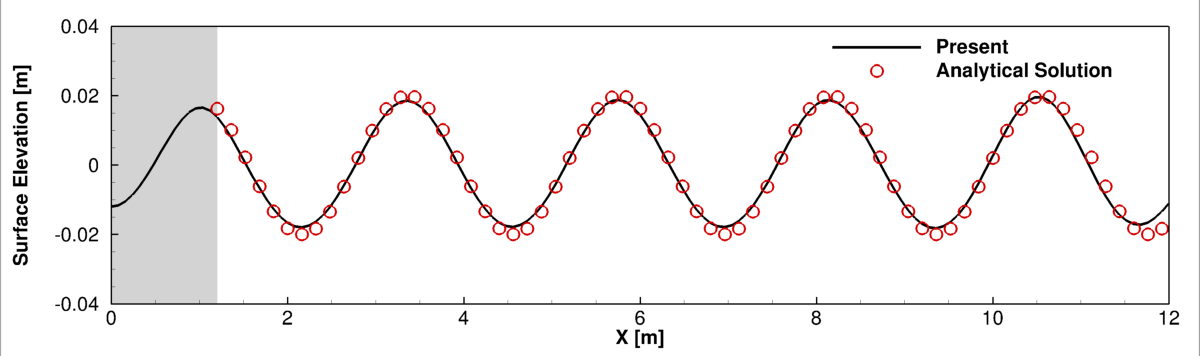}
		\label{subfig::monowavesnum_c}
	}
	\caption{	Generation of monochromatic waves. Computed and analytical free surface elevation of several monochromatic wave cases with different wavenumbers but maintaining a fixed wave slope ($Ak=0.01$). The results correspond to time $t=14s$ and the time step used is $0.002s$.}\label{fig:monowavesnum}
\end{figure}
\begin{figure}[h!bt]
	\centering
	\subfigure[Wave case 4: $Ak=0.01 m$]{
		\includegraphics[trim=0.1cm .3cm 0.1cm 0.5cm,clip,width=.7\textwidth]{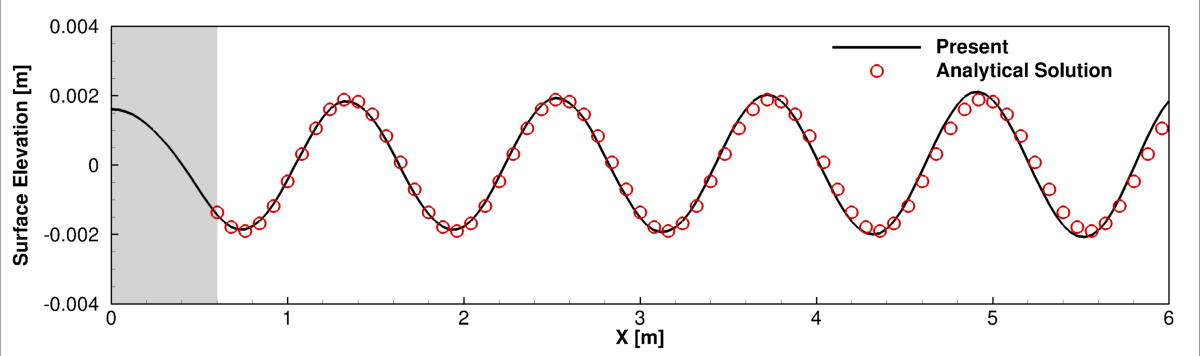}
		\label{subfig::monowavesslope_a}
	}
	\subfigure[Wave case 5: $Ak=0.05 m$]{
		\includegraphics[trim=0.1cm 0.3cm 0.1cm 0.5cm,clip,width=.7\textwidth]{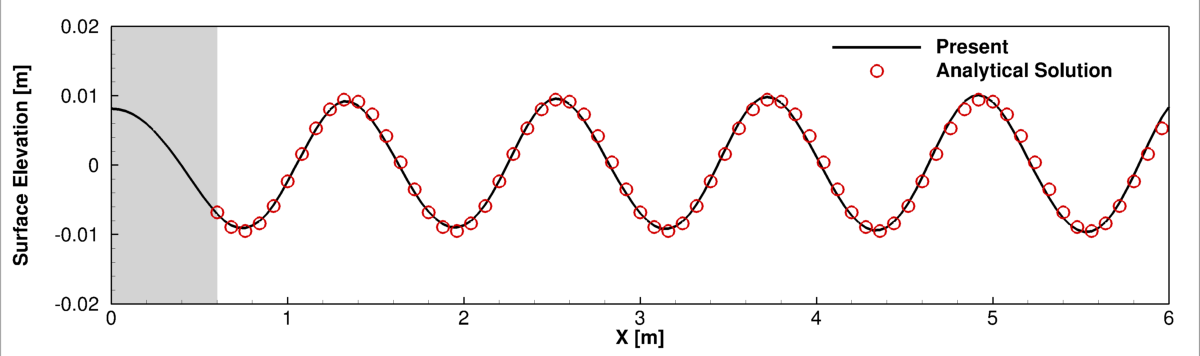}
		\label{subfig::monowavesslope_b}
	}
	\subfigure[Wave case 6: $Ak=0.10 m$]{
		\includegraphics[trim=0.1cm 0.3cm 0.1cm 0.5cm,clip,width=.7\textwidth]{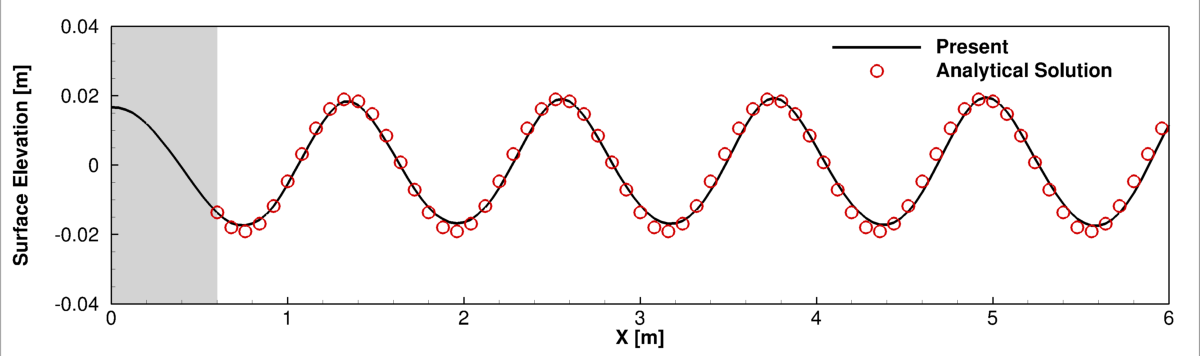}
		\label{subfig::monowavesslope_c}
	}
	\caption{Generation of monochromatic waves. Computed and analytical free surface elevation of several monochromatic wave cases with fixed wavelength ($L=1.2 m$) but different slope. The results correspond to time $t=14s$ and the time step used is $0.002s$.}\label{fig:monowavesslope}
\end{figure}
To validate and demonstrate the accuracy of the wave forcing method described in Section \ref{sec:WaveGeneration} we first consider a simple test case of linear monochromatic waves in a \gls{2d} rectangular channel of constant depth. To analyze the sensitivity of the method when generating waves with different wavelengths and wave slopes, we consider six different wave cases, summarized in Table \ref{tab:monowaves}. In the three first cases, case 1, case 2, and case 3, the wavelength is, respectively, $L=0.6 m$, $L=1.2 m$, and $L=2.4 m$ while keeping a constant wave slope of $A k_x = 0.05$. In the remaining three wave  cases, case 4, case 5, and case 6, the value that is maintained constant is the wavelength $L=1.2 m$ and the slope varies as follows, $A k_x=0.01$ for case 4, $A k_x=0.05$ for case 5, and $Ak_x=0.1$ for case 6. Note that the wave slope, also known as wave steepness, can be defined as the wave amplitude $A$ times the wavenumber $k$. For a given linear wave of amplitude $A$ and length $L$ the analytical solution is known through the linear wave theory provided in equation (\ref{eq:elevation2d}). 

For all of the wave cases, we consider herein a \gls{2d} domain of length equal to $40L$, water depth of $2 m$, air column above the water of height $1 m$, and a gravitational acceleration of $g=9.81m/s^2$. The large domain length was taken with the purpose of preventing, or at least minimizing, possible wave reflections at the side walls that are also treated with the sponge layer method.    

We employ five different non-uniform meshes: grid 1 with $301\times180$, grid 2 with $601\times180$, grid 3 with $1201\times180$, grid 4 with $601\times 140$, and grid 5 with $601\times 100$ nodes in the horizontal and vertical directions, respectively. While the horizontal grid spacing is constant throughout the domain, the vertical spacing is only constant along a rectangular region centered on the undisturbed free surface defined by $z=[-0.1m, 0.1m]$. Within this region the vertical grid spacing is $0.005 m$ for grid 1, grid 2, and grid 3, $0.01 m$ for grid 4, and $0.02m$ for grid 5. Outside of this region the vertical grid spacing increases progressively with a maximum stretching ratio of $1.05$. The horizontal grid spacing is $L/7.5$ for grid 1, $L/15$ for grid 2, grid 4, and grid 5, and $L/30$ for grid 3. The description of the five meshes is summarized in Table \ref{tab:monowaves_grid}. 

The time step of the simulation for all cases is $0.002s$, and the thickness $\epsilon$ of the interface is four times the vertical grid spacing. The source region is centered on the origin, its length $\epsilon_{x}$ is half the wavelength, $L/2$, and its thickness $\epsilon_{\phi}$ is equal to the interface thickness $\epsilon$. The sponge layer method with length equal to $L$ is applied at the two ends of the computational domain. The initial velocity field is zero, and the initial pressure in the air phase is zero. The density and dynamic viscosity are respectively set to $1{,}000kg/m^3$ and $1.0\times10^{-3}Pa s$ for water and  $1.2kg/m^3$ and $1.8\times10^{-5}Pa s$ for air. The free-slip boundary condition is applied at all the four boundaries of the domain.

The free surface elevation for wave case 2 under grid 2 is presented at different instances in time in Fig.\ \ref{fig:monowavestime}, which shows the formation of the wave fields propagating symmetrically with respect to the origin $x=0$. As it is observed in the figure the resulting surface elevations are nearly identical to the theoretical solution, except in the source region and in the wave front region where the simulated results are not expected to follow the analytical free surface pattern (\ref{eq:elevation2d}). For the same wave case (case 2) computed on grid 2, Fig.\ \ref{fig:monowavesvel} presents the velocity profiles at several streamwise locations confirming the accuracy of the free surface forcing method of generating monochromatic waves. The analytical velocity profiles that have been used in the figure for comparison are the following
\begin{equation}\label{eq:wave_vel_theo_u}
u_{lin}(x,z,t)=A \omega\frac{\cosh(k_x h + k_x z)}{\sinh(k_x h)}\cos(k_x x - \omega t) ,
\end{equation}
\begin{equation}\label{eq:wave_vel_theo_v}
v_{lin}(x,z,t)=A \omega\frac{\sinh(k_x h + k_x z)}{\sinh(k_x h)}\sin(k_x x - \omega t) .
\end{equation}

To investigate the grid sensitivity effects in the application of the forcing method, we present in Fig.\ \ref{fig:monochrom_refin} the free surface elevation of wave case 2 using the five aforementioned grids (see Table \ref{tab:monowaves_grid}). In particular, grid 1, grid 2, and grid 3 are successively refined only in the streamwise direction, while grid 5, grid 4, and grid 2 are refined only in the vertical direction. The objective of refining separately along the streamwise and vertical directions is to better analyze: (1) the number of grid points required along a wavelength, and (2) the vertical grid spacing requirement to properly resolve the diffused interface. The same time step of $0.002s$ and interface thickness of $\epsilon=4\Delta y_{min}$ has been used for all grids.
As one would expect, the surface elevation converges monotonically to the analytical solution, either when refining in the streamwise direction (Fig.\ \ref{fig:monochrom_refin}a) or when refining in the vertical direction (Fig.\ \ref{fig:monochrom_refin}b). 
With reference to Fig.\ \ref{fig:monochrom_refin}a,  grid 2 and grid 3, with respectively 15 and 30 cells across the wavelength, can accurately capture the wave frequency and wave amplitude. In contrast, grid 1, with only 7.5 cells across the wavelength, slightly over-predicts the frequency and under-predicts the amplitude. A first conclusion that can be drawn from these results is that, in order to obtain accurate wave fields, the mesh needs to be constructed with a number of nodes per wavelength larger than $7.5$ and ideally approaching $15$. 
Fig.\ \ref{fig:monochrom_refin}b shows that the vertical grid spacing in the vicinity of the interface also has an important role for obtaining accurate results. While grid 2 and grid 4, with a near interface vertical spacing of $0.005 m$ and $0.01 m$, respectively, accurately predict the free surface, grid 5, with spacing of $0.02 m$, fails considerably.  
 
To analyze the performance of the wave forcing method for different wavelengths while maintaining a fixed wave slope, we compare in Fig.\ \ref{fig:monowavesnum} the computed surface elevation resulting from case 1, case 2, and case 3, all employing a grid with $15$ nodes per wavelength (grid 2). As seen in the figure, the free surface agrees very well with the theoretical solution for the three cases.

Finally, we present in Fig.\ \ref{fig:monowavesslope} the surface elevation of case 4, case 5, and case 6, all computed on grid 2, to analyze the ability of the proposed wave maker to generate waves of different slopes. In these three cases, the wavelength has been kept constant to $L=1.2m$, while the wave amplitude has been varied such that the wave slope considered were $Ak=0.01$ for case 4, $Ak=0.05$ for case 5, and $Ak=0.10$ for case 6. As shown in the figure, the computed waves compare well with the linear theoretical solution. Note however that for the high slope case (case 6) the wave troughs are slightly raised compared to the linear solution, due to the fact that the minor non-linear effects start taking part in the simulation.

\subsection{Forcing method validation case: directional waves}
\label{sec:directionalwaves}
\begin{table}[t]
  \caption{Description of the parameters used in the directional wave cases.}\
\centering
\begin{tabular}{c c c c c c}
\hline
\textbf{Wave case} & \textbf{Direction [deg]} & \textbf{$k_x$ [rad/m]} & \textbf{$k_y$ [rad/m]} & \textbf{$\omega$ [rad/s]}\\
\hline
1 & 15 & 5.0576  & 1.3552   &   7.1669 \\
2 & 30 & 4.5345  & 2.6180   &   7.1669 \\
\hline
\end{tabular}
\label{tab:MonochWaves}
\end{table}
\begin{figure}[h!bt]
\centering
\subfigure[Free surface profile at plane $y=0m$]{%
\label{subfig:dirwave_case1_a}%
\includegraphics[trim=0.1cm 0.0cm 0.0cm 0.0cm,clip,width=0.765 \textwidth]{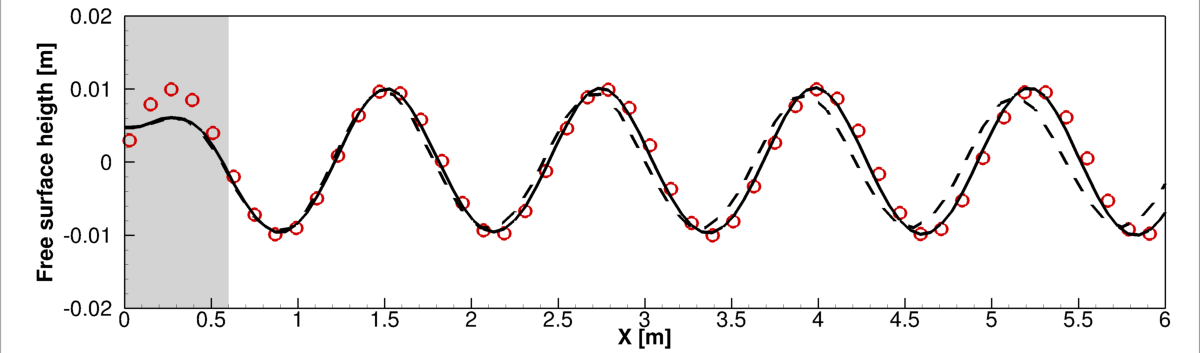}%
\includegraphics[trim=0.3cm 0.3cm 0.0cm 0.0cm,clip,width=0.232 \textwidth]{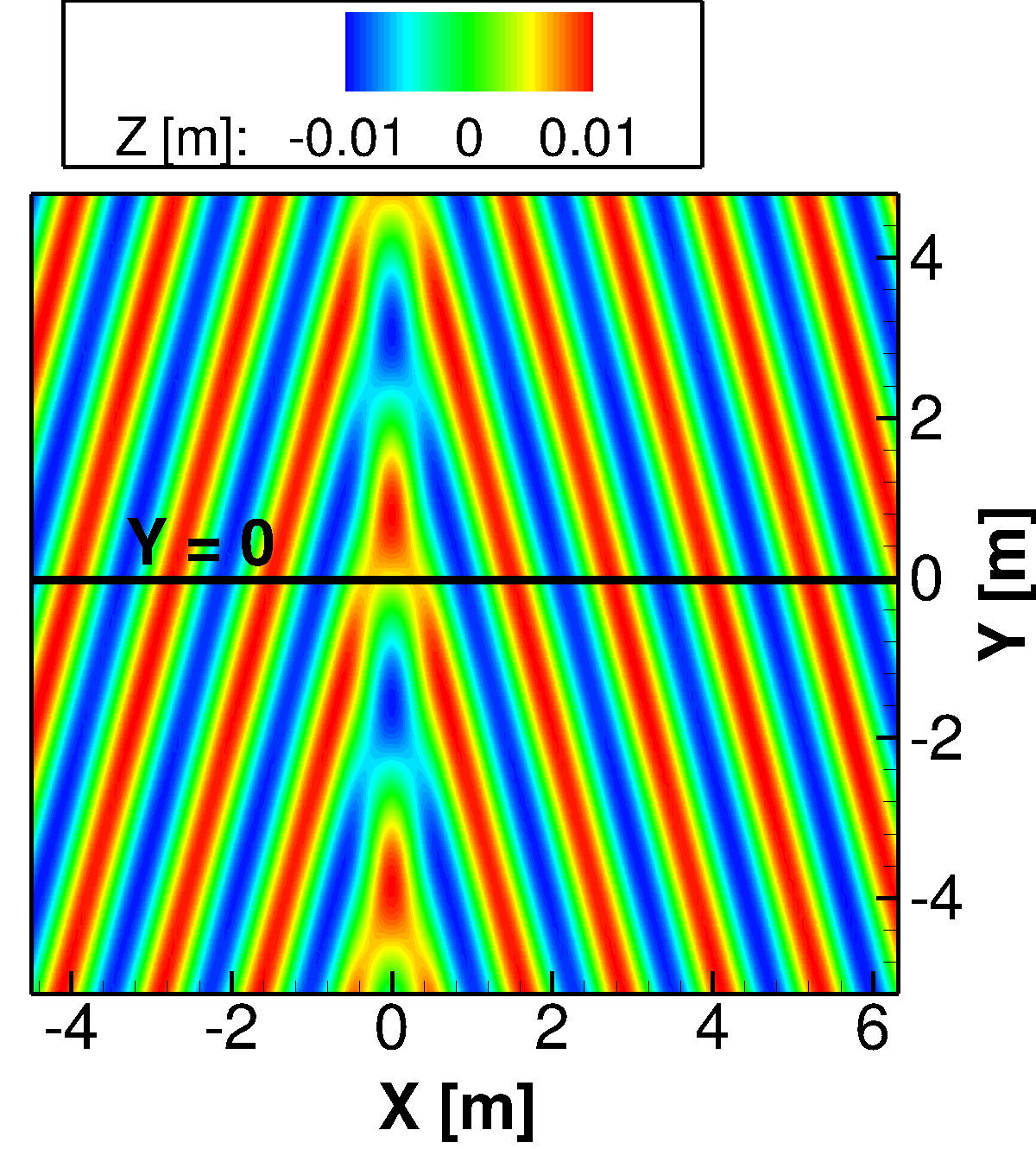}%
}
\subfigure[Free surface profile at plane $x=1.5 m$]{%
\label{subfig:dirwave_case1_b}%
\includegraphics[trim=0.1cm 0.0cm 0.0cm 0.0cm,clip,width=0.765 \textwidth]{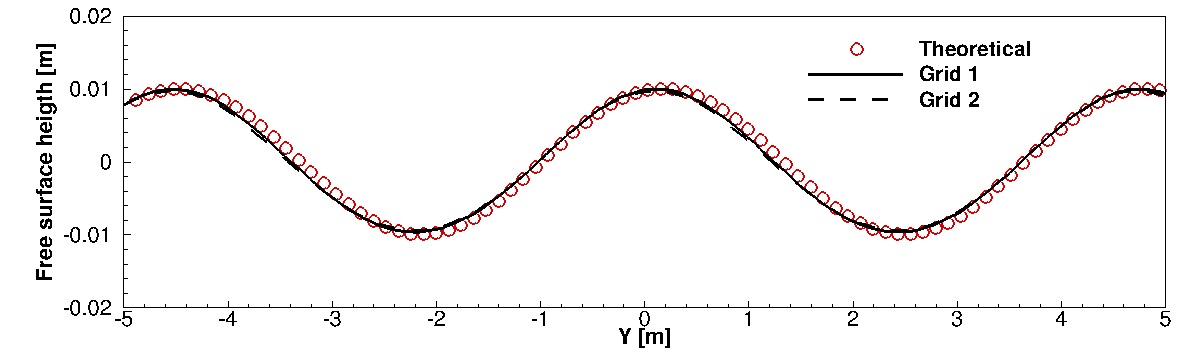}%
\includegraphics[trim=0.3cm 0.3cm 0.0cm 0cm,clip,width=0.232 \textwidth]{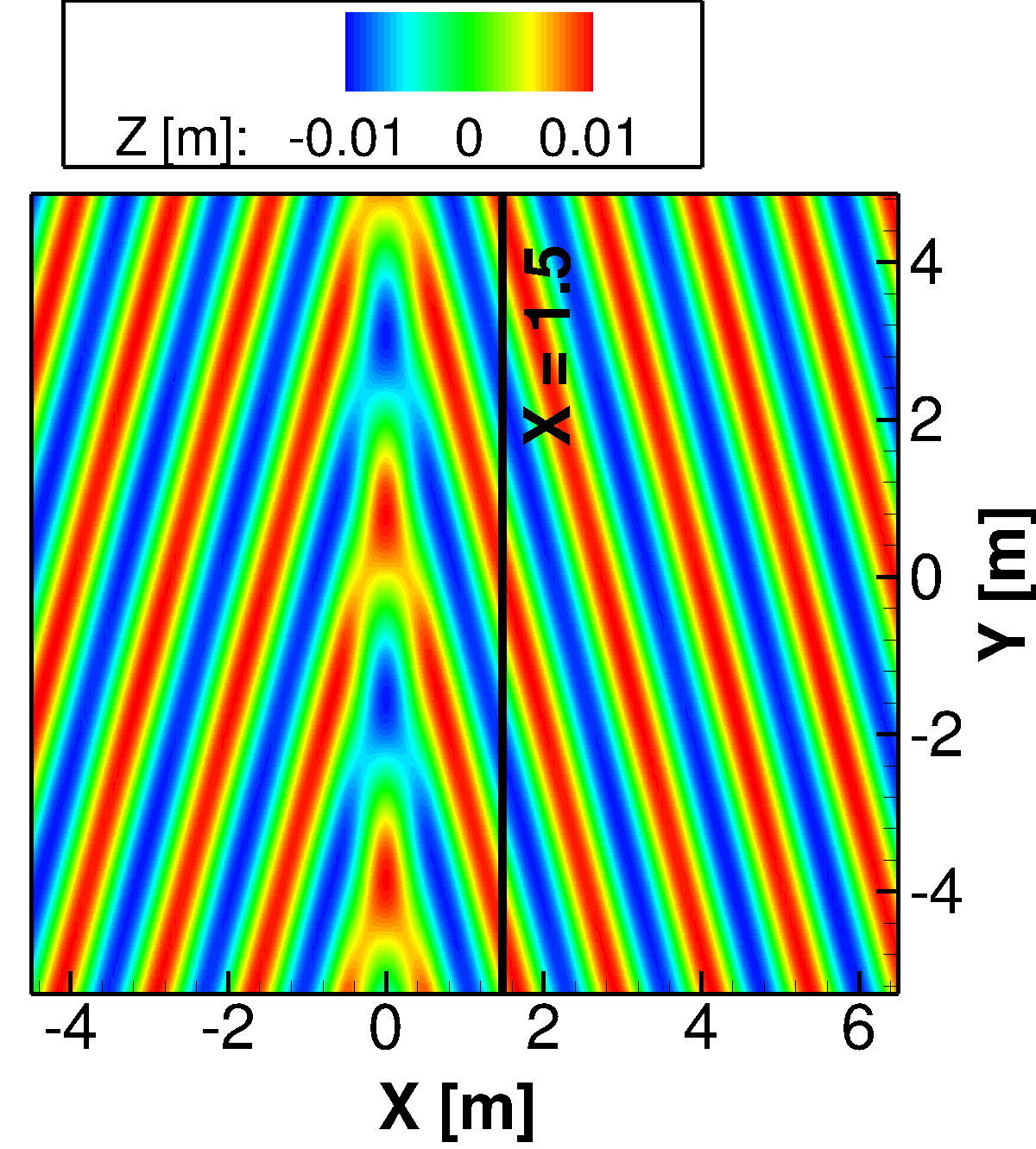}%
}%
\caption{Generation of \gls{3d} directional waves. The left figures show the free surface elevation profiles computed on grid 1 and grid 2 and the theoretical solution from linear wave theory. The right figures show the surface elevation contours (in meters) with a horizontal line in (a) and a vertical line in (b) to indicate the positions of the planes shown in the left figures. The results correspond to time $t=14 s$ and the time step used in the simulation is $0.001 s$. The grey shaded area represents the source region.}
\label{fig:dirwave_case1}
\end{figure}
\begin{figure}[h!bt]
\centering
\subfigure[Free surface profile at plane $y=0m$]{%
\includegraphics[trim=0.0cm 0.0cm 0.0cm 0.0cm,clip,width=0.772\textwidth]{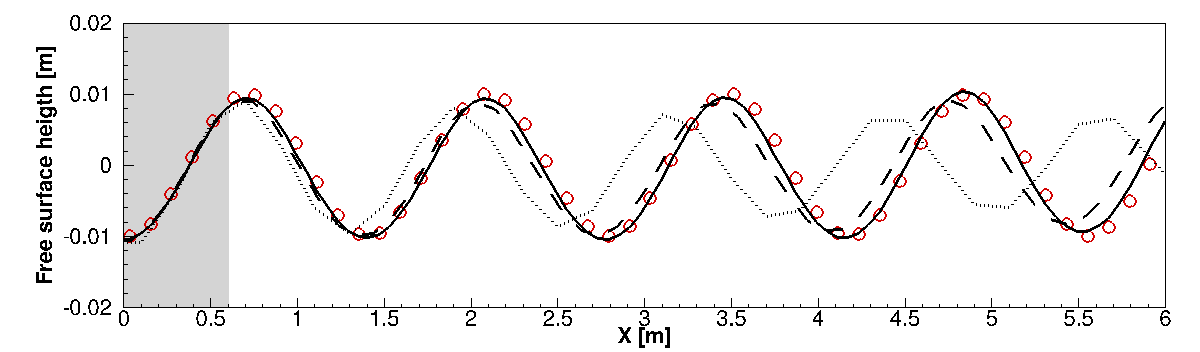}%
\includegraphics[trim=0.0cm 0.0cm 0.0cm 0.0cm,clip,width=0.230 \textwidth]{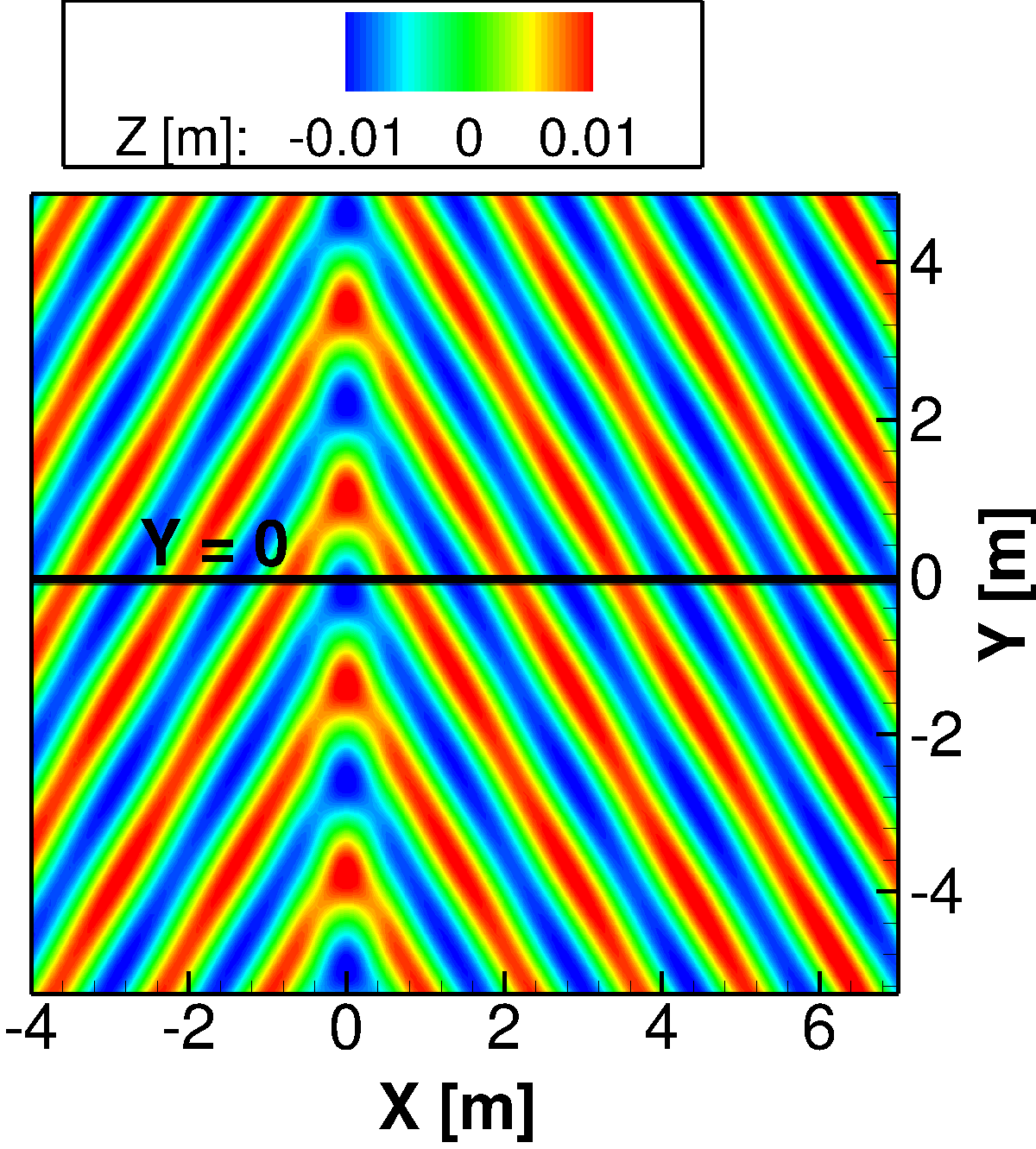}%
\label{subfig:dirwave_case2_a}%
}
\subfigure[Free surface profile at plane $x=1.5 m$]{%
\includegraphics[trim=0.0cm 0.0cm 0.0cm 0.0cm,clip,width=0.772\textwidth]{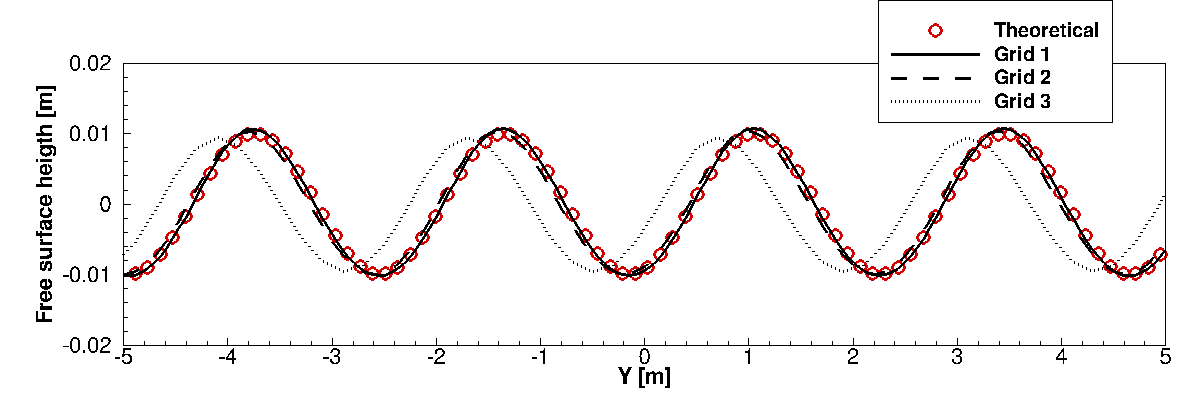}%
\includegraphics[trim=0.0cm 0.0cm 0.0cm 0.0cm,clip,width=0.230 \textwidth]{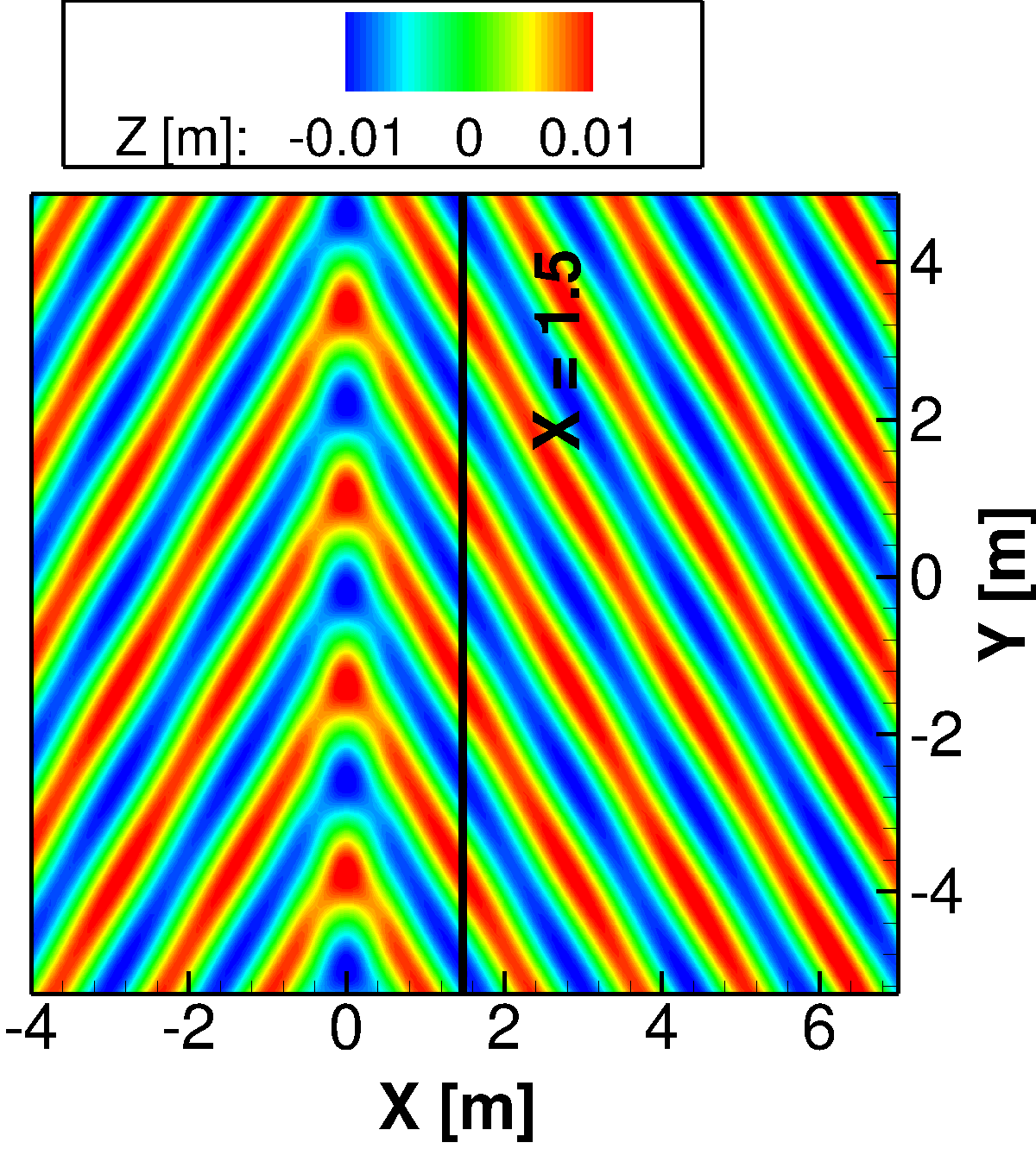}%
\label{subfig:dirwave_case2_b}%
}
\caption{Generation of \gls{3d} directional waves. The left figures show the free surface elevation profiles computed on grid 1, grid 2, and grid 3 and the theoretical solution from linear wave theory. The right figures show the surface elevation contours (in meters) with a horizontal line in (a) and a vertical line in (b) to indicate the positions of the planes shown in the left figures. The results correspond to time $t=16 s$ and the time step used in the simulation is $0.001 s$. The grey shaded area represents the source region.}\label{fig:dirwave_case2}
\end{figure}

In this section we validate the forcing method for wave generation by simulating, in a \gls{3d} basin of constant depth, two cases involving a linear directional wave field. The wave we consider in case 1 has an amplitude of $A=0.01 m$, a wavelength of $L=1.2 m$, and a propagating direction $\beta=15^{\circ}$, with the direction $\beta$ defined as the angle between the wave propagation direction and the $x$ axis. The wave in case 2 has the same amplitude and wavelength as in case 1, but a propagating direction $\beta=30^{\circ}$. 

The domain length for case 1 is $24 m$ ($x=[-12 m, 12 m]$) in the longitudinal direction and $13.91m$ ($y=[-6.96 m,\allowbreak 6.96 m]$) in the span-wise direction, and the depths of water and air are $2 m$ and $1 m$, respectively. For case 2, the only difference in the domain size with respect to case 1 is the span-wise dimension taken as $12m$ ($y=[-6 m, 6 m]$), due to the use of periodic boundary conditions, which requires sufficient space to accommodate an integer number of wavelengths.  

Three non-uniform Cartesian grids with successively refined resolution are considered to perform a mesh convergence study. The three meshes follow a common pattern consisting of an inner rectangular region with uniform grid spacing and an outer region within which the mesh is gradually stretched towards the boundaries using a hyperbolic function and limiting the stretching ratio to $1.05$. The inner region, which contains the source region, is the same for the three meshes and spans $x=[-6 m, 6m]$ in the stream-wise direction, the whole length in the span-wise direction, and $z=[-0.1 m, 0.1 m]$ in the vertical direction. The grid spacing in the inner region for grid 1, which is the finest of the three meshes, is $\Delta x =\Delta y=0.06 m$ and $\Delta z=0.005 m$, for grid 2 is $\Delta x =\Delta y=0.1 m$ and $\Delta z=0.01 m$, and for grid 3 is $\Delta x =\Delta y=0.2 m$ and $\Delta z=0.02 m$. The total number of nodes per each mesh is $301\times201 (233)\times199$, $201\times121 (141)\times139$, $109\times61\times92$, for grids 1 to 3, respectively. The numbers in the parentheses in the span-wise direction refer to the numbers of grid nodes for case 1, which slightly differs from the value in case 2 as mentioned above. 

The source region is centered at $x=0$ with horizontal thickness of $\epsilon_{x}=L/2=0.6m$. Its vertical thickness is equal to the thickness of the smoothed free surface interface $\epsilon_{\phi}=\epsilon$, which is taken as four times the vertical grid spacing. That is  $\epsilon=0.02m$ for grid 1, $\epsilon=0.04m$ for grid 2, and $\epsilon=0.08m$ for grid 3. The time step used is $0.001s$ and the gravitational acceleration is $g=9.81m/s^2$. The sponge layer method with length equal to $1.2m$ is applied at the stream-wise boundaries, periodic boundary conditions are applied in the span-wise direction, and the free-slip boundary condition is applied at the top and bottom walls. The simulation is started with an undisturbed free surface and a zero velocity field, pressure field of zero in the air phase, and hydrostatic pressure at the water phase. The density and dynamic viscosity are the same as in the previous wave cases, $1{,}000kg/m^3$ and $1.0\cdot10^{-3}Pa s$, respectively, for water and  $1.2kg/m^3$ and $1.8\cdot10^{-5}Pa s$ for air. 

The results of the free surface elevation with the grid sensitivity study and the comparison with the analytical solution are presented in Fig.\ \ref{fig:dirwave_case1} for case 1 and in Fig.\ \ref{fig:dirwave_case2} for case 2. The two figures show that the proposed wave generation method can successfully simulate \gls{3d} directional waves of different propagation angle $\beta$ with results converging monotonically to the expected theoretical solution as the grid resolution is refined. Analyzing the effect of the grid resolution on the accuracy of the computed waves, we observe that while grids 1 and 2 can represent the wave with an accurate amplitude and frequency, specifically grid 1 provides excellent accuracy, grid 3 fails to provide reasonable results. 
To see the number of grid nodes per wavelength in a \gls{3d} directional wave case, we first need to compute the projected wavelength to the $x$ and $y$ axes, which is done with the following expressions: $L_x=2\pi/k_x$ and $L_y=2\pi/k_y$. In case 1, $L_x=1.25 m$ and $L_y=4.6 m$, which respectively correspond to $25$ and $92$ grid nodes for grid 1 and $12.4$ and $46$ grid nodes for grid 2. In case 2, $L_x=1.39 m$ and $L_y=2.4 m$ and the number of grid nodes per wavelength is respectively $28$ and $48$ for grid 1, $14$ and $24$ for grid 2, and $7$ and $12$ for grid 3. Looking at the $x$ direction values, which are the critical ones, the minimum number of grid nodes per wavelength to get accurate free surface results is approximately 12, corresponding to grid 2. The 7 nodes used for grid 3 are clearly insufficient. The above grid resolution requirement is in line with those discussed in the previous section for monochromatic waves.

\subsection{Far-field/near-field coupling validation cases}

In this section we present three wave cases aimed to validate the far-field/near-field wave coupling algorithm described in Section \ref{sec:WaveGeneration}. We also want to demonstrate the ability of the forcing method to generate complex wave fields composed of various superposed frequencies. 

In all of the three cases we present, the simulation is started in the far-field domain with the HOS code, by setting the initial velocity potential and free surface elevation to that of the given wave case at time zero. A fast Fourier transform of the free surface elevation is then applied at every time step to extract the wave frequencies and amplitudes, which are then incorporated to the near-field solver with the proposed surface forcing method. After the simulation is advanced for sufficient time of a few wave periods, the wave field of the near-field simulation in the region comprised between the source region ($x>\epsilon_x$) and the sponge layer is expected to match the wave field from the overlapped sub-region of the far-field simulation. In the present cases we do not consider any mean airflow in order to reduce the complexity of the problem.

\subsubsection{Simulation of \gls{2d} superposed monochromatic waves}
\begin{table}[t]
  \caption{Description of the three wave components in the simulation of \gls{2d} superposed monochromatic waves case}\
\centering
\begin{tabular}{c c c c c}
\hline
\textbf{Wave component} & \textbf{$L$ [m]} & \textbf{$k_x$ [rad/m]} & \textbf{$A$ [m]} & \textbf{$\omega$ [rad/s]}\\
\hline
1   &   12.57  &   0.5  & 0.040  & 2.24\\
2   &   6.28   &   1.0  & 0.010  & 3.16\\
3   &   3.14   &   2.0  & 0.005 & 4.47\\
\hline
\end{tabular}
\label{tab:SPosed_monoch}
\end{table}
\begin{figure}[h!bt]
	\centering
	\includegraphics[trim=0.0cm 0.0cm 0.0cm 0.0cm,clip,width=1.\textwidth]{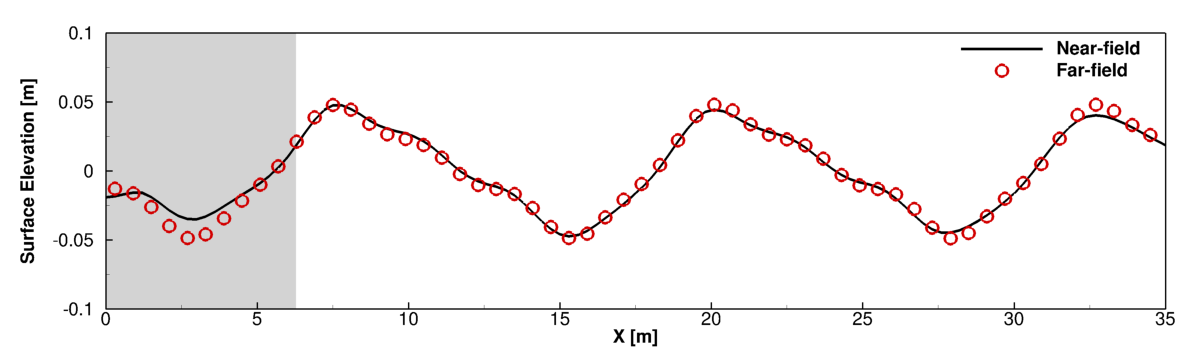}
	\caption{Far-field/near-field coupling case 2: simulation of superposed monochromatic waves. Comparison of the computed free surface elevation in the near-field domain between using the far-field/near-field coupling and letting the far-field solver to continue to carry out the simulation.
}\label{fig:coupling_case1}
\end{figure}
In this test case, a set of three superposed monochromatic waves are simulated in a \gls{2d} rectangular channel. The three wave components have the amplitudes of $A_1=0.040m$, $A_2=0.009m$, and $A_3=0.005m$, respectively, and wavenumbers of $k_{x1}=0.5rad/m$, $k_{x2}=1.0rad/m$, and $k_{x3}=2.0rad/m$ (as summarized in Table \ref{tab:SPosed_monoch}).

The computational domain is a \gls{2d} rectangular channel of length $120 m$ ($x=[-30 m ,90 m]$), the water depth is $10 m$, and the air domain height is $1 m$. A grid of size $300\times 200$ is constructed following the same dual inner/outer region pattern described in Section \ref{sec:monoch} for the monochromatic wave case.  The region of uniform spacing is defined respectively in the stream-wise and vertical directions as $x=[-7 m , 60 m ]$ and $z=[-0.1m ,0.1m ]$, and the corresponding grid spacing is $0.3m$ and $0.005m$, respectively. The time step is $0.00125s$ and the gravity is $g=10m/s^2$. Free-slip condition is considered at all the four boundaries in combination with a sponge layer of length $9 m $ applied at the side boundaries.   

The computed free surface elevation in the near field resulting from the far-field/near-field coupling process at time $t=30s$ is presented in Fig.\ \ref{fig:coupling_case1}. As shown in the figure, the surface elevation generated in the near-field domain, in the area beyond the source region, agrees very well with the elevation computed in the corresponding region with the far-field solver when it continues the simulation alone. With the stream-wise spacing of $0.3 m$ that we used in the simulation, the first wave component ($L=12.57m$) is resolved with 42 grid nodes, the second component ($L=4.44m$) with 15 grid nodes, and the third component ($L=2.51m$) with 8.4 grid nodes. Using the conclusions obtained from the grid sensitivity analysis in Section \ref{sec:monoch}, we can say that the number of cells per wavelength used for wave components 1 and 2 is sufficient for resolving them with high accuracy. However, for wave component 3 the number of cells per wavelenth of 8.4 is low, which may explain the minor discrepancies that can be seen in the figure.    

\subsubsection{Simulation of three \gls{3d} superposed directional waves}
\label{sec:FF-NF_coupling_case2}
\begin{table}[t]
  \caption{Description of the three wave components in the simulation of \gls{3d} superposed directional waves case}\
\centering
\begin{tabular}{c c c c c c c c}
\hline
\textbf{Wave component} & \textbf{$L$ [m]} & \textbf{$k_x$ [rad/m]}&\textbf{$k_y$ [rad/m]} & \textbf{$L_x$ [m]} & \textbf{$L_y$ [m]} & \textbf{$A$ [m]} & \textbf{$\omega$ [rad/s]}\\
\hline
1   &   12.57  &   0.5 & 0.0 & 12.57 & 0 & 0.040  & 2.24\\
2   &   4.44   &   1.0 & 1.0 & 6.28 & 6.28 & 0.014  & 3.76\\
3   &   2.51   &   2.0 & 1.5 & 3.14 & 4.20 & 0.008 & 5.00\\
\hline
\end{tabular}
\label{tab:coupling_case2}
\end{table}
\begin{figure}[h!bt]
	\centering
	\subfigure[Section $Y=2$]{%
		\includegraphics[trim=0.0cm 0.0cm 0.0cm 0.0cm,clip,width=0.77311\textwidth]{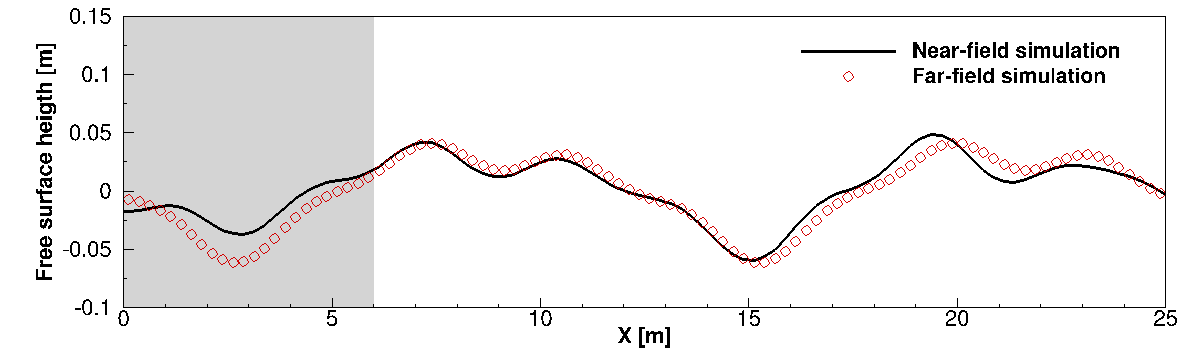}%
		\includegraphics[trim=0.0cm 0.0cm 0.0cm 0.0cm,clip,width=0.234\textwidth]{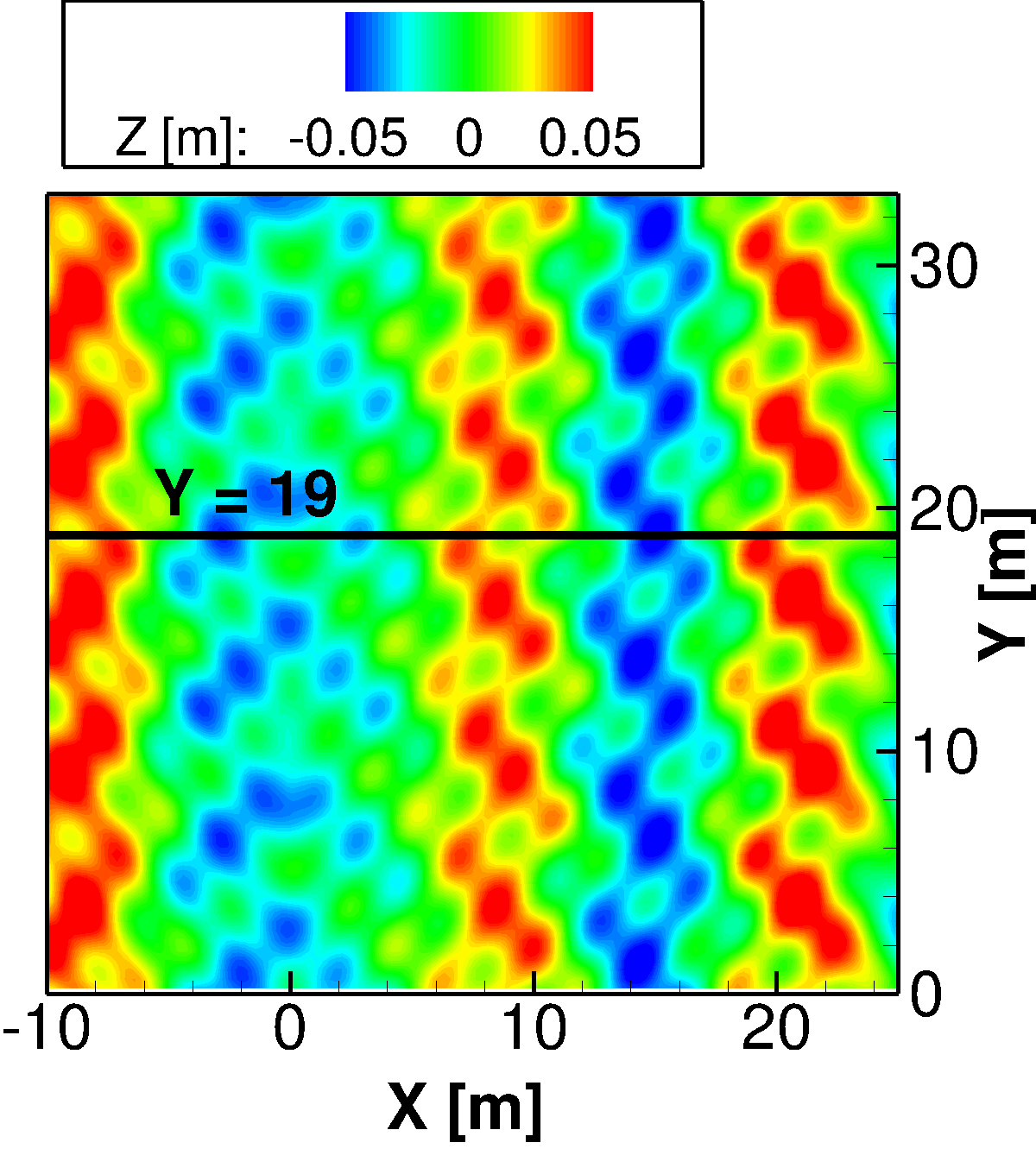}%
		\label{subfig:coupling_case2_a}%
	}
	\subfigure[Section $X=8$]{%
    \includegraphics[trim=0.0cm 0.0cm 0.0cm 0.0cm,clip,width=0.77311\textwidth]{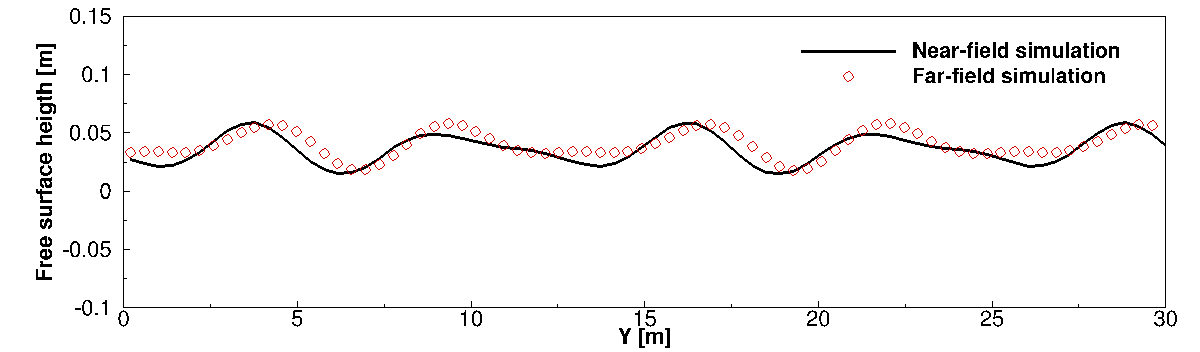}%
    \includegraphics[trim=0.0cm 0.0cm 0.0cm 0.0cm,clip,width=0.234\textwidth]{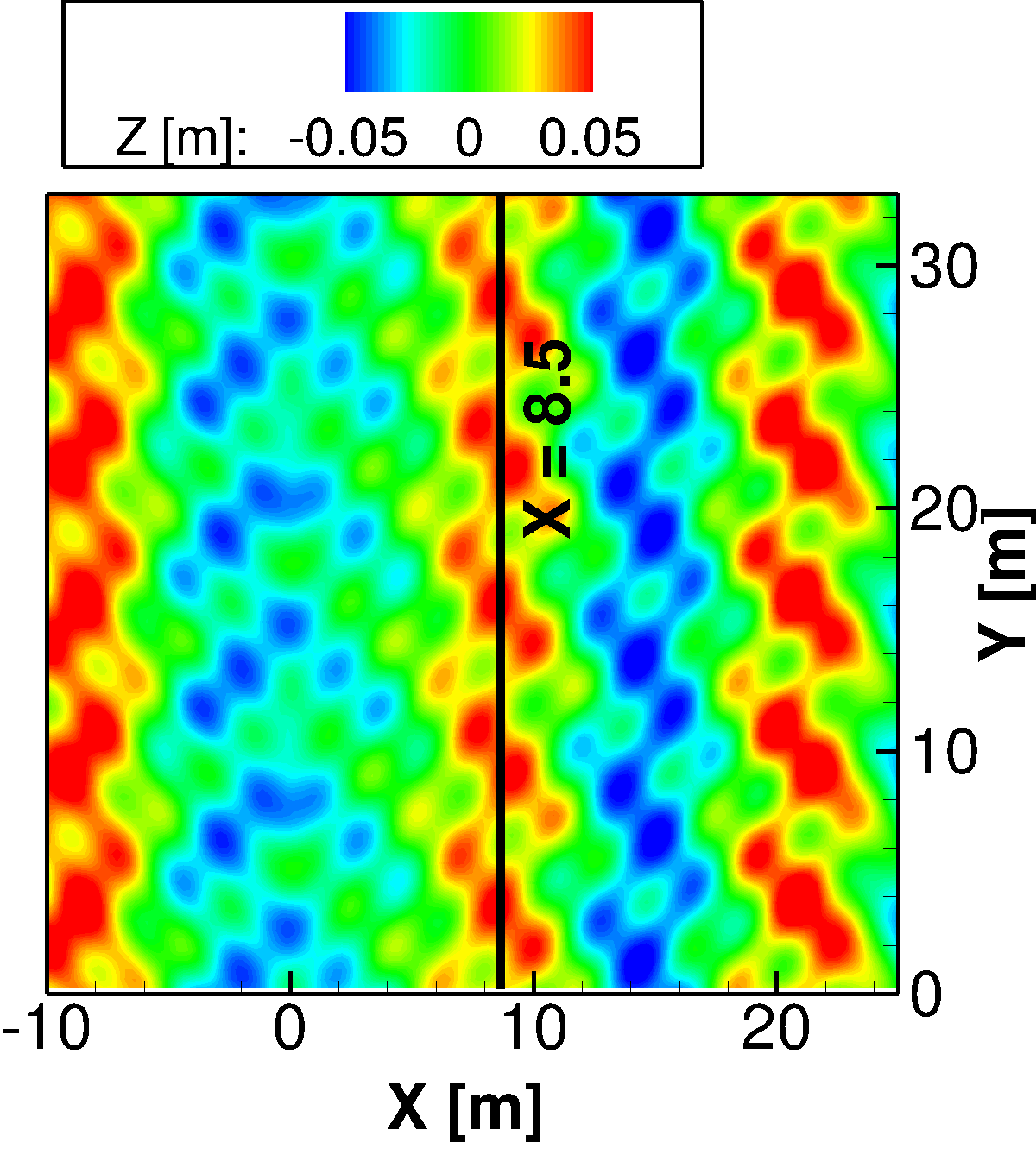}%
		\label{subfig:coupling_case2_b}%
	}
	\caption{Far-field/near-field coupling case 2: simulation of three superposed directional waves. The left figures show computed free surface elevation profiles in the near-field domain between using the far-field/near-field coupling and letting the far-field solver to continue carry out the simulation. The right figures show the surface elevation contours (in meters) computed with the near-field solver with a horizontal line in (a) and a vertical line in (b) to indicate the position of the planes. The results correspond to time $t=30 s$ and the time step used in the simulation is $0.0025 s$. The grey shaded area represents the source region.}\label{fig:coupling_case2}
\end{figure}
In this test case, a wave field composed of three directional wave components are simulated in a \gls{3d} rectangular basin. As already mentioned, the wave field is initialized in the far-field domain and transferred with the pressure forcing method to the near-field solver by following the coupling algorithm described in Section \ref{sec:WaveGeneration}. 

The three wave components of the present case have an amplitude of $A_1=0.040m$, $A_2=0.014m$, and $A_3=0.008m$, a wavenumber in the $x$ direction of $k_{x1}=0.5rad/m$, $k_{x2}=1.0rad/m$, and $k_{x3}=2.0rad/m$, and a wavenumber in the $y$ direction of $k_{y1}=0.0rad/m$, $k_{y2}=1.0rad/m$, and $k_{y3}=1.5rad/m$, respectively. A summary of the three wave component parameters including wavelength and wave frequency is presented in Table \ref{tab:coupling_case2}.

The near-field computational domain is a rectangular basin of length $93.5 m$ ($x=[-31.16m, 62.34m]$) in  the stream-wise direction, 31.16m ($y=[0 m , 31.16 m ]$) in the span-wise direction, water depth of $10 m$, and air column height of $1m$. 
The grid is non-uniform with the same structure used in the directional wave cases in Section \ref{sec:directionalwaves}. The region with uniform spacing is the following: $x=[-7.5m, 30m ]$, $y=[0m, 31.16m]$, and $z=[-0.2m,0.2m]$, and the grid spacing within this area is, respectively, $0.4m$, $0.25m$, and $0.02m$. The outer region is stretched using a hyperbolic function with ratios limited to $1.05$. The time step is $0.0025s$ and the gravity $g=10m/s^2$. The two simulations, far-field and near-field, are run for $30 s$. 

The results of the surface elevation at time $t=30s$ from both the near-field simulation using the far-field/near-field coupling and the corresponding region of the far-field simulation alone are presented in Fig.\ \ref{fig:coupling_case2}. As shown in the figure, the computed near-field solution agrees well with the far-field results. In this case, the wave component that is resolved with the least number of nodes per wavelength is wave component 3 ($L=2.51$) with 12 nodes per $L_x$ and 10.5 nodes per $L_y$.

\subsubsection{Simulation of a broadband wave spectrum}
\begin{figure}[h!bt]
	\centering
	\includegraphics[trim=0.0cm 6.5cm 0.0cm 5.0cm,clip,width=.6\textwidth]{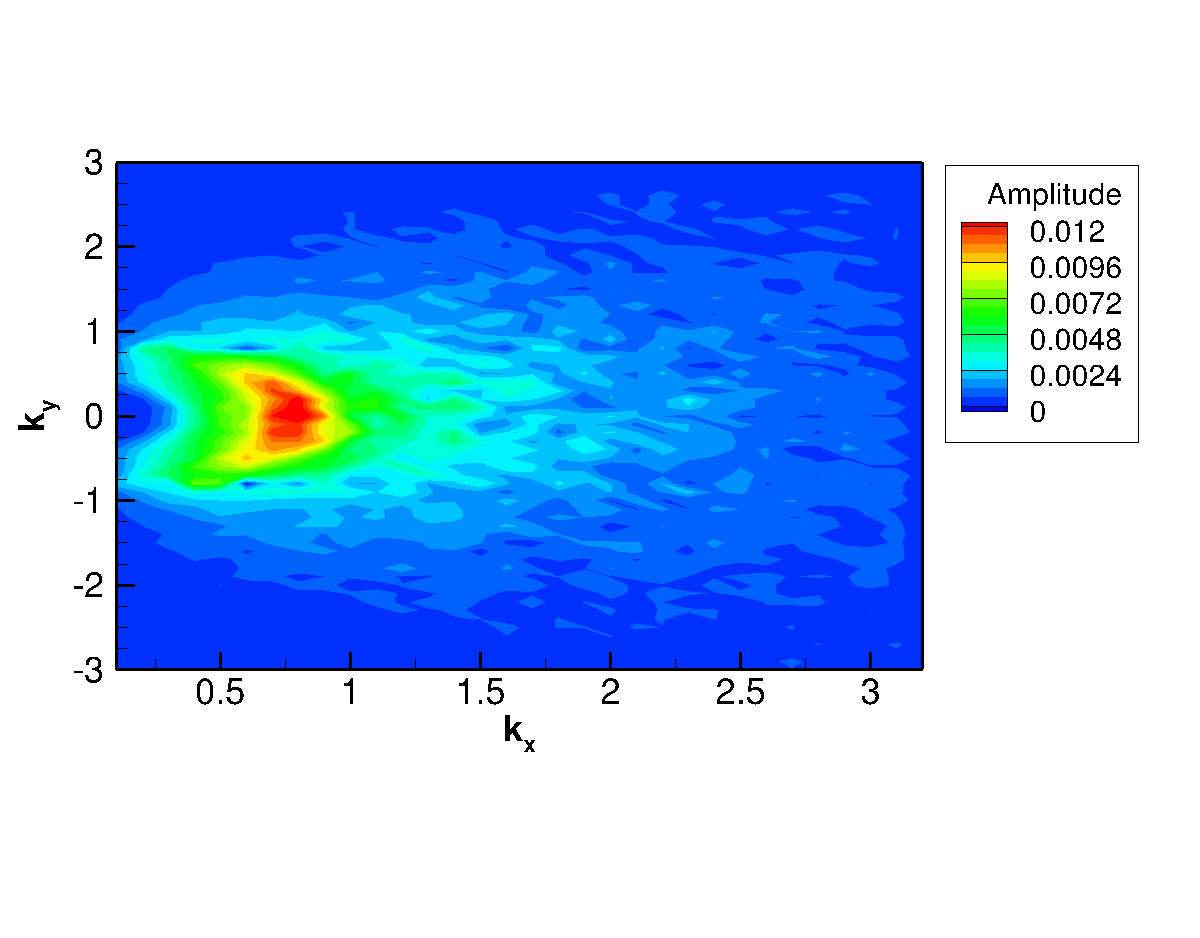}
	\caption{Far-field/near-field coupling case 3: simulation of a broadband spectrum of waves. Plotted is the \gls{2d} wave spectrum obtained from the far-field simulation at time $t=30s$. 
}\label{fig:coupling_case3_freq}
\end{figure}
\begin{figure}[h!bt]
	\centering
	\subfigure[Near-field simulation]{%
		\includegraphics[trim=0.0cm 0.0cm 0.0cm 0.0cm,clip,width=0.49\textwidth]{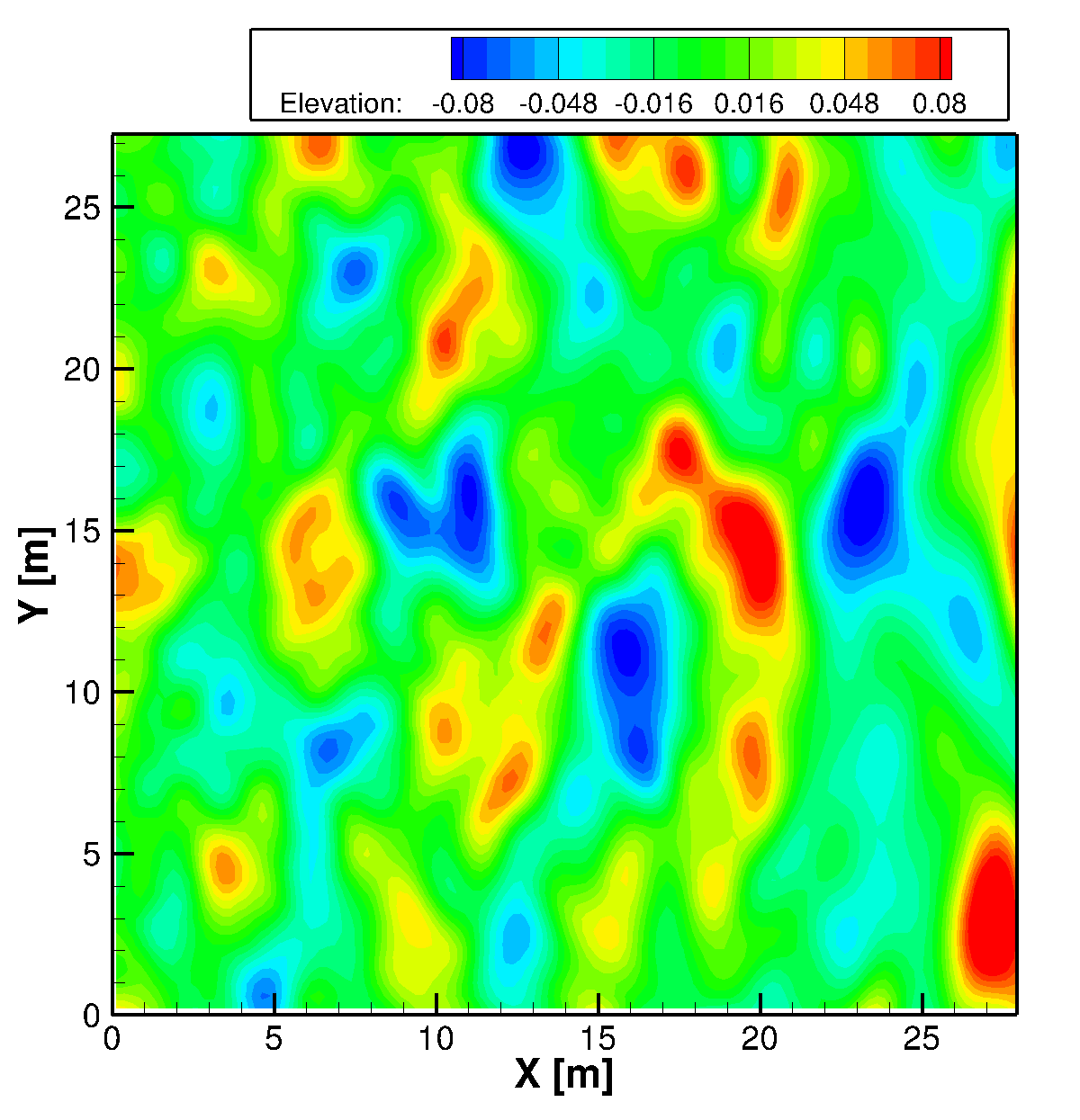}%
		\label{subfig:coupling_case3_cont_a}%
	}
	\subfigure[Far-field simulation]{%
    \includegraphics[trim=0.0cm 0.0cm 0.0cm 0.0cm,clip,width=0.49\textwidth]{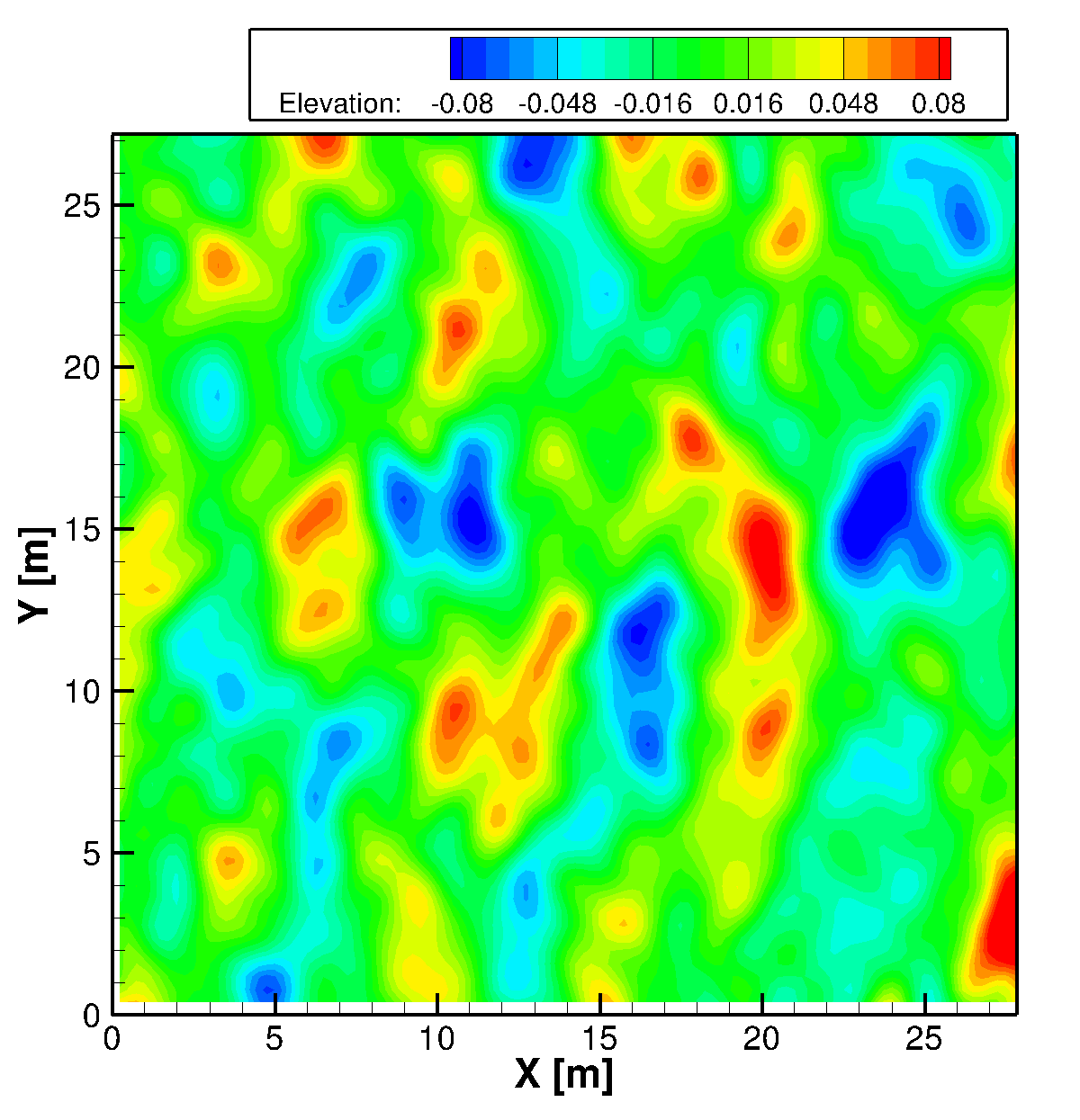}%
		\label{subfig:coupling_case3_cont_b}%
	}
	\caption{Far-field/near-field coupling case 3: simulation of a broadband spectrum of waves. Computed free surface elevation (in meters) resulting from the near-field simulation using the far-field/near-field coupling and the corresponding region of the far-field simulations alone. The results correspond to time $t=30 s$.}\label{fig:coupling_case3_cont}
\end{figure}
\begin{figure}[h!bt]
	\centering
	\subfigure[$y=5.0 m$]{%
		\includegraphics[trim=0.0cm 0.2cm 0.0cm 0.7cm,clip,width=0.8\textwidth]{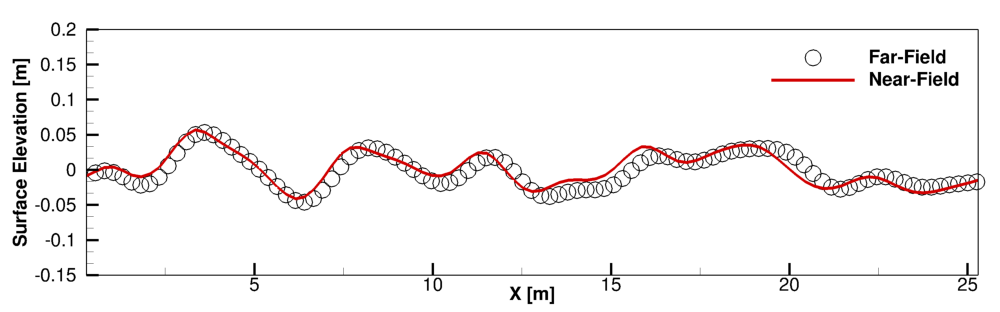}%
		\label{subfig:coupling_case3_profiles_a}%
	}
	\subfigure[$y=10.0 m$]{%
    \includegraphics[trim=0.0cm 0.2cm 0.0cm 0.7cm,clip,width=0.8\textwidth]{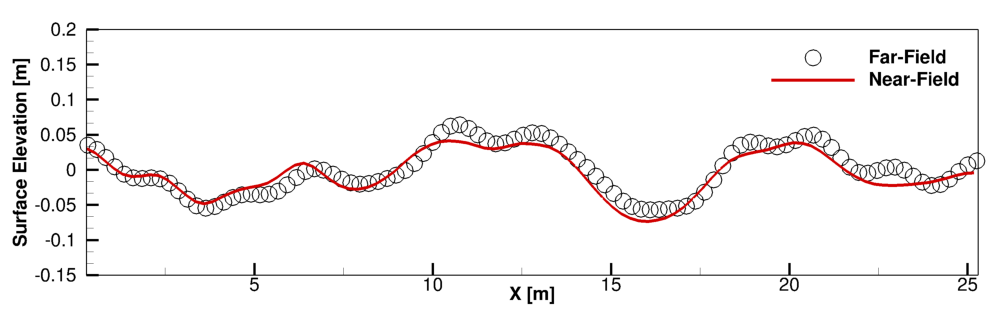}%
		\label{subfig:coupling_case3_profiles_b}%
	}
  \subfigure[$y=15.0 m$]{%
    \includegraphics[trim=0.0cm 0.2cm 0.0cm 0.7cm,clip,width=0.8\textwidth]{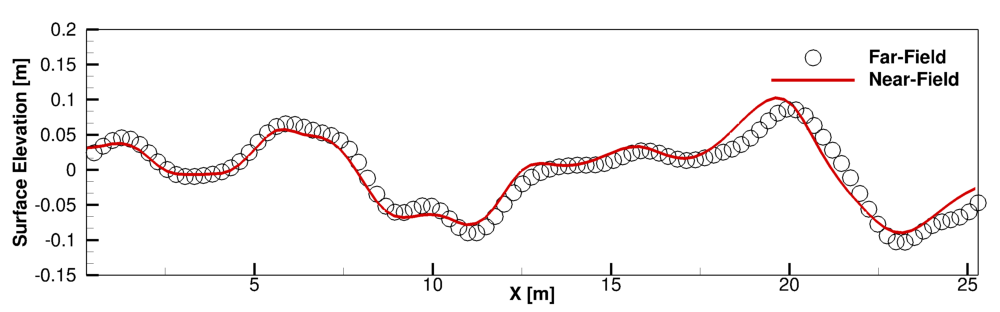}%
		\label{subfig:coupling_case3_profiles_c}%
	}
  \subfigure[$y=20.0 m$]{%
    \includegraphics[trim=0.0cm 0.2cm 0.0cm 0.7cm,clip,width=0.8\textwidth]{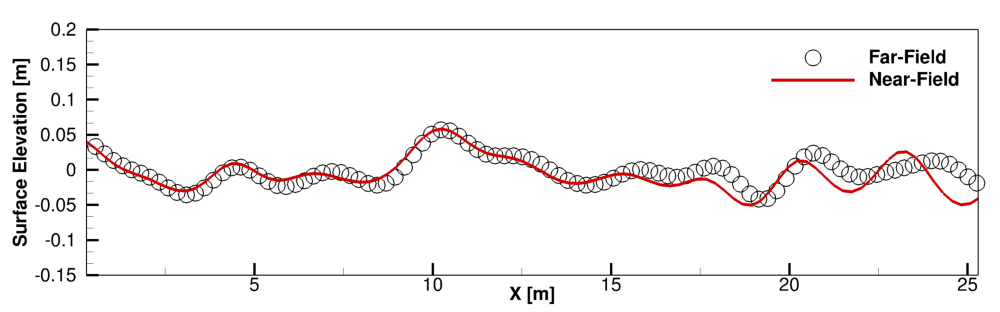}%
		\label{subfig:coupling_case3_profiles_d}%
	}
	\caption{Far-field/near-field coupling case 3: simulation of a broadband spectrum of waves. Computed free-surface elevation profiles from both the near-field simulation using the far-field/near-field coupling and the corresponding region of the far-field simulations alone at different $y$ planes. The results correspond to time $t=30 s$.}
\label{fig:coupling_case3_profiles}
\end{figure}

The final case for validating the far-field/near-field coupling algorithm and the ability to simulate a wave field with multiple frequency components is the simulation of a broadband spectrum of waves with the spectrum peak approximately at the wavelength $L\approx8 m$. The \gls{2d} wavenumber distribution of the energy spectrum extracted from the far-field simulation at time $t=30s$ is shown in Fig.\ \ref{fig:coupling_case3_freq}.
For this test case the domain size, the mesh dimensions, and all the case parameters such as the fluid properties, the gravity, the boundary conditions, and the time step are the same as in the previous simulation of three \gls{3d} superposed directional waves case in Section \ref{sec:FF-NF_coupling_case2}.  

The far-field simulation, which was initialized with a broadband spectrum of waves, was advanced for $30s$. From the beginning of the simulation, we extracted at every time step the far-field spectrum of waves by calculating the \gls{2d} Fourier transform of the surface elevation as described in Section \ref{sec:WaveGeneration}. We specified the extracted data into the source region of the near-field domain using the pressure forcing method, also at every time step, during the $30s$ for which data from the far-field was available. In Figs.\ \ref{fig:coupling_case3_cont} and \ref{fig:coupling_case3_profiles}, the free surface elevation results computed with the near-field solver in the area beyond the source region are compared with those computed with the far-field solver when it continues the simulation alone in the corresponding area at time $t=30s$. In particular, Fig.\ \ref{fig:coupling_case3_cont} shows the elevation contours and Fig.\ \ref{fig:coupling_case3_profiles} shows several elevation profiles at different $y$ planes. As demonstrated in the two figures, the wave field resulting from the near-field simulation using the coupled approach agrees well with that from the far-field simulation alone. There are some discrepancies that can be explained by the fact that not all wave frequency components are resolved with the adequate number of grid nodes per wavelength.

\subsection{Approach 2 validation case: generation of a broadband spectrum of waves}
\label{sec:approach2_results}
\begin{figure}[h!bt]
	\centering
    \includegraphics[trim=0.0cm 0.0cm 0.0cm 0.0cm,clip,width=1.0\textwidth]{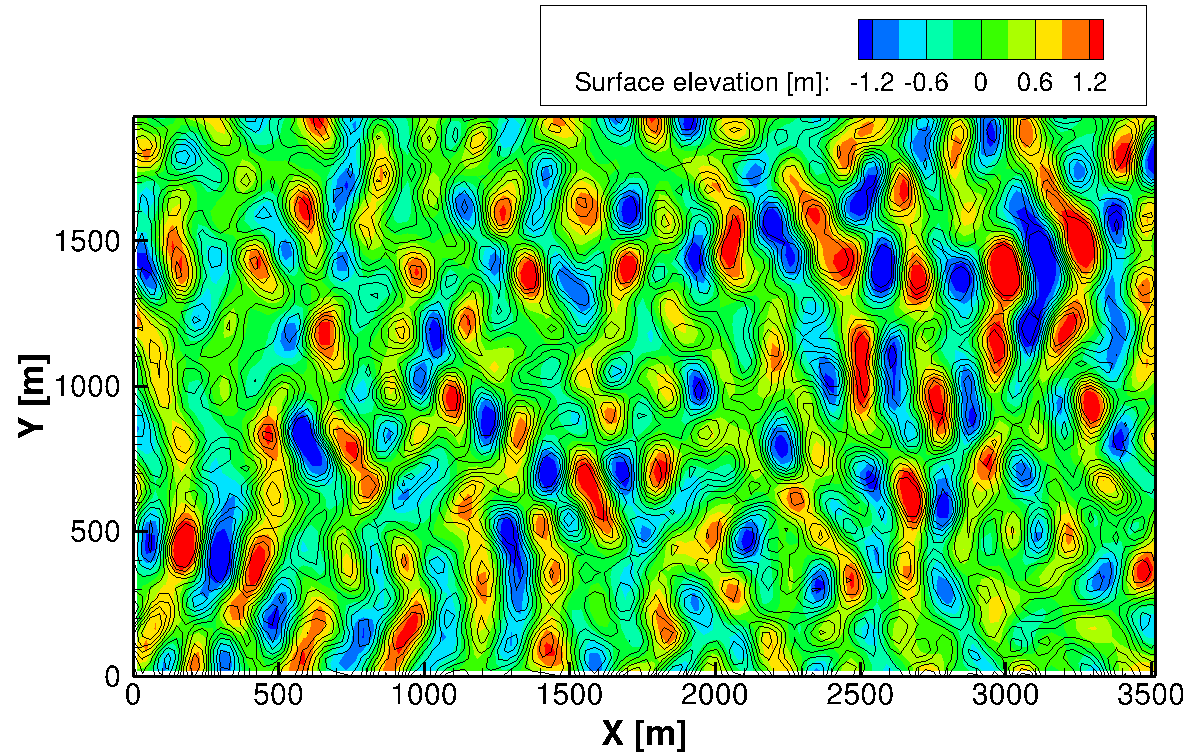}%
	\caption{Approach 2 validation case. Instantaneous free surface elevation $\eta_{FFR}$ reconstructed from the spectrum of waves extracted from the HOS far-field simulation  (contour lines) and the free surface elevation $\eta_{NF}$ from the CURVIB-level set simulation (colored contours).}\label{fig:approach2_vali}
\end{figure}
\begin{figure}[h!bt]
	\centering
	\subfigure[$y=500 m$]{%
		\includegraphics[trim=0.0cm 0.0cm 0.0cm 0.0cm,clip,width=0.8\textwidth]{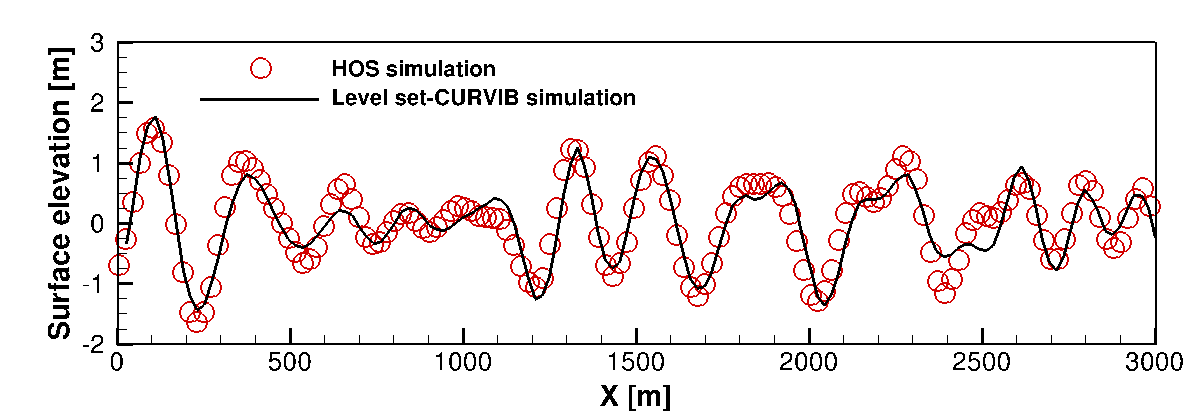}%
		\label{subfig:approach2_vali2_a}%
	}
	\subfigure[$y=1500 m$]{%
    \includegraphics[trim=0.0cm 0.0cm 0.0cm 0.0cm,clip,width=0.8\textwidth]{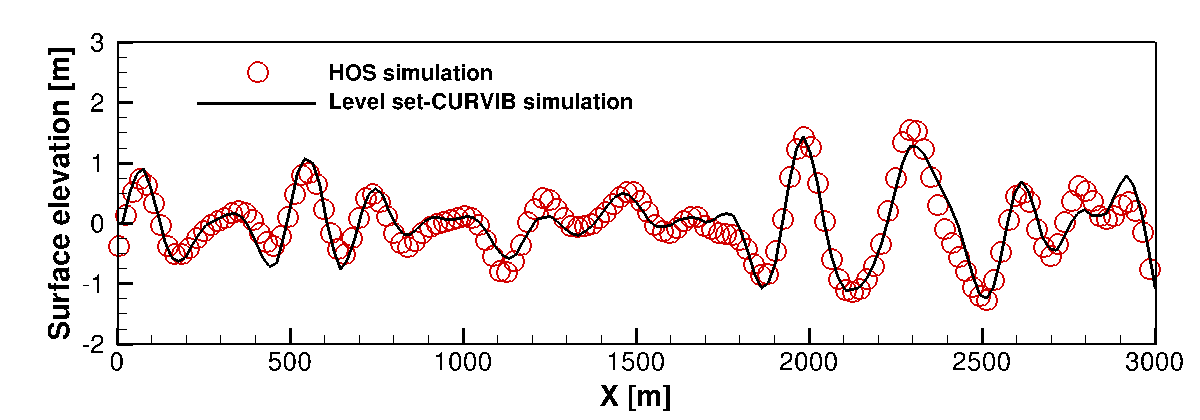}%
		\label{subfig:approach2_vali2_b}%
	}
	\caption{Approach 2 validation case. Profiles at $y$-planes of the instantaneous free surface elevation $\eta_{FFR}$ reconstructed from the spectrum of waves extracted from the HOS far-field simulation and the free surface elevation $\eta_{NF}$ from the CURVIB-level set simulation.}\label{fig:approach2_vali2}
\end{figure}

This test case is aimed to validate the coupling algorithm in Approach 2 by incorporating as initial condition for the level set-CURVIB simulation a complex wave field extracted from the LES-HOS simulation. 
To accomplish this goal, we first perform a wind-wave simulation using the LES-HOS code till the flow conditions are fully developed. Then, the wave data is extracted from the HOS simulation using the 2D Fourier transform, and it is transferred to the level set-CURVIB code by applying the forcing method described in Section \ref{sec:approach2_method}. In Approach 2, the pressure is applied at the whole free surface for a short period of time ($2\Delta$). 

\begin{figure}[h!bt]
	\centering
	\includegraphics[trim=0.0cm 6.5cm 0.0cm 4.5cm,clip,width=.6\textwidth]{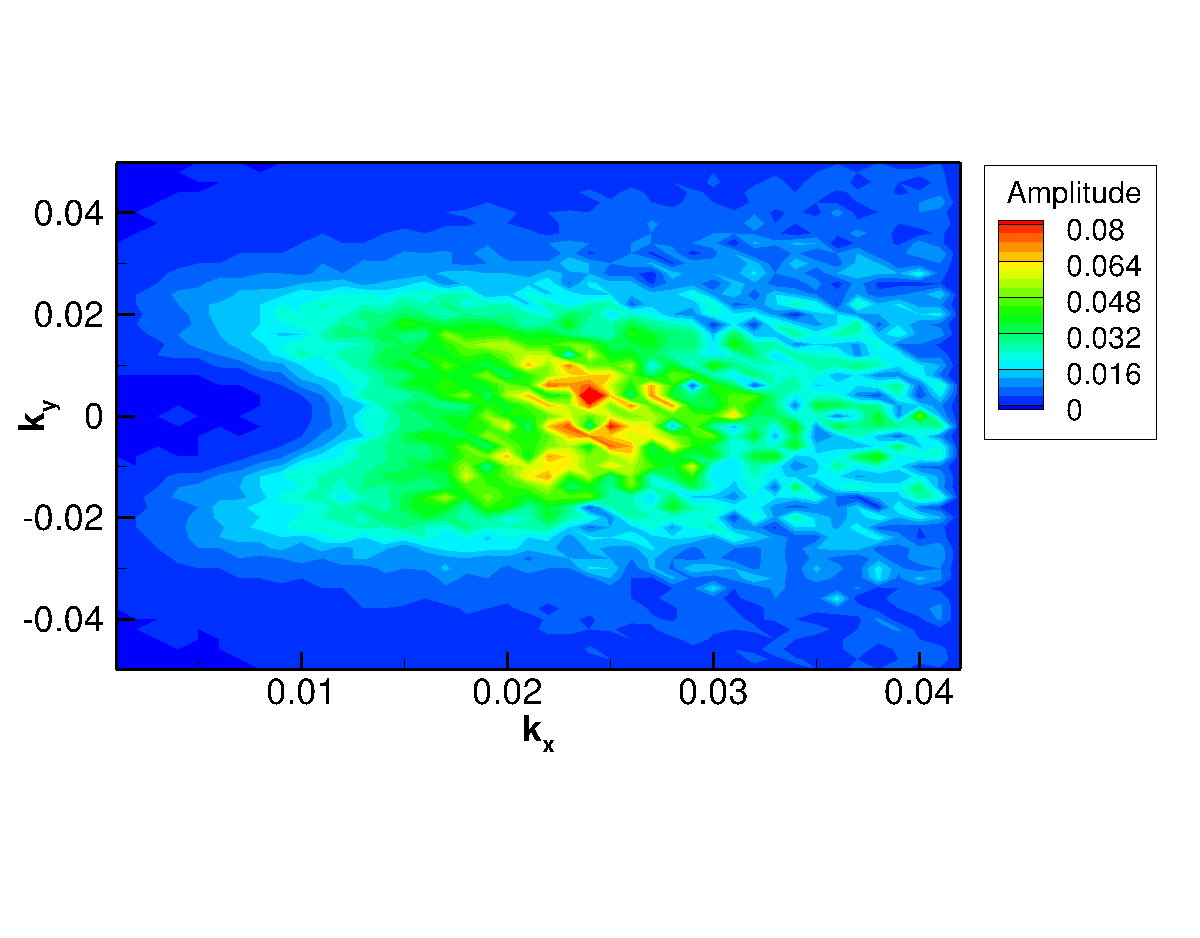}
	\caption{The broadband waves represented by wave amplitude contours as a function of the directional wavenumbers. Computed in the HOS domain at a time for which the flow is fully developed ($t=163{,}500s$).  
    }\label{fig:turbinecase_wavefreq}
\end{figure}

For the LES-HOS simulation we consider a domain size of $6{,}280m$, $3{,}140m$, and $1{,}000m$, in the stream-wise, span-wise, and vertical direction, respectively. The LES domain is discretized by a non-uniform mesh of size $128\times128\times128$. While in the stream-wise and span-wise directions the spacing is constant, in the vertical direction the grid cells are clustered near the free surface. The mesh of the HOS for simulating the wave field is uniform and of size $769\times769$. The free surface initial condition is a broadband wave spectrum of type \gls{jonswap} with wave peak period $T_{peak}=12.75s$. The wind field, which is driven by a constant pressure gradient such that the velocity at height $5m$ is $6m/s$, has been solved in a coupled manner with the wave field as described in Section \ref{sec:leshos}. The LES-HOS simulation has been advanced for a long time of $300{,}000$ time steps, with a time step size of $0.545s$ to ensure that fully developed wind and wave conditions were achieved. The wave spectrum for the developed wave field is shown in Fig.\ \ref{fig:turbinecase_wavefreq}.

The level set-CURVIB computational domain considered is a $6{,}283m$ long and $3{,}141m$ wide basin with $280m$ of water depth and $500m$ of air height. The mesh is uniform in the horizontal directions with 211 and 106 grid points in the stream-wise and span-wise directions, respectively. In the vertical direction the grid is uniform from $z=-12m$ to $z=12m$ with a spacing $\Delta z=2m$, and then the spacing progressively increases towards the top and bottom boundaries. The fluid properties and the gravity are the same as in the previous wave cases. Slip-wall boundary conditions are used at the top and bottom boundaries and periodic boundary conditions at the lateral boundaries. The time step is set to $0.02s$ and the interface thickness $\epsilon$ is $4m$. The time smoothing parameter $\Delta$ has been chosen to be $1s$. 

Let $t_0$ be the time for which the spectrum of waves is extracted from the HOS solver. In the forcing method of Approach 2, the application of the surface pressure is smeared in time, starting at time $t_{-1}=t_0-\Delta$ and finalizing at time $t_1=t_0+\Delta$. As a result, the level set-CURVIB surface elevation starts from a flat surface at time $t_{-1}$ and it progressively transitions to the wave field given by the provided spectrum of waves at time $t_1$. In order to validate the level set-CURVIB surface elevation $\eta_{NF}$ at time $t_2\ge t_0+\Delta$, it would not be entirely consistent to compare it with the free surface elevation of the HOS simulation $\eta_{FF}$ at the corresponding time $t_2$, as in this case the wind field is not transferred into the level set-CURVIB simulation, and thus the two simulations evolve slightly differently in time. Alternatively, we can reconstruct the surface elevation $\eta_{FFR}$ at the corresponding time $t_2$ by applying Equation (\ref{eq:surfacepressure2}) to all the frequencies of the spectrum that were imposed.
Figs.\ \ref{fig:approach2_vali} and \ref{fig:approach2_vali2} show the computed instantaneous free surface elevation $\eta_{NF}$ and the reconstructed free surface elevation $\eta_{FFR}$.
The agreement seen in these figures demonstrates the ability of the pressure forcing method to initialize a complex wave fields in a Navier-Stokes based two-phase flow solver such as the presented \gls{curvib}-level set  method. 
\subsection{\Gls{fsi} simulation of an offshore floating wind turbine}
\label{sec:turbinecase}
The aim of this simulation case is to demonstrate the full capabilities of the proposed far-field/near-field computational framework by simulating an offshore floating wind turbine under realistic wind and wave conditions representative of a site-specific environment such as the \gls{pnw}.

\subsubsection{Definition of the offshore environmental conditions}
To determine the wind velocity and the wave field that is representative of the environmental conditions for offshore floating wind turbine deployment, we use the data measurements taken from the Station 46041 of the National Data Buoy Center. In particular, Ref. \cite{DataBuouy} provides an annual \gls{jpd} of the most commonly occurring waves elaborated with several decades of wave measurements. According to the \gls{jpd}, the most common waves occurring more than $10\%$ of the time have a dominant period of about $10.0$ to $12.9s$, and a wave height between $1.5$ and $2.4m$. Based on that, we choose for the wave field a broadband wave spectrum of dominant peak period $T_{peak}=12.75s$, which corresponds to a peak wavelength of $L_{peak}=251m$ according to the wave dispersion relation.

With regards to the wind field, \cite{DataBuouy} provides monthly averaged values of wind speed taken at $5m$ above the \gls{msl}. The values show that the averaged wind speed is considerably lower during the warmer season (August and September), approximately $3.8m/s$, and increases to averaged values of about $7m/s$ during the colder season (December and January). We thus adopt an intermediate wind speed of $5m/s$ at $5m$ above the \gls{msl}. Using the law of the wall, the extrapolated wind speed at the turbine hub height (133m) is approximately $9m/s$.

\subsubsection{The floating wind turbine}
We consider a large floating wind turbine system, consisting of a $13.2MW$ wind turbine installed on a tri-column triangular platform. The turbine rotor, which is a design by the Sandia National Laboratories, has a $200m$ long diameter and a three SNL100-00 blades (see Griffith and Ashwill \cite{griffith2011sandia} for details). The hub height with respect to the \gls{msl} of the turbine is $133.5m$. 

\begin{figure}[h!bt]
\label{fig:turbinecase_ibmesh}
\centering
\includegraphics[trim=0.1cm 0.1cm 0.1cm 0.1cm,clip,width=.4\textwidth]{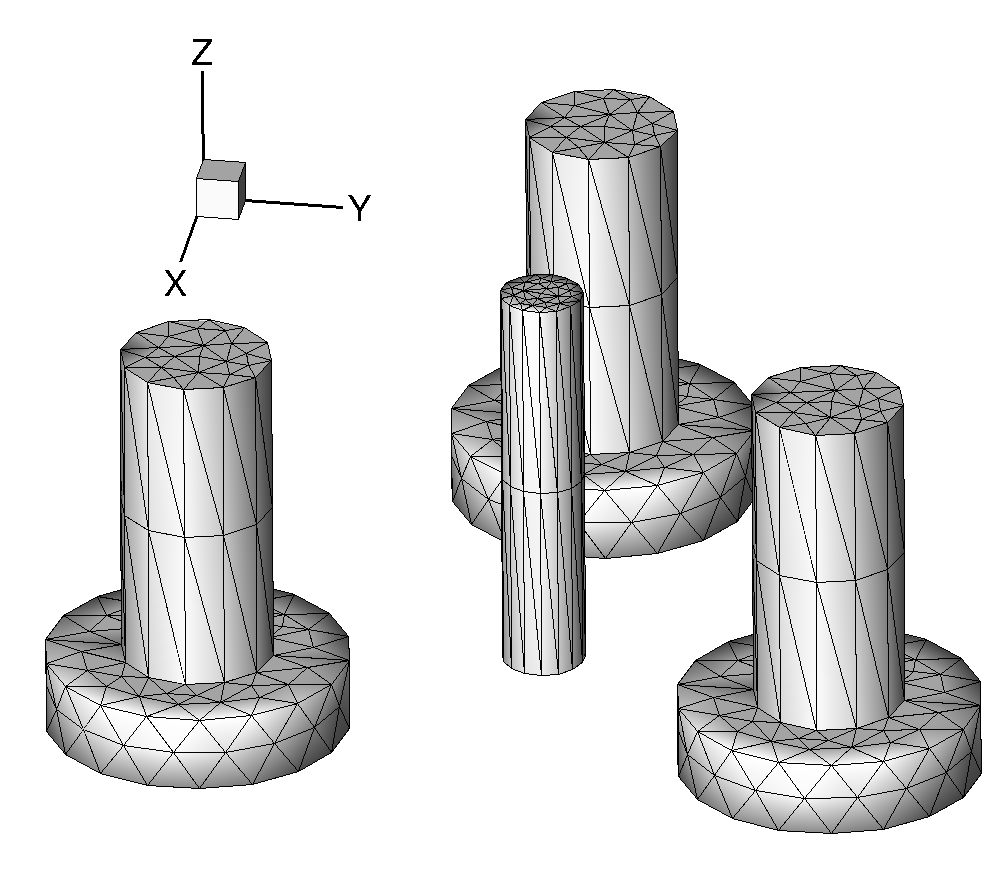}
\caption{Mesh of the floating platform used in the \Gls{ib} method in the offshore floating wind turbine case.}
\end{figure}

The floating platform that supports the turbine, designed by Principle Power, is a scaled up version (using Froude number similitude and a scale factor of $\lambda=1.4$) of the the OC4 semi-submersible design presented in Robertson et al. \cite{robertson2012definition}. It is composed of a main central column of $9.1m$ diameter and $42.0m$ height and three offset cylindrical columns of $16.8m$ diameter and $36.4m$ height that are interconnected through pontoons and cross braces. The offset columns, which are spaced $70m$ apart from each other, have a base of $33.6m$ diameter and $8.4m$ height. The geometry of the four columns is illustrated in Fig.\ \ref{fig:turbinecase_ibmesh}.

\subsubsection{Hydrodynamic properties of the floating turbine}
In order to introduce all the floating turbine parameters required for computing the 6 \gls{dof} dynamics of the floating wind turbine, we first rewrite the rigid body \gls{eom} given by equations (\ref{eq:EOM1}) and (\ref{eq:EOM2}) in matrix form, including case specific external forces. The 6 \gls{dof} \gls{eom} can be written in the inertial frame of reference and in principal axis as follows: 
\begin{equation}\label{eq:new_eom}
\textbf{M}\frac{\partial^2 \textbf{q}}{\partial t^2}=\textbf{F}_{fluid}+\textbf{F}_{gyro}+\textbf{F}_{mooring}+\textbf{F}_{actline}
\end{equation}
where $\textbf{M}$ is the $6\times6$ mass matrix, $\textbf{q}$ the 6 \gls{dof} position vector (includes 3 displacements $q_x$, $q_y$, and $q_z$, and 3 rotations $q_{\theta_x}$, $q_{\theta_y}$, and $q_{\theta_z}$), and $\textbf{F}_{fluid}$, $\textbf{F}_{actline}$, $\textbf{F}_{gyro}$, and $\textbf{F}_{mooring}$ are the vectors containing forces and moments due to the fluid, the rotor thrust force, the gyroscopic effects of the spinning rotor, and the mooring system, respectively. 

The mass matrix when the equations are in principal axis have zero non-diagonal terms, and reads as follows:
\begin{equation}\label{eq:mass_matrix}
\textbf{M}=\left\{ 
\begin{array}{cccccc}
m&0&0&0&0&0\\
0&m&0&0&0&0\\
0&0&m&0&0&0\\
0&0&0&I_x&0&0\\
0&0&0&0&I_y&0\\
0&0&0&0&0&I_z\\  
\end{array}\right\}.
\end{equation}

In the present case, we take the mass of the floating turbine system to be $m=3.7\cdot10^7 kg$, and the inertia $I_x=3.64\cdot10^{10} kg\cdot m^2$, $I_y=3.64\cdot10^{10} kg\cdot m^2$, and $I_z=6.39\cdot10^{10} kg\cdot m^2$. The center of gravity is located $18.84m$ below the \gls{msl} and the platform draft is $28m$.

$\textbf{F}_{fluid}$ is computed by integrating the fluid pressure and shear stresses on the body surface as described in \cite{calderer_fsi_2014}. The thrust force, when using the actuator line model described in Section \ref{sec:actuatorline}, is computed in the following manner:
\begin{equation}\label{eq:new_al}
\textbf{F}_{actline}= \sum_{n_l}\left(\textbf{n}_t \textbf{L} a  + \textbf{n}_t \textbf{D} a\right)
\end{equation}
where $\textbf{L}$ and $\textbf{D}$ are the lift and drag coefficients computed at each line element using equations (\ref{eq:act_line1}) and (\ref{eq:act_line2}), respectively, $n_l$ is the total number of elements (including all blades), $a$ is the length of each element, and $\textbf{n}_t$ is the rotor normal direction pointing towards the stream-wise direction. 
The moments due to the gyroscopic effects of the spinning rotor are computed using the method described in \cite{jensen2011numerical}. The method, which is based on the assumption that the turbine system undergoes small rotations, reads as follows:
\begin{equation}\label{eq:gyro}
\textbf{F}_{gyro}=\left\{ \begin{matrix}
  0\\
  0\\
  0\\
  I_p \Omega \frac{\partial q_{\theta_y}}{\partial t}\\
  -I_p \Omega \frac{\partial q_{\theta_x}}{\partial t}\\
  0
 \end{matrix}\right\}
\end{equation}
where $I_p$ is the rotational inertia of the rotor, $6.471\cdot10^8kg\cdot m^2$ for the present turbine, and $\Omega$ is the angular velocity of the rotor determined by the \gls{tsr} and the inflow velocity. 
The calculation of the last necessary term for the turbine simulation, $\textbf{F}_{mooring}$,  is described in the subsequent section.

\subsubsection{The mooring system}
\begin{figure}[h!bt]
	\centering
	\includegraphics[trim=0.cm 0.cm 0.cm 0.cm,clip,width=.4\textwidth]{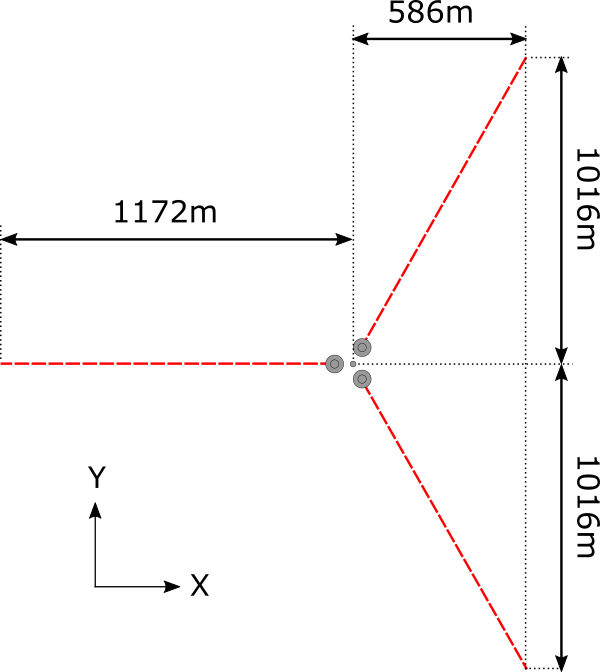}
	\caption{Schematic description of the mooring system composed of three catenary lines employed in the floating turbine case.   
}\label{fig:turbinecase_catenary}
\end{figure}

 The floating platform is secured in place using a mooring system consisting of three catenary lines distributed symmetrically with respect to the platform vertical axis as illustrated in Fig. \ref{fig:turbinecase_catenary}. The design is taken from Robertson et al. \cite{robertson2012definition}, with the difference that it has been properly scaled by a factor of $\lambda=1.4$ to accommodate the larger dimensions of the present turbine design. The mooring cables are attached to the upper part of the base columns at a location corresponding to a water depth of $19.6m$ and a radial distance of $57.2m$. The other end of the cables is attached to the sea bottom at a radial distance of $1{,}172m$ and a depth of $280m$.

To incorporate the mooring system into the computational framework we opt for the linearized model of \cite{robertson2012definition}. In the linearized model, the forces induced by the entire mooring system, $F^{mooring}$, are estimated in the following manner:
\begin{equation}\label{eq:lin_mooring}
\textbf{F}^{mooring}(\textbf{q})=\textbf{F}^{mooring,0}-\textbf{C}^{mooring}\textbf{q},
\end{equation}
where $\textbf{F}^{mooring,0}$ is the force vector of the mooring system when the system is in equilibrium, and $\textbf{C}^{mooring}$ is the $6\times6$ linearized restoring matrix. Robertson et al. \cite{robertson2012definition} provide the values of $\textbf{F}^{mooring,0}$ and $\textbf{C}^{mooring}$ for the present mooring system, estimated using a linearized perturbation analysis in FAST. Using an scaling factor of $\lambda^3$ for the forces, a factor of $\lambda^4$ for the moments, a factor of $\lambda$ for the lengths, and a unit factor for the angles, the scaled $\textbf{F}^{mooring,0}$ and $\textbf{C}^{mooring}$ read as follows:
\begin{equation}\label{eq:lin_mooring_force1}
\textbf{F}^{mooring,0}=\left\{ \begin{matrix}
  0 \\
  0 \\
  -5.05\cdot 10^6 N\\
  0 \\
  0 \\
  0
 \end{matrix}\right\},
\end{equation}
and
\begin{eqnarray}\label{eq:lin_mooring_force2}
\textbf{C}^{mooring}&=&\left\{ \begin{array}{ccc}
  1.39\cdot 10^5 N/m & 0  &0    \\                
  0&1.39\cdot 10^5 N/m&0        \\                
  0&0&3.74\cdot10^4N/m            \\              
  0&2.94\cdot 10^6 Nm/m&0        \\              
  -2.94\cdot 10^6 Nm/m&0&0         \\              
  0&0&0                                          
 \end{array}\right.\nonumber\\
&&\left. \begin{array}{ccc}
0&-2.96\cdot 10^5 N/rad&0\\
2.96\cdot 10^5N/rad&0&0\\
0&0&0\\
3.35\cdot10^8Nm/m&0&0\\
0&3.35\cdot10^8Nm/m&0\\
0&0&4.49\cdot 10^8 Nm/m
\end{array}\right\}. 
\end{eqnarray}

We note that the linearized mooring model assumes small motions. This assumption is reasonable given that the objective of the present simulation is to demonstrate the capabilities of the framework. In future simulation we will consider more elaborate models, treating individual lines and considering non-linear effects.

\subsubsection{Simulation results}
\begin{figure}[h!bt]
	\centering
	\includegraphics[trim=0.1cm 0.1cm 0.1cm 0.1cm,clip,width=1.0\textwidth]{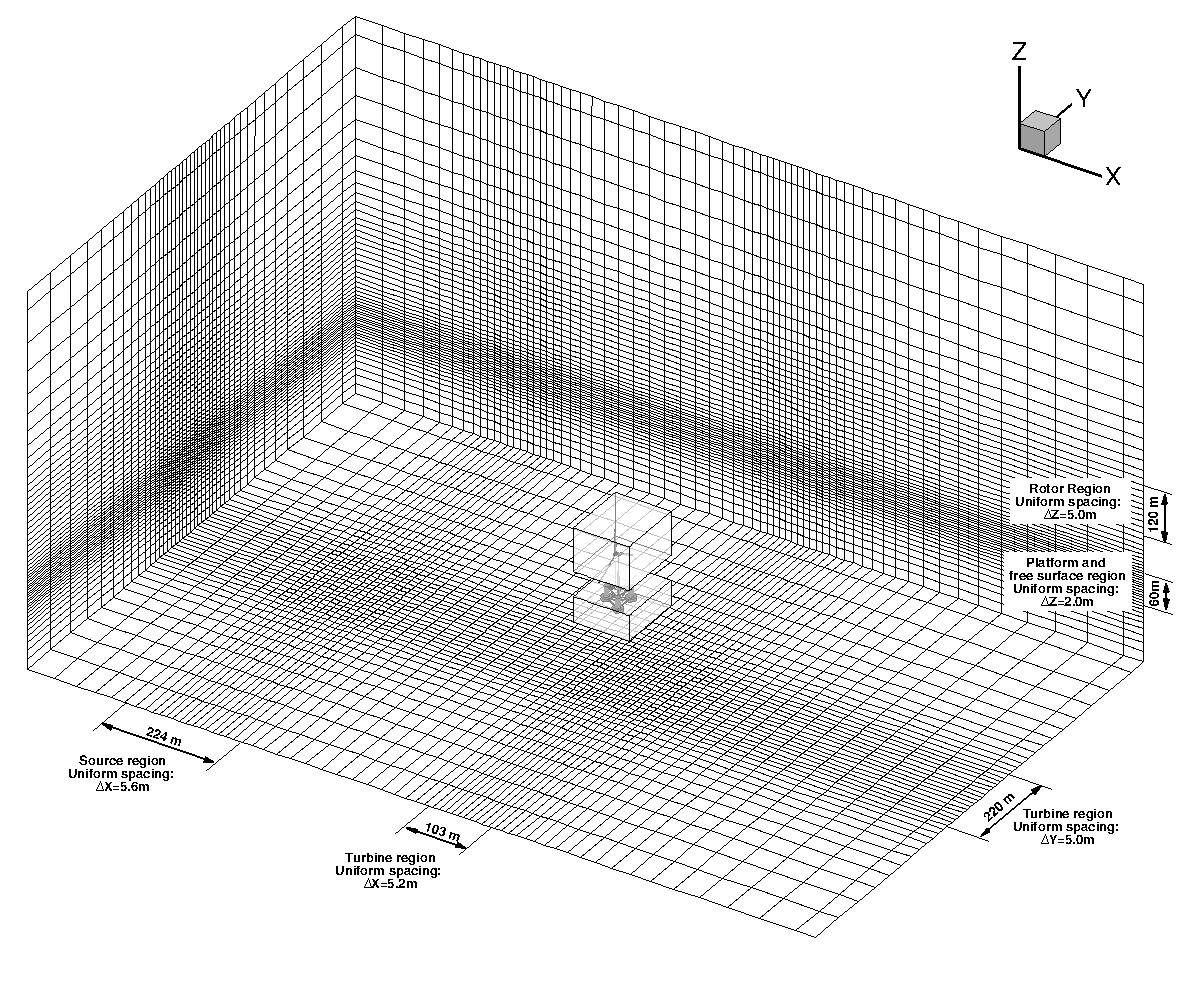}
	\caption{Schematic description of the fluid mesh used in the near-field domain of the offshore floating wind turbine case. The rectangular boxes indicate the two regions of constant grid spacing where the floating turbine is located. In this figure, for every grid line shown four are skipped.   
}\label{fig:turbinecase_fluidmesh}
\end{figure}

In this simulation we employ the far-field/near-field approach developed herein for studying floating structures under ocean wind and waves. The near-field computational domain is $2{,}675m$ long in the stream-wise direction ($x=[-550m, 2125m]$) and $1{,}750m$ wide in the span-wise direction ($y=[-875m,875m]$), the water depth is $280m$ and the air column above the \gls{msl} is $1{,}000m$. The source region has a length $\epsilon_x$ of $224m$ and is centered on $x=0$. 
The floating turbine is positioned downstream of the sponge layer at $x=900m$ and centered on $y=0m$. 
We use a non-uniform mesh of size $163\times207\times259$ that is schematically described in Fig.\ \ref{fig:turbinecase_fluidmesh}. 
In the stream-wise direction, the spacing $\Delta x$ is constant in the following two regions: (1) in the source region ($x=[-112m, 112m]$) where the spacing is $\Delta x=5.6m$; and (2) in the region near the floating structure (defined by $x=[848.2m, 951.8m]$) where the spacing is $\Delta x=5.18m$. 
From the end of the first region ($x=112m$) to the beginning of the second region ($x=848.2m$), $\Delta x$ varies smoothly across the two values, and outside of these two regions the spacing increases progressively towards the inlet and outlet boundaries. In the vertical direction, the spacing $\Delta z$ follows the same spacing pattern also with two regions of constant spacing: (1) the region from $z=-40m$ to $z=20m$ which has spacing $\Delta z$ equal to $2m$ and comprises the floating platform and the free surface; and (2) the region from $z=130m$ to $z=250m$ which has spacing $\Delta z$ equal to $5m$ and comprises most of the rotor. 
Finally, in the span-wise direction the spacing $\Delta y$ is constant and equal to $5m$ in a single region spanning from $y=-110m$ to $y=110m$ and enclosing the floating turbine. The stretching ratio used in all directions is always limited to $1.05$ and its variation follows a hyperbolic function. The thickness of the interface is set to $\epsilon=4m$, the gravity to $g=9.81m/s^2$, and the density and dynamic viscosity for the water to $1{,}000kg/m^3$ and $1.0\cdot10^{-3}Pa s$, respectively, and for the air $1.2kg/m^3$ and $1.8\cdot10^{-5}Pa s$. Slip-wall boundary conditions are adopted at the $y$ and $z$ boundaries, and sponge layers with thickness of $200m$ are applied near the lateral boundaries. 

 The geometrical parts of the platform considered in the simulation and its dimensions are presented in Fig.\ \ref{fig:turbinecase_ibmesh}, which also shows the triangular mesh used for discretizing the structure in the \gls{curvib} method. The structural elements interconnecting the four columns have been neglected due to their small size in comparison to the large dimensions of the columns.

The turbine rotor, in the present simulation, is treated with the actuator line model described in Section \ref{sec:actuatorline}.  The turbine is simulated with a constant \gls{tsr} of 8, which given a hub height incoming velocity of about $9m/s$, is estimated to be close to the optimal value for performance.

The wind and wave conditions are incorporated from the far-field precursor simulation already presented in Section \ref{sec:approach2_results} for the Approach 2 validation case. It consists of a wind-wave coupled simulation started with an initial condition of the wave field defined by the \gls{jonswap} spectrum of peak period $T_{peak}=12.75s$. Once the precursor simulation was advanced for about $300{,}000$ time steps, with a time step size of $0.545s$, reaching a fully developed stage, the wind and wave conditions are both fed to the near-field turbine simulation domain using the Approach 1 coupling method (Section \ref{sec:WaveGeneration}).


%
\begin{figure}[h!bt]
	\centering
	\subfigure[Time $t=273 s$]{%
		\includegraphics[trim=0.0cm 0.0cm 0.0cm 0.0cm,clip,width=0.49\textwidth]{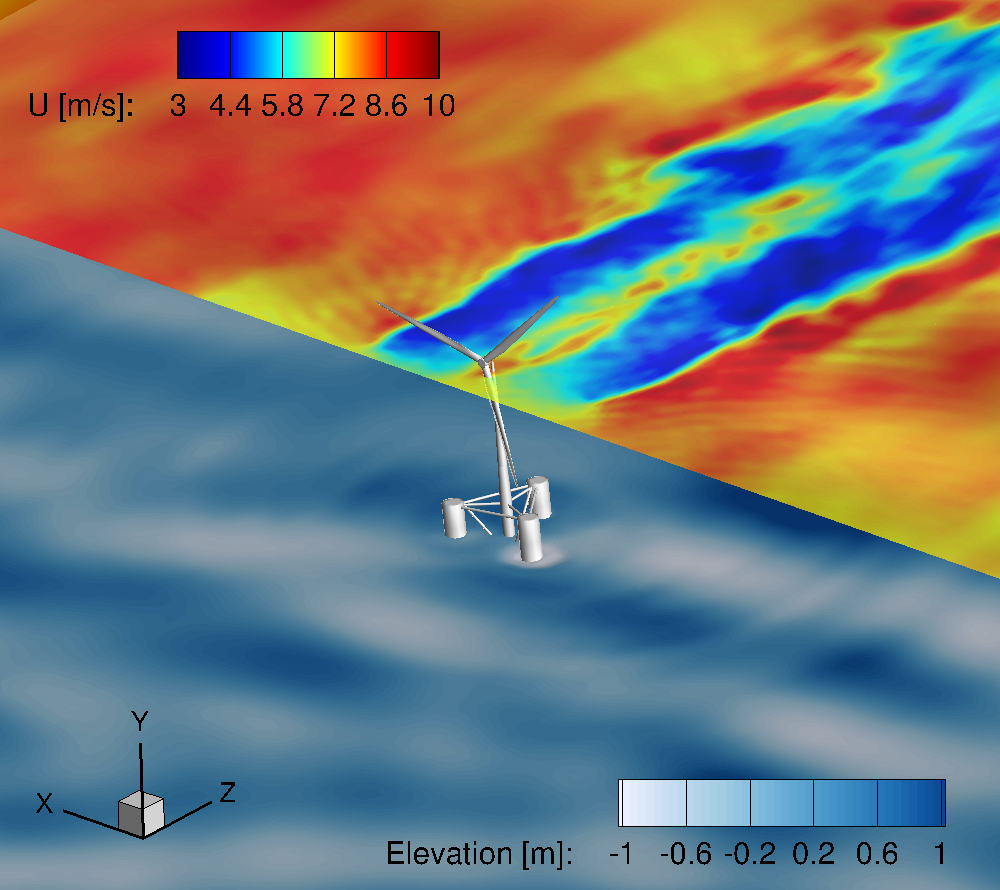}%
		\label{subfig:turbinecase_vel_wave_a}%
	}
	\subfigure[Time $t=685 s$]{%
        \includegraphics[trim=0.0cm 0.0cm 0.0cm 0.0cm,clip,width=0.49\textwidth]{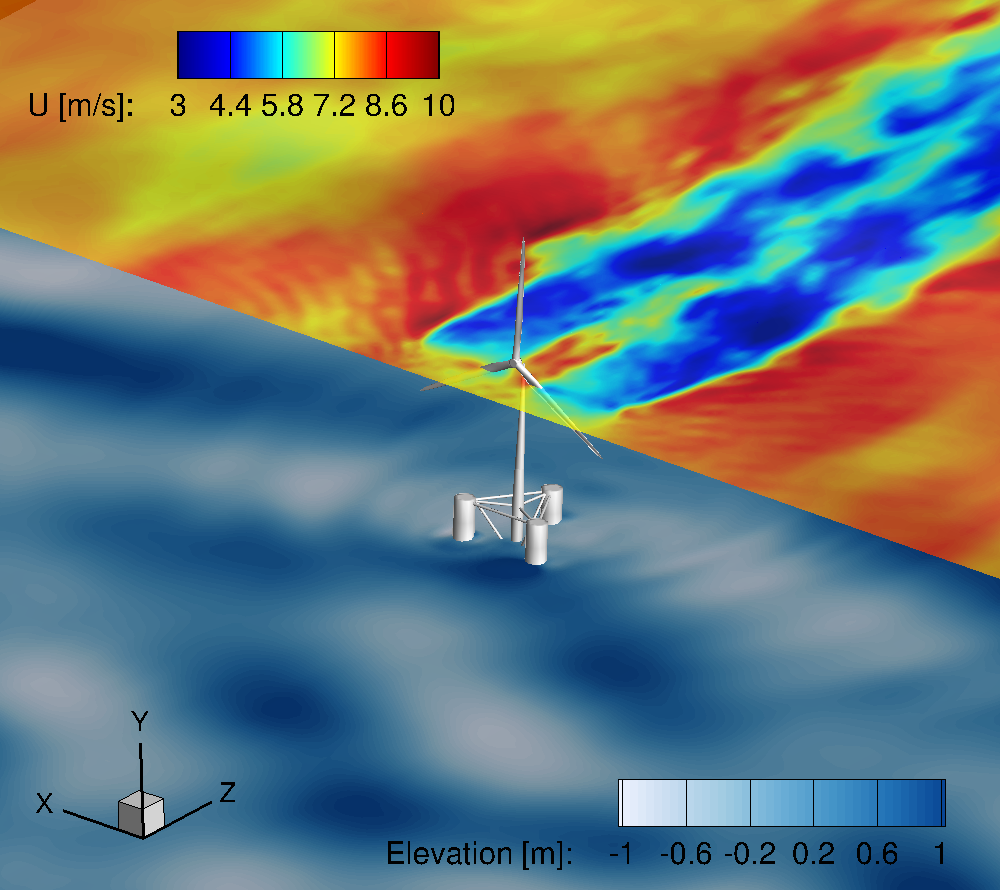}%
		\label{subfig:turbinecase_vel_wave_b}%
	}
	\caption{Offshore floating wind turbine case. \gls{3d} view of the floating wind turbine with the free surface colored with elevation contours and a horitzonal plane at hub heigth of stream-wise velocity.}\label{fig:turbinecase_vel_wave}
\end{figure}
\begin{figure}[h!bt]
	\centering
	\subfigure[Response in Surge]{%
		\includegraphics[trim=0.0cm 0.0cm 0.0cm 0.0cm,clip,width=0.49\textwidth]{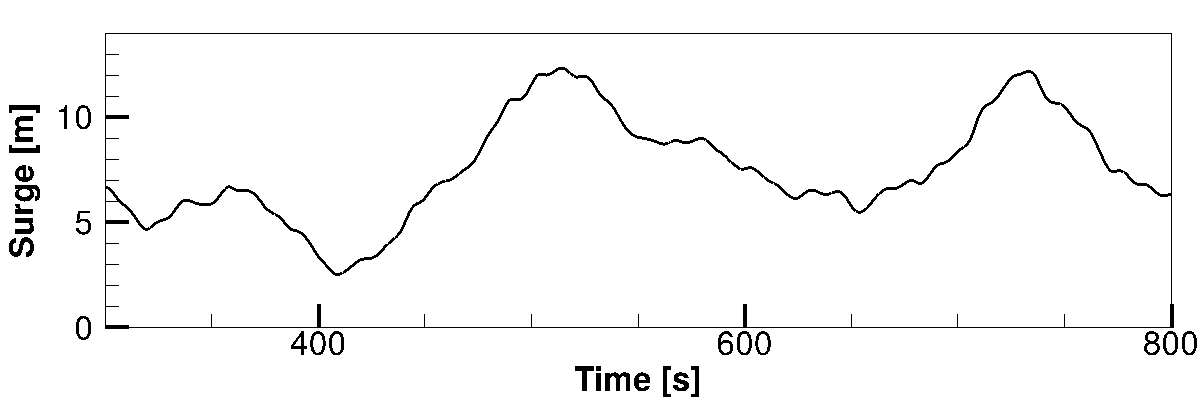}%
		\label{subfig:turbinecase_response_a}%
	}
	\subfigure[Response in Roll]{%
        \includegraphics[trim=0.0cm 0.0cm 0.0cm 0.0cm,clip,width=0.49\textwidth]{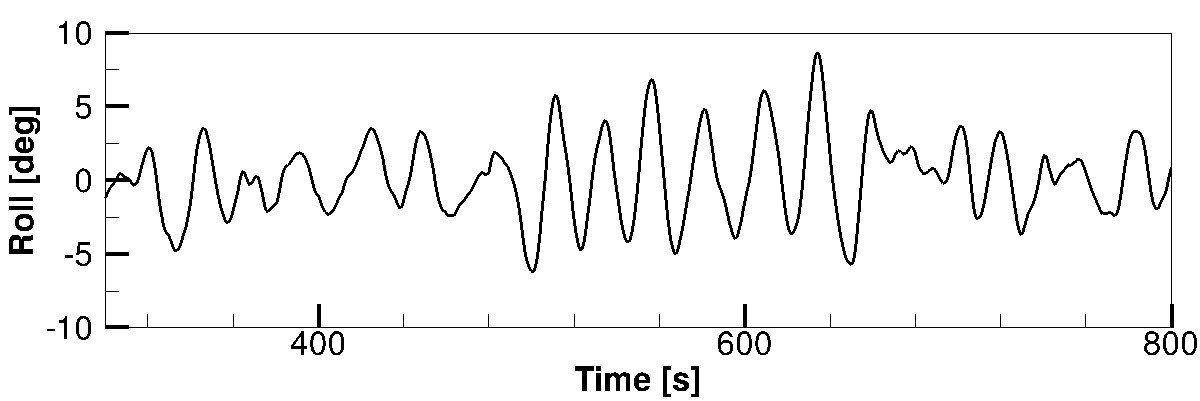}%
		\label{subfig:turbinecase_response_b}%
	}
	\subfigure[Response in Sway]{%
		\includegraphics[trim=0.0cm 0.0cm 0.0cm 0.0cm,clip,width=0.49\textwidth]{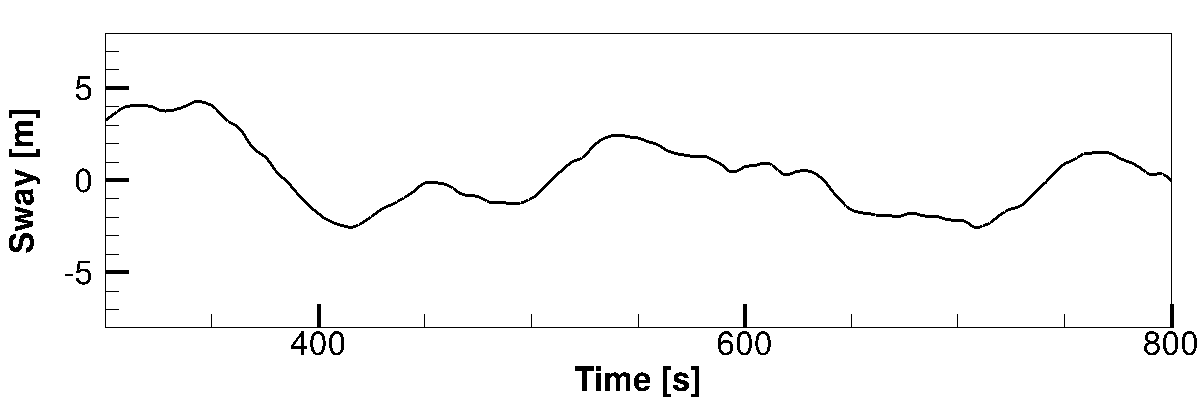}%
		\label{subfig:turbinecase_response_c}%
	}
	\subfigure[Response in Pitch]{%
        \includegraphics[trim=0.0cm 0.0cm 0.0cm 0.0cm,clip,width=0.49\textwidth]{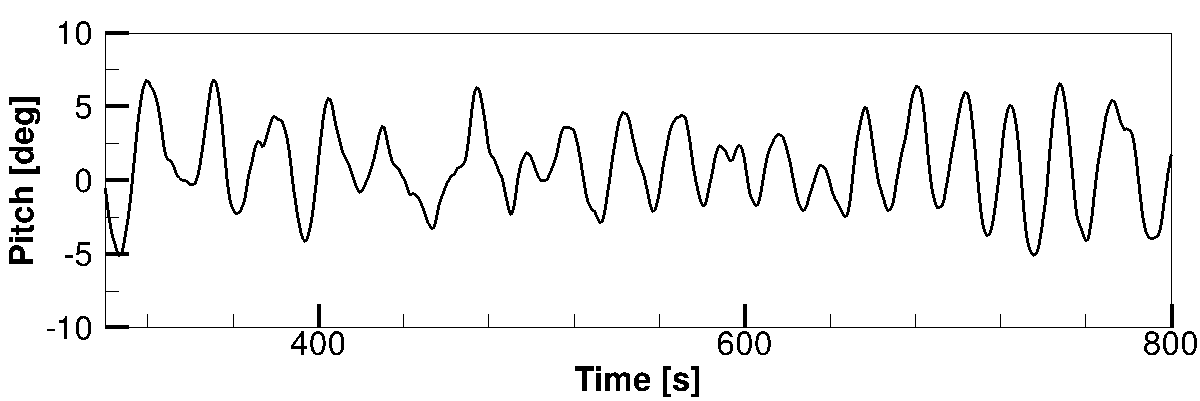}%
		\label{subfig:turbinecase_response_d}%
	}
	\subfigure[Response in Heave]{%
		\includegraphics[trim=0.0cm 0.0cm 0.0cm 0.0cm,clip,width=0.49\textwidth]{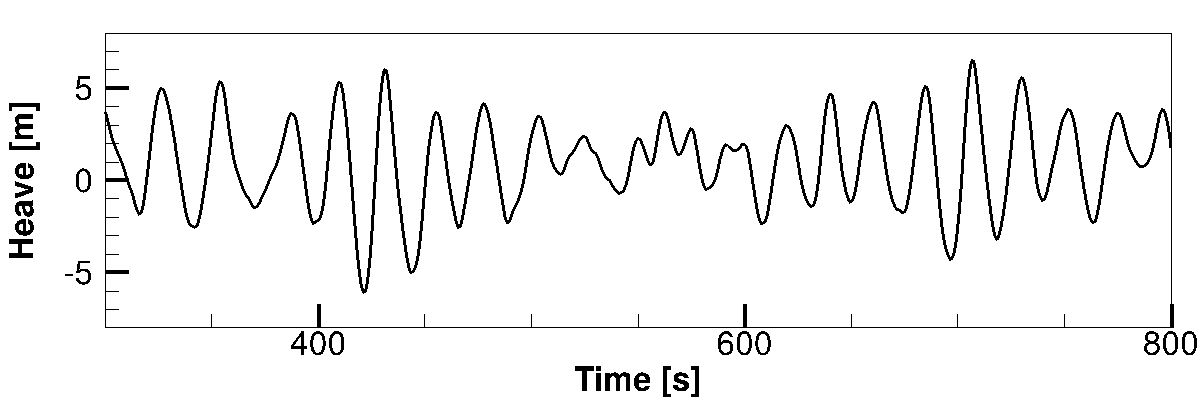}%
		\label{subfig:turbinecase_response_e}%
	}
	\subfigure[Response in Yaw]{%
        \includegraphics[trim=0.0cm 0.0cm 0.0cm 0.0cm,clip,width=0.49\textwidth]{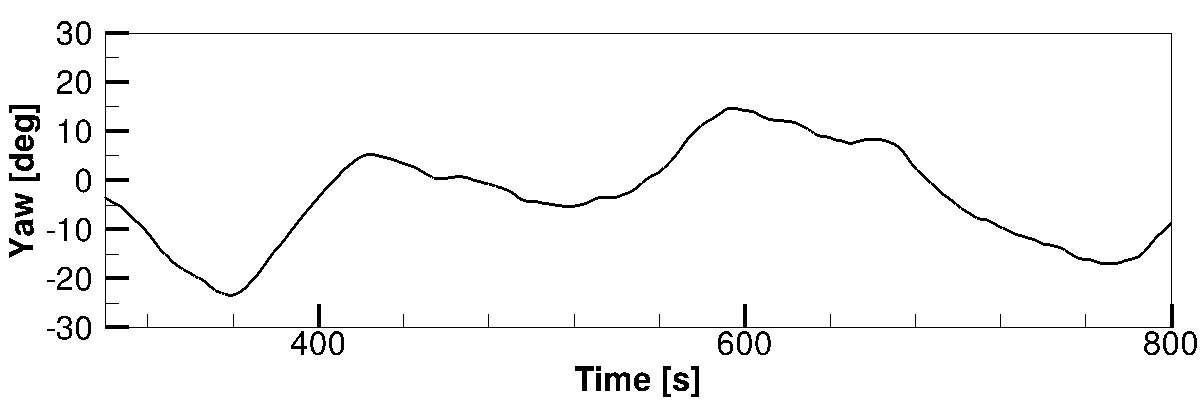}%
		\label{subfig:turbinecase_response_f}%
	}
	\caption{Offshore floating wind turbine case. Structural response of the floating turbine system in the six \gls{dof}.}\label{fig:turbinecase_response}
\end{figure}

In Fig.\ \ref{fig:turbinecase_vel_wave} we present the near-field results of the floating wind turbine including the free surface and the stream-wise velocity on a horizontal plane at hub height. As seen in the figure, the wave field clearly shows the formation of radiated waves induced by the motion of the platform. The structural response of the floating structure in the six \gls{dof} is given in Fig.\ \ref{fig:turbinecase_response}. Note that surge, sway, and heave correspond to the translational \gls{dof} in the stream-wise, span-wise, and vertical directions, respectively, and roll, pitch, and yaw correspond to the rotations with respect to the stream-wise, span-wise, and vertical axes, respectively. Looking at the pitch response in Fig.\ \ref{subfig:turbinecase_response_b}, the turbine is slightly inclined towards positive angles as a result of the wind effect. 

This test case clearly illustrates the ability of the present method to simulate wind-wave-body interactions using real environmental conditions and a complex floating structure. Some of the models that have been integrated in the framework for this simulation, such as the mooring system model and the computation of the gyroscopic forces, are relatively simple models that assume low amplitude motions. The flexibility of the code, however, allows easy implementations of more sophisticated methods for individual aspects of the application.   

More in depth results analyzing the water and air flows around this and other floating turbines are certainly necessary to get better insights into the problem. The computational framework we present in this paper is a unique tool that allows, for the first time, to study offshore applications considering as many coupled physical phenomena.  

\section{Conclusions}\label{sec:conc}
The objective of the present work was to develop a computational framework that can simulate real life complex floating structures and its interaction with realistic ocean wave and wind fields. To efficiently deal with the computational challenge of large disparity of scales associated with such type of problems, we adopted a partitioned far-field/near-field approach. The two-fluid method of Yang and Shen \cite{yang_simulation_2011,yang_simulation_2011_1} applied in the far-field domain allows to obtain fully developed wind and wave conditions taking advantage of the expedience of the HOS method for wave simulation. In the near-field domain, the \gls{fsi}-level set method of Calderer et al. \cite{calderer_fsi_2014} allows to study the interaction of complex floating structures under the wind and wave conditions developed in the far-field domain.

The validity and performance of two proposed far-field/near-field coupling algorithms based on the surface forcing method of Guo and Shen \cite{guo_generation_2009} were systematically verified with a number of wave cases of increasing complexity, including broadband wave fields. The computed wave fields in the near-field domain were seen to agree very well with either the theoretical solutions or the corresponding computed results from the far-field simulation. 

To demonstrate the potential of the framework we applied it to simulate a $13.2MW$ offshore floating wind turbine under realistic site-specific ocean wind and wave conditions. The method was able to capture the turbine response in the 6 \gls{dof}, considering the platform-wave interactions, the effect of the turbulent wind on the turbine, and the gyroscopic effect of the rotor.   

As a future work, we will further validate the present near-field solver using a set of experiments carried out at the  St.\ Anthony Falls Laboratory of the University of Minnesota consisting of a model floating platform interacting with different wave cases. Another aspect to consider as a future work is the validation of the framework with measurements of an operational floating wind turbine, which was not possible in this work due to the lack of data.
\section*{Acknowledgements}
This work has been supported by the US Department of Energy (DE-EE 0005482), the US National Science Foundation (CBET-1341062 and CBET-1622314), the Office of Naval Research (N00244-14-2-008), and the University of Minnesota Initiative for Renewable Energy and the Environment. The computational resources were provided by the Minnesota Supercomputing Institute and Sandia National Laboratories.
%
\printglossary[type=\acronymtype]
%
%
\bibliographystyle{unsrt}
\bibliography{references}
\end{document}